\documentclass[twocolumn, twocolappendix]{aastex631}

\begin{document}

\title{A Deep uGMRT view of the ultra steep spectrum radio halo in Abell 521}

\author[0009-0002-0373-570X]{R. Santra}
\affiliation{National Centre for Radio Astrophysics, Tata Institute of Fundamental Research, Pune 411007, India}

\author[0000-0003-1449-3718]{R. Kale}
\affiliation{National Centre for Radio Astrophysics, Tata Institute of Fundamental Research, Pune 411007, India}

\author[0000-0002-1634-9886]{S. Giacintucci}
\affiliation{Naval Research Laboratory, 4555 Overlook Avenue SW, Code 7213, Washington, DC 20375, USA}

\author[0000-0003-0144-4052]{M. Markevitch}
\affiliation{NASA/Goddard Space Flight Center, Greenbelt, MD 20771, USA}

\author[0000-0001-5573-726X]{F. De Luca}
\affiliation{Dipartimento di Fisica, Università di Roma Tor Vergata, Via della Ricerca Scientifica 1, I-00133 Roma, Italy}
\affiliation{INFN, Sezione di Roma 2, Università di Roma Tor Vergata, Via della Ricerca Scientifica, 1, Roma, Italy}

\author{H. Bourdin}
\affiliation{Dipartimento di Fisica, Università di Roma Tor Vergata, Via della Ricerca Scientifica 1, I-00133 Roma, Italy}
\affiliation{INFN, Sezione di Roma 2, Università di Roma Tor Vergata, Via della Ricerca Scientifica, 1, Roma, Italy}

\author[0000-0002-8476-6307]{T. Venturi}
\affiliation{INAF - IRA, Via Gobetti 101, I-40129 Bologna, Italy; IRA - INAF, via P. Gobetti 101, I-40129 Bologna, Italy}

\author[0000-0003-1246-6492]{D. Dallacasa}
\affiliation{Dipartimento di Fisica e Astronomia, Università di Bologna, via P. Gobetti 93/2, 40129, Bologna, Italy}

\author[0000-0003-4046-0637]{R. Cassano}
\affiliation{INAF - IRA, Via Gobetti 101, I-40129 Bologna, Italy; IRA - INAF, via P. Gobetti 101, I-40129 Bologna, Italy}

\author[0000-0003-4195-8613]{G. Brunetti}
\affiliation{INAF - IRA, Via Gobetti 101, I-40129 Bologna, Italy; IRA - INAF, via P. Gobetti 101, I-40129 Bologna, Italy}

\author[0000-0003-3445-4521]{K.D.Buch}
\affiliation{Digital Backend Group, Giant Metrewave Radio Telescope, NCRA-TIFR, Pune, 410504, India}




\begin{abstract}

We present the first detailed analysis of the ultra-steep spectrum radio halo in the merging galaxy cluster Abell 521, based on upgraded Giant Metrewave Radio telescope (uGMRT) observations. The combination of radio observations (300-850 MHz) and archival X-ray data provide a new window into the complex physics occurring in this system. When compared to all previous analyses, our sensitive radio images detected the centrally located radio halo emission to a greater extent of $\sim$ 1.3 Mpc. A faint extension of the southeastern radio relic has been discovered. We detected another relic, recently discovered by MeerKAT, and coincident with a possible shock front in the X-rays, at the northwest position of the center. We find that the integrated spectrum of the radio halo is well-fitted with a spectral index of $-1.86 \pm 0.12$. A spatially resolved spectral index map revealed the spectral index fluctuations, as well as an outward radial steepening of the average spectral index. The radio and X-ray surface brightness are well correlated for the entire and different sub-parts of the halo, with sub-linear correlation slopes (0.50$-$0.65). We also found a mild anti-correlation between the spectral index and X-ray surface brightness. Newly detected extensions of the SE relic and the counter relic are consistent with the merger in the plane of the sky.
\end{abstract}




\section{Introduction} \label{sec:1}

Galaxy clusters accrete mass via mergers of galaxy groups and other clusters. These processes result in turbulent motions in the Intra-cluster medium (ICM), the largest baryonic pool in the universe. Consequently, a notable amount of energy is released during a merger, accelerates the cosmic ray (CR) population, and amplifies the ICM magnetic field. The observational footprints of these events are seen via diffuse radio emission in galaxy clusters \citep[for reviews;][]{2014IJMPD..2330007B, 2019SSRv..215...16V}. These radio sources have a steep spectra\footnote{S$_{\nu}$ $\propto$  $\nu ^{\alpha}$, where S$_{\nu}$ is the flux density at the frequency at $\nu$ and $\alpha$ stands for spectral index} ($\alpha$ \textless -1). The origin of these synchrotron radio-emitting electrons, discovered to fill the entire cluster volume is still poorly understood \citep{2014IJMPD..2330007B, 2022arXiv221101493B}.

Diffuse radio sources in clusters are mainly classified into two classes: radio halos and radio relics \citep{2012A&ARv..20...54F}. Radio halos are the centrally located $\sim$ Mpc scale diffuse radio sources and are unpolarized \citep[][]{2009A&A...507.1257G, 2010PhDT.......342B}. Radio relics are the diffuse sources located at the cluster peripheries. They are thought to be the tracers of the outgoing merger shock waves \citep[e.g.][]{2019SSRv..215...16V}. Radio relics show filamentary structures on kpc scales, a high degree of linear polarization, and aligned magnetic field distribution \citep[e.g.][]{2010Sci...330..347V, 2017A&A...600A..18K,2021ApJ...911....3D,2021A&A...646A.135R}. Advancement of both the theoretical models and observations that resulted in the discovery of many sources \citep[e. g.][]{2022A&A...657A..56K, 2022A&A...660A..78B} provide the tools to explore the possible physical mechanism behind the origin of radio halos and relics.

The currently favored scenario for the generation of the radio halo involves the re-acceleration of the pre-existing relativistic electrons, due to turbulence, originating in cluster mergers \citep[re-acceleration models][]{2001MNRAS.320..365B,petrosian2001nonthermal,2006AN....327..557C, 2007MNRAS.378..245B, 2013ApJ...771..131B, 2016MNRAS.458.2584B}. Observational signatures of the radio halos match quite well with the predictions of the turbulent re-acceleration model \citep{2021A&A...647A..51C,2023A&A...672A..43C}. Contribution from secondary electrons from hadronic collisions \citep[e.g][]{1999NuPhS..70..495B} have been constrained to be subdominant from Fermi-LAT observations and by arguments based on the energetics of CRs \citep[e.g][]{2012MNRAS.426...40B, 2017AIPC.1792b0009B, 2021A&A...648A..60A}, yet secondary electrons may contribute significantly if they are re-accelerated by turbulence during mergers \citep[e.g.][]{2011MNRAS.410..127B, 2022ApJ...934..182N}.

It is widely accepted that the fraction of kinetic energy from the shock fuels the radio emission of the relic by the Diffusive Shock Acceleration \citep[DSA;][]{1998A&A...332..395E,2007MNRAS.375...77H} of the cosmic ray electrons. The ongoing debate on the acceleration of the electrons is, whether they are accelerated from a thermal pool (keV) to relativistic energy (GeV) \citep[standard scenario;][]{1998A&A...332..395E,2007MNRAS.375...77H} or from a pre-existing relativistic electron population \citep[re-acceleration;][]{2005ApJ...627..733M,2011ApJ...734...18K,2016ApJ...823...13K}. 
There are a few examples that show some connection between the relic emission to an Active Galactic Nucleus (AGN) and radio galaxy \citep[e.g.][]{2014ApJ...785....1B,2015MNRAS.449.1486S,2017NatAs...1E...5V,2018ApJ...865...24D,2019MNRAS.489.3905S}. The standard scenarios have explained many observable from the relics. But still, some disputes remain regarding the acceleration of the CRe from the thermal pool via weak shock \citep{2020A&A...634A..64B}.

Radio halos have been observed to have a broad range of spectral indices, varying from -2.1 to -1.0 \citep[see][for reviews]{2012A&ARv..20...54F, 2019SSRv..215...16V}. In particular, there are radio halos, detected over the years, that have shown an integrated spectral index of \textless -1.5. This population is known as Ultra-Steep Spectrum Radio Halos (USSRH), which are thought to be produced during less energetic mergers (involving clusters with masses \textless 10$^{15}$M$_{\odot}$), is the key expectation of the re-acceleration model \citep[e.g.][]{2006AN....327..557C, 2010ApJ...721L..82C}. The prime example is Abell 521 \citep{2008Natur.455..944B, 2009ApJ...699.1288D, 2013A&A...551A.141M}, the cluster with first discovered a USSRH. The number of steep spectrum radio halos has increased thanks to low-frequency observations from LOFAR, MWA, GMRT \citep[e. g.][]{2018MNRAS.473.3536W,2021A&A...646A.135R, 2021PASA...38...31D,2021NatAs...5..268D,2021A&A...650A..44B,2022arXiv221101493B}.

The integrated radio spectrum for the radio halos generally follows a single power law, although there are a few examples of halos with curved spectrum (Coma \citealt{2003A&A...397...53T}, MACSJ017.5+3745 \citealt{2021A&A...646A.135R}). A radio spectral index is a key tool in the understanding of the shape of the energy distribution of the relativistic electrons  and properties of the turbulence in ICM. Only for very few clusters, high-resolution resolved spectral maps, and robust estimations of the spectral index, are available. Some halos show a uniform spectral index all over the spatial scales, whereas some show small-scale fluctuations in the resolved spectral maps \citep{2017ApJ...845...81P, 2020ApJ...897...93B,2022arXiv220903288R}. Resolved spectral maps for the steep spectrum radio halos are not explored significantly, which can throw light on the self-similarity behavior (if any) of the global spectral index at local kpc scales. 

Here, we present the deep upgraded Giant Metrewave Radio telescope (uGMRT) radio observations of the prototype USSRH in Abell 521. For a better understanding of the diffuse emission, we have also used archival \textit{\textit{Chandra}} and XMM-\textit{Newton} data sets. For the \textit{Chandra} analysis, we use the results presented by \citealt{2019PhDT........98W}. The \textit{Chandra} data include four archival \textit{\textit{Chandra}} datasets (ObsIDs 430, 901, 12880, 13190) with a total exposure of 146 ks on the cluster (after cleaning for background flares; see \citealt{2019PhDT........98W} for details \footnote{\url{http://hdl.handle.net/1903/22163}}). The XMM-\textit{Newton} observations were published by \cite{2013ApJ...764...82B}. 

The paper is organized as follows. In Sec.~\ref{sec:2} we provide a brief overview of Abell 521. The uGMRT observations and data analysis procedures are explained in Sec.~\ref{sec:3}. In Sec.~\ref{sec:4}, we show the uGMRT continuum images at different resolutions and discuss newly discovered features from our study. The results obtained from spectral analysis and radio vs X-ray correlations are described in Sec.~\ref{sec:5}$-$\ref{sec:7}. Then we summarised our results and findings in Sec.~\ref{sec:8}. Throughout this paper, we have adopted a flat $\Lambda$CDM cosmology with H$_{0}$ = 70 km s$^{-1}$, $\Omega$ $_{m}$ = 0.3 , $\Omega_{\Lambda} =0.7$. At the redshift of Abell 521, 1$''$ corresponds to a linear scale of 4 kpc.

\section{The galaxy cluster Abell 521} \label{sec:2}

The galaxy cluster Abell 521 (hereafter A521) is a very massive and merging cluster, located at a redshift of 0.247 \citep[e.g.][]{2000A&A...355..461A}, having a center at RA = 04:54:6.9, DEC = -10:13:26.2 \citep{2020ApJ...903..151Y}. The mass (M$_{200}$) of the cluster is 1.3$_{-1.3}^{+1.0}$ $\times$ 10$^{15}$ M$_{\odot}$ \citep{2020ApJ...903..151Y}, with a very high X-ray luminosity of L$_{\rm X [0.1 - 2.4 \rm keV]}$ $\sim$ 8 $\times$ 10$^{44}$ erg s$^{-1}$, and a temperature of 5.9 keV \citep{2013ApJ...764...82B}.

Optical and X-ray observations of this cluster have shown high dynamical activity at the center. \textit{Chandra} X-ray observations show that this cluster constitutes two main clumps along the north-west and south-east direction \citep{2006A&A...446..417F}. The southern gas clump is more diffuse compared to the northern one. Using the XMM-\textit{Newton} observations, \citet{2013ApJ...764...82B} have detected two shock fronts (one along the south-east and the other one along the south-west direction) and two cold fronts at the intersection region of these two gas clumps. Temperature maps from the XMM-\textit{Newton} observations suggest that the cluster has a temperature of $\sim$ 4.5 keV in the northern part, whereas in the southern it is $\sim$ 4 keV. The interacting region between these two sub-clumps has a temperature of \textgreater 7 keV.

Optical spectroscopic observations of 113 member galaxies of A521 have shown a complex galaxy distribution \citep{2003A&A...399..813F}. There are 7 different galaxy groups situated in the NW-SE region. Additionally, the velocity analysis suggests that this cluster is very complex along the line of sight. Recent HST weak lensing studies by \citet{2020ApJ...903..151Y} have shown that the cluster is composed of mainly three distinct substructures named North-West (NW), South-East (SE), and Central (C). Using the available multi-wavelength observations, they revisited the merging scenario of A521 via numerical simulations. However, their study did not get a conclusive picture of the merging scenario in the central region.

\subsection{Previous radio studies}\label{sec:2.1}

 A521 is widely observed in radio wavelengths. \citet{2008Natur.455..944B} reported the discovery of a radio halo with the legacy GMRT (240, 325, 610 MHz). They estimated an integrated spectral index of the halo of $\sim$ -1.9, which is very steep compared to the typical observed spectral index of a radio halo. \citet{2009ApJ...699.1288D} have reported the detection of the radio halo at 1.4 GHz, using the VLA (D-array and BnC array) and obtained a very steep spectral index of $\sim$ -1.86 $\pm$ 0.08. \citet{2013A&A...551A.141M} had also reported the spectral index of the halo to be very steep $\sim$ -1.81 $\pm$ 0.02. 
 
\citet{2000A&A...355..461A} had first presented the image of the radio relic using the archival National Radio Astronomy Observatory Very Large Array Sky Survey (NVSS) \citep{1998AJ....115.1693C}. Since then, the relic has been observed at radio frequencies ranging from 150 to 4800 MHz with legacy GMRT and VLA. The radio relic is detected up to 8.4 GHz, with  a good significance \citep{2008A&A...486..347G}. The measured integrated spectral index of the relic over a frequency range of 325 to 4800 MHz is -1.48 $\pm$ 0.01. \textit{Chandra} observations of A521 at 0.5$-$4 keV also suggested the presence of an X-ray surface brightness edge, a shock front at that position of the relic \citep{2008A&A...486..347G}. The shock front was later confirmed by \citet{2013ApJ...764...82B}.

\section{Observations and Data analysis} \label{sec:3}

\begin{table}
  \centering
  \caption{Summary of uGMRT observations.}
  \begin{tabular}{@{}lcc@{}}
    \hline
      &  Band 3    & Band 4 \\
    \hline
    Frequency range (MHz) & 300-500 & 550-750  \\ 
 
 No. of channels & 2048  & 2048 \\

 Channel width (kHz)& 97.65 & 97.65   \\
 
 Bandwidth (MHz) & 200 &  200 \\
 
 On source time (Hr) & 9 + 9  & 9 + 9   \\
 
 Polarisation & RR, LL & RR, LL  \\

 Flux calibrator & 3C147 & 3C147 \\ 
 
 Largest angular scale & 1920$''$  & 1020$''$  \\
   \hline
  \end{tabular}
  \label{table1}
    
    \tablecomments{The observations reported in the table are recorded in two sessions: one using the uGMRT real-time RFI-Filter and the other session without the filter. }
\end{table}

\begin{table*}
  \centering
  \caption{Summary of the radio images.}
  \begin{tabular}{@{}ccccccc@{}}
    \hline\hline
    & Name & Beam Size ($''$) & Robust & \textit{uv} range & \textit{uv}taper ($''$) & Map RMS ($\mu$Jybeam$^{-1}$)  \\
    \hline
    uGMRT band3 & IMG1 & 8$''$ $\times$ 7$''$  & 0.5 & None & None& 24.9   \\ 
 
 & IMG2  & 6$''$ $\times$ 6$''$  & -0.5  & \textgreater 0.2k$\lambda$  & 5$''$ & 28.7   \\
 
 & IMG3 & 10$''$ $\times$ 10$''$  & -0.5 & \textgreater0.2k$\lambda$& 10$''$ & 31.7\\

 & IMG4 & 15$''$ $\times$ 15$''$ & 0 & \textgreater 0.2k$\lambda$& 15$''$ & 34.0  \\
 
  & IMG5 & 20$''$ $\times$ 20$''$  & -0.5 & \textgreater0.2 k$\lambda$& 20$''$ & 38.6\\

 \hline

 uGMRT band 4  & IMG6 & 5$''$ $\times$ 5$''$ & 0.5& None& None& 8.5  \\ 
   
  & IMG7 & 6$''$ $\times$ 6$''$ & 0.0 & \textgreater 0.2k$\lambda$ &5$''$ & 11.0   \\
 
 & IMG8 & 10$''$ $\times$ 10$''$ & -0.5 & \textgreater 0.2k$\lambda$ &10$''$ &16.7  \\

 & IMG9 & 15$''$ $\times$ 15$''$ & 0 & \textgreater 0.2 k$\lambda$ &  15$''$&21.6 \\
 
 & IMG10 & 20$''$ $\times$ 20$''$ &0  & \textgreater 0.2 k$\lambda$ &20$''$ &25.6  \\
  \hline
  \end{tabular}
  \label{table2}
\end{table*}

\subsection{uGMRT} \label{sec:3.1}

A521 was observed with the uGMRT at bands 3 and 4 using the GMRT Wideband Backend (GWB) \citep{2017JAI.....641011R}. The observations in each band were taken in two sessions, placed on consecutive days. We used real-time RFI filtering \citep[e.g.][]{2019JAI.....840006B,2022JAI....1150008B,2023JApA...44...37B} in GWB, targeted for mitigating broadband RFI. We used it in the mode where the RFI samples (those that exceeded the 3$\sigma$ threshold in voltages) were replaced with digital noise. Band 3 observations were done on July 6 (with filter) and July 7 (without a filter) and band 4 observations were done on July 8 (with filter) and July 9 (without a filter) in 2018. In Table~\ref{table1} we have summarized the observational parameters. Further details on the effect of the RFI filtering are provided in Appendix~\ref{app:A}.  

We processed the uGMRT data with \texttt{CASA} using the  \texttt{CAPTURE\footnote{\url{https://github.com/ruta-k/CAPTURE-CASA6}}} pipeline \citep{2021ExA....51...95K}. After this initial flagging, the flux density of the primary calibrator was set according to the flux scale of Perley - Butler 2017 \citep[][]{2017ApJS..230....7P}. After following the standard calibration routines, the calibrated target source data were split and were further flagged using the automated flagging modes. To reduce the volume of the data but still prevent bandwidth smearing, 10 frequency channels ($\sim$ 1 MHz) were averaged. The procedure was identical for bands 3 and 4. The target visibilities were imaged using the CASA task \texttt{tclean} with wide-field and wide-band imaging algorithms. At the stage of imaging and self-calibration, the target source measurement set was divided into 8 frequency sub-bands, each sub-band containing 20 channels. It divides the whole 200 MHz bandwidth into 8 spectral windows and does imaging and self-calibration on the sub-banded data file. 

We assume the flux density uncertainty is $10\%$ for both the uGMRT frequencies \citep{2017ApJ...846..111C}. The flux density uncertainty ($\Delta f$) is defined by:
\begin{equation}
    \Delta f = \sqrt{(0.1f)^{2} + (N_{\rm beam}(\sigma_{\rm rms})^{2}}),
    \label{eq:1}
\end{equation}
where $f$ is the flux density, $N_{\rm beam}$ is the number of beams, and $\sigma_{\rm rms}$ is the rms noise. We have used the task \texttt{ugmrtpb}\footnote{\url{https://github.com/ruta-k/uGMRTprimarybeam-CASA6}} for the primary beam correction of all the images.

\begin{figure*}
    \centering
    \includegraphics[width=0.90\textwidth]{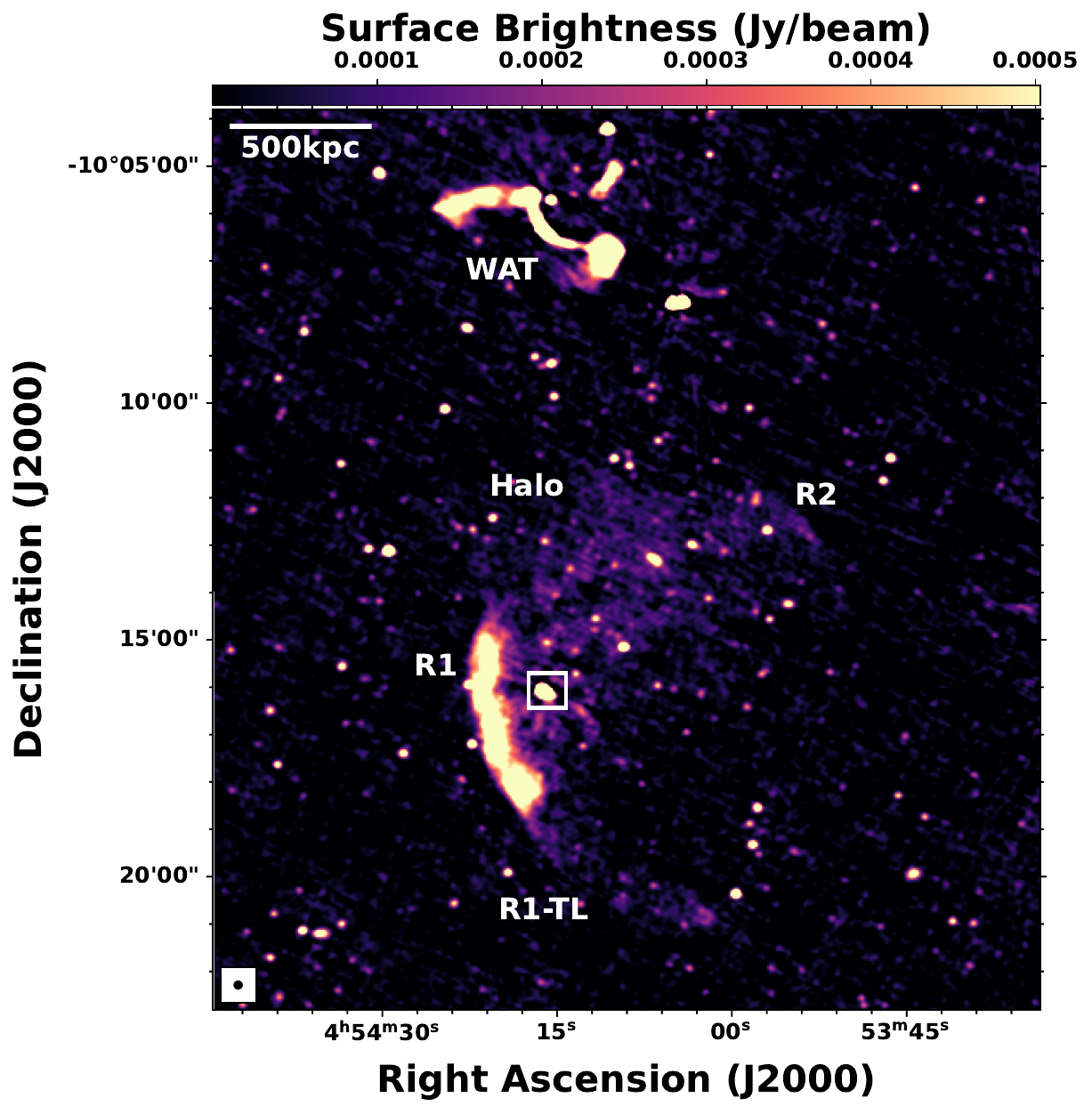}
    \caption{uGMRT band 3 full resolution image IMG1 obtained with robust$=0.5$ is shown here. The rms is $\sigma_{\rm rms}$ $=$ 24.9 $\mu$Jybeam$^{-1}$ and the beam is 8$''$ $\times$ 7$''$ . We have labeled the extended sources; R1 the relic situated at the South-east position of the cluster, R2 the newly discovered relic, ``R1-TL'' the faint extension of the R1 relic, and the Wide Angle-Tailed galaxy (WAT). At the central region, the low surface brightness radio halo emission is seen. The boxed region indicates the radio galaxy J0454-1016a.}
    \label{img:1}
\end{figure*}

\begin{figure*}
    \centering
    \includegraphics[width=13cm, height=12.5cm]{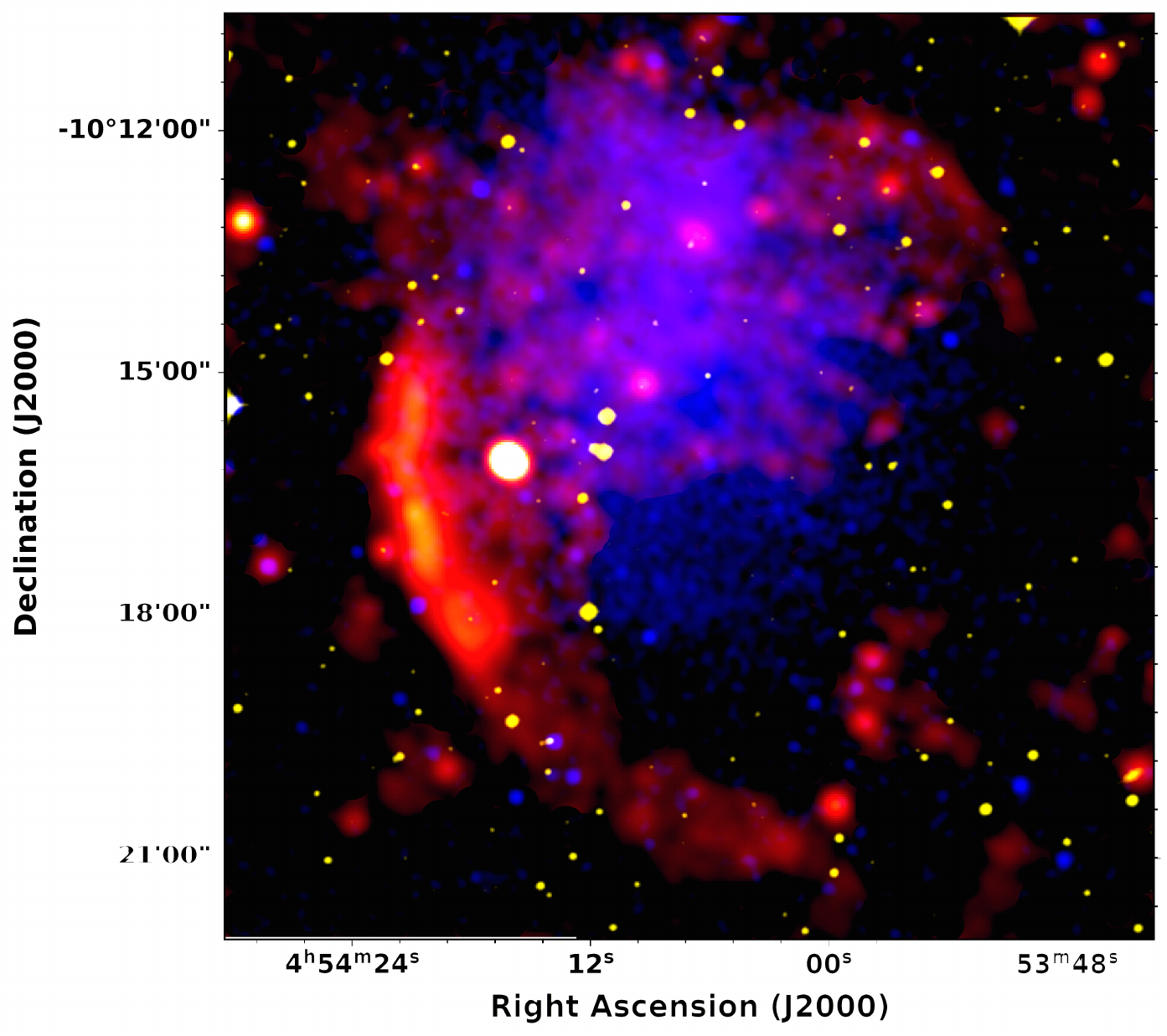}
    \caption{Multi-wavelength view for the central region of the A521 is shown. The yellow intensity shows the image of the A521 field in the DSS2 red. The intensity in blue shows the \textit{Chandra} image at 0.5$-$4 keV from \citet{2019PhDT........98W}, which is smoothed up to 5$''$ using a Gaussian kernel. The radio emission from the uGMRT band 3 image is shown in red.}
    \label{img:2}
\end{figure*}

\subsection{Removal of the discrete sources}\label{sec:3.2}

The diffuse radio emission in A521 is contaminated by weak/faint discrete sources like unresolved point sources and radio galaxies in the images which bring complications in the measurement of the flux densities of the diffuse sources. For the removal of the discrete sources at band 3 and band 4, their models were created by applying a \textit{uv}-cut of \textgreater 5 k$\lambda$ (angular scale of 40$''$). In both frequencies, we have kept the \textit{uv} cut-off range the same to match the spatial scales. In Appendix~\ref{app:b}, we have presented the high-resolution point source images for the A521 cluster field in both uGMRT bands 3 and 4 frequencies.

After creating the model image for the point sources, the clean components corresponding to the ``point'' sources were subtracted from the observed data.  After subtraction, the data file was then imaged again using \texttt{tclean} with a \textit{uv} baseline of \textless 10 k$\lambda$, to highlight the extended emission. We used the \texttt{multi-scale} deconvolver to properly map the large-scale diffuse emission. We have checked the consistency of the point source subtraction process, by calculating the flux density of the radio halo with that obtained by algebraically subtracting the flux density of the embedded sources from the total (halo $+$ sources), finding a good agreement (5\% for uGMRT band 3 and band 4). During the measurement of the errors regarding the flux density estimation of the extended sources, we used the equation~\ref{eq:1}, with one extra term for the point source subtraction error ($\sigma_{\rm sub}$), 5\% for the uGMRT, added in quadrature.

\subsection{X-ray observations}

Using the XMM-\textit{Newton} pointing ObsID 0603890101, we generated a surface brightness and projected temperature map. XMM-\textit{Newton} data have been pre-processed in the framework of the CHEX-MATE project \citep{2021A&A...650A.104C}, as detailed in De Luca et al., (in prep). This pre-processing includes, in particular, spatial and spectral modeling of the sky and instrumental backgrounds, together with the detection and masking of point sources. The XMM-\textit{Newton} surface brightness map is derived from wavelet analysis of the photon image that we adapted for spatial variations in the effective area and a background noise model. The photon image has been denoised via the 4-$\sigma$ soft-thresholding of a variance-stabilized wavelet transform \citep{2009A&A...504..641S, 4531116}, that is especially suited to process low photon counts. The XMM-\textit{Newton} temperature map has been computed using a spectral-imaging algorithm that combines spatially weighted likelihood estimates of the projected intra-cluster temperature with a curvelet analysis, that is especially suited to recover curvilinear features expected near shocks or cold fronts. This algorithm can be seen as an adaptation of the spectral-imaging algorithm presented in \cite{2015ApJ...815...92B} and recently applied also in \citet{2022arXiv220909601O}. Briefly, temperature log-likelihoods are first computed from spectral analyses in each pixel of the maps, then spatially weighted with kernels that correspond to the negative and positive parts of B3-spline wavelets functions. This allows us to derive wavelet coefficients of the temperature features and their expected fluctuation, which we derive from spatially weighted Fisher information. We use these wavelet coefficients to compute a curvelet transform that typically analyses features of apparent size in the range of [3.5, 60] arcsec. We finally reconstruct the temperature map from a de-noised curvelet transform that we inferred from a 4-sigma soft thresholding of the curvelet coefficients. For each pixel of the temperature map, we associate a temperature uncertainty that results from a parametric bootstrap re-sampling. Specifically, combining the best-fit temperature and surface brightness maps with the effective area and background noise models, we generate mock XMM-\textit{Newton} event lists whose statistical properties mimic the real pointing and derive bootstrap temperature maps from curvelet analyses of these mock event lists. The temperature uncertainty map derives from the standard deviations of temperature values registered in each pixel of 100 bootstrap maps. It can be noticed that temperature uncertainties increase at the same time as the best-fit temperature values increase and the X-ray surface brightness decreases.

\begin{figure*}
    \includegraphics[width=10cm, height = 10cm]{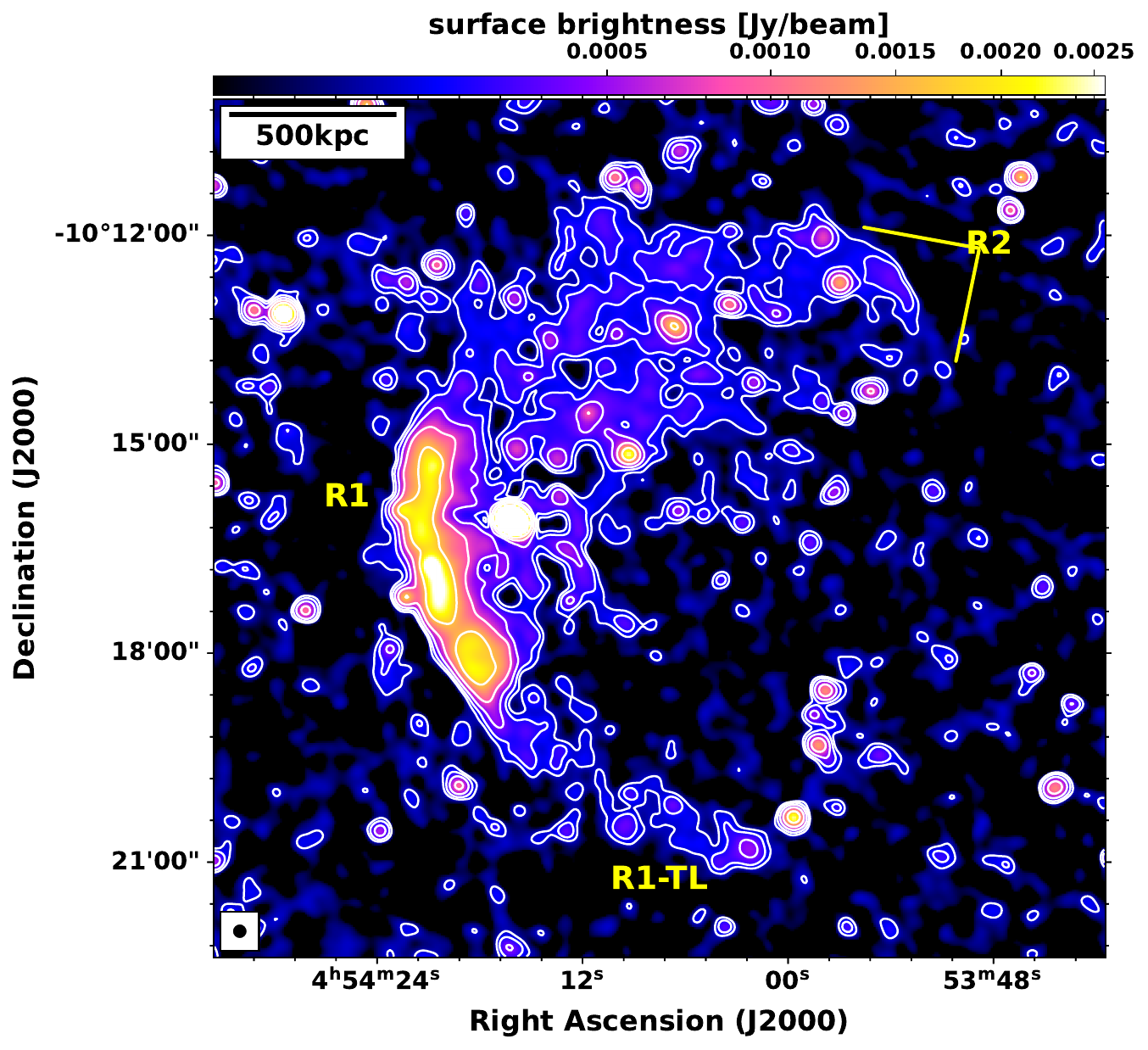}
    \includegraphics[width=8.2cm, height=10cm]{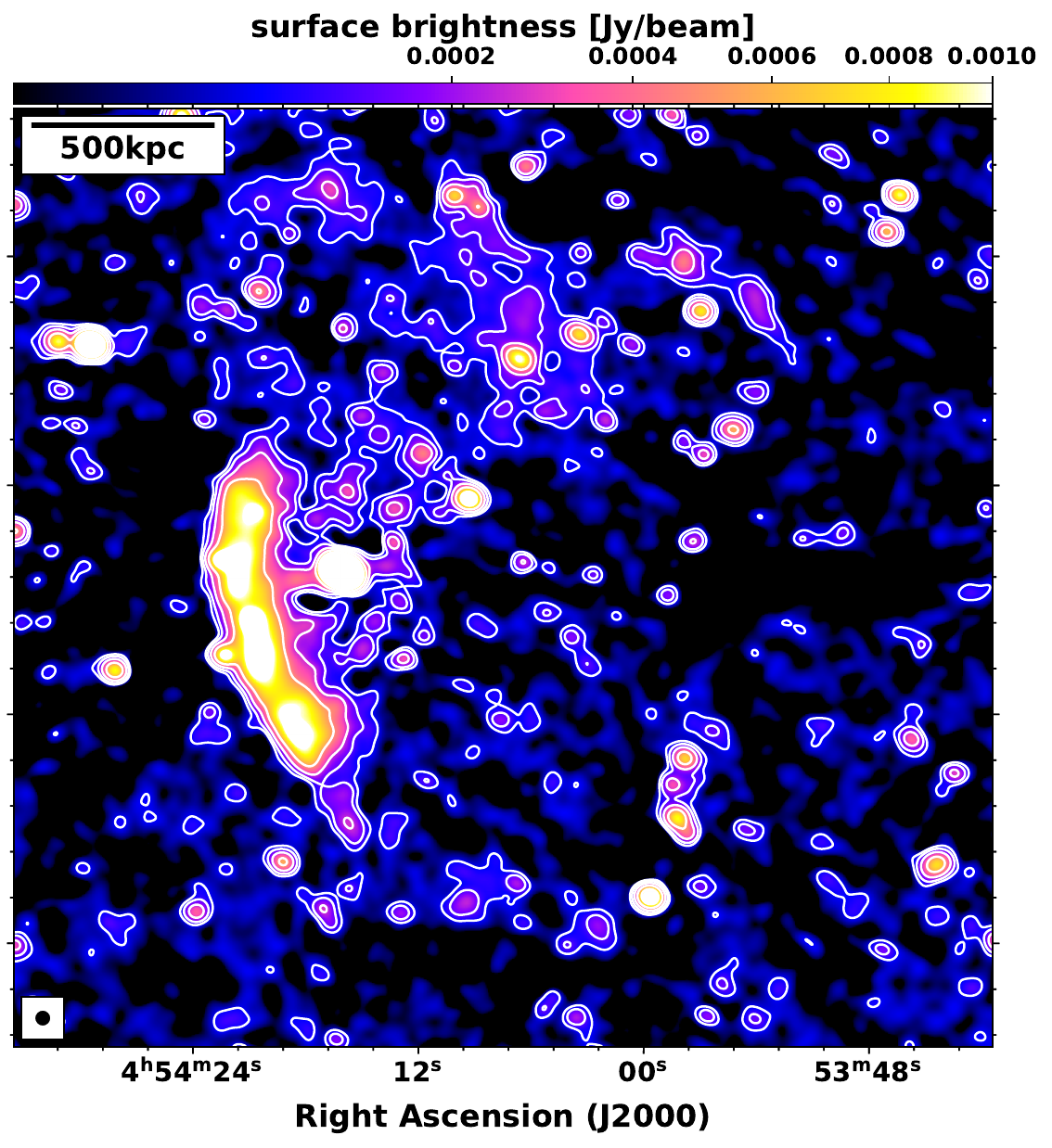}
    \includegraphics[width=10cm, height =10cm]{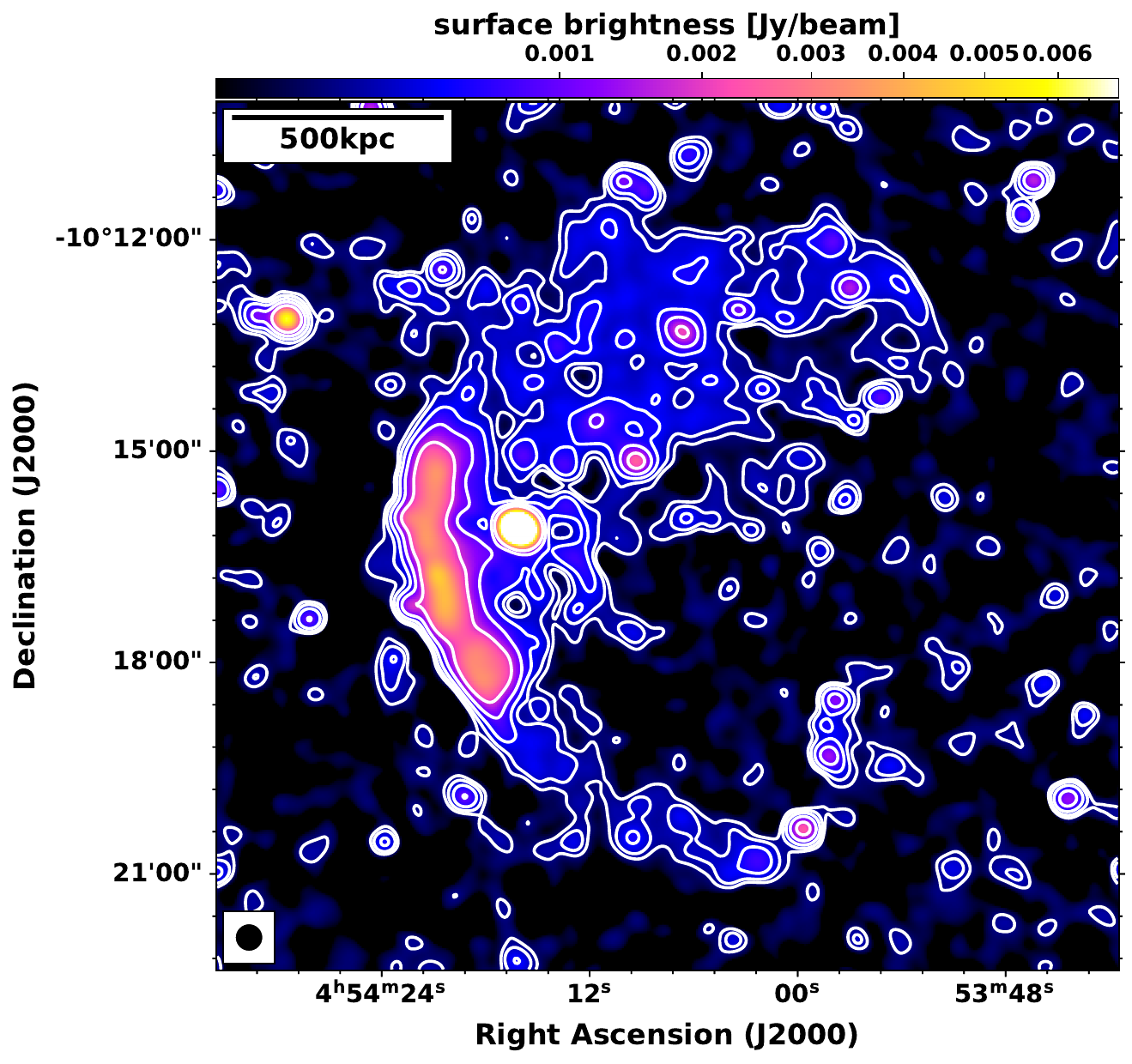}
    \includegraphics[width=8.2cm, height=10cm]{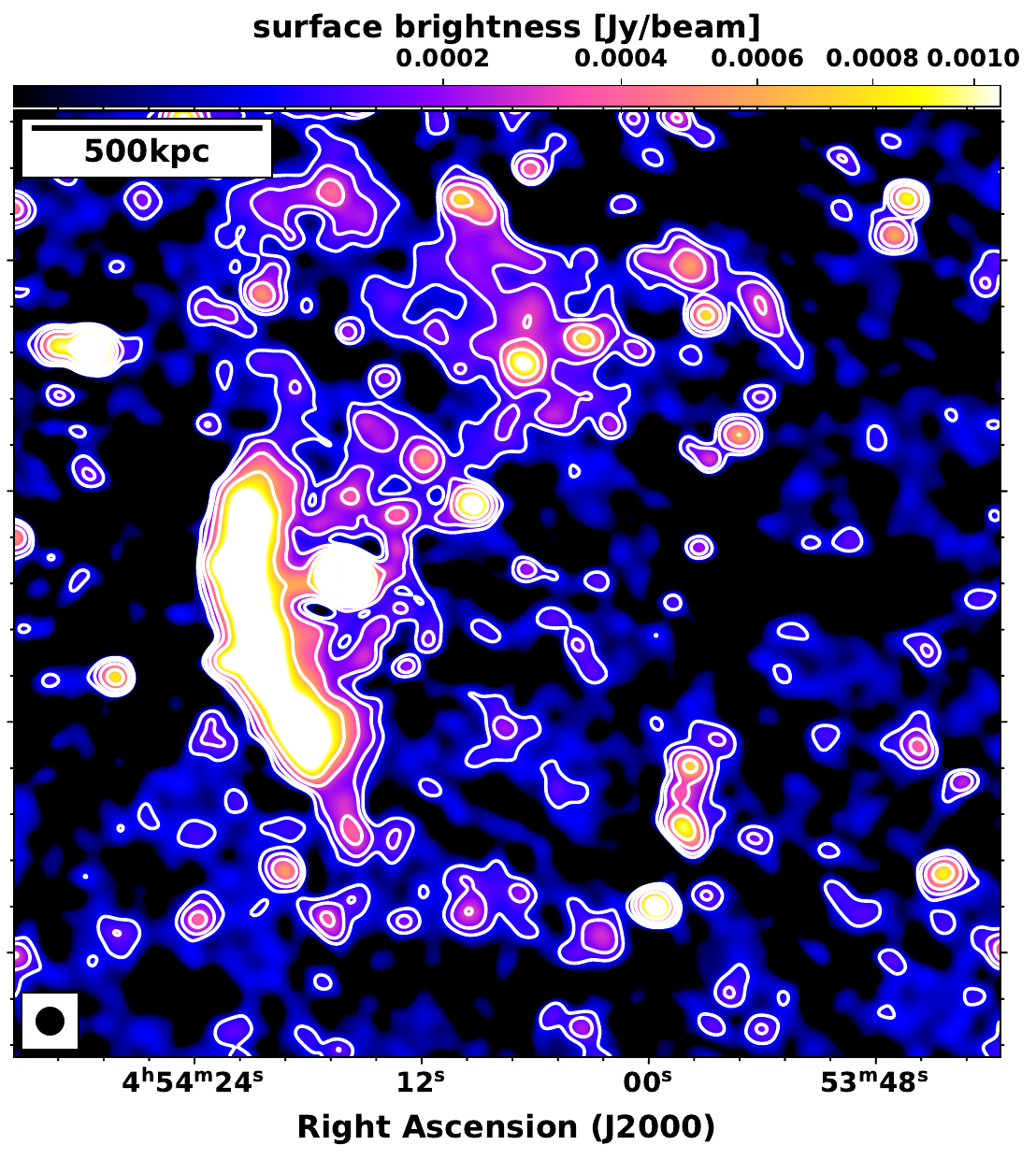}
     
    \caption{ \textit{Top left:} Total power uGMRT band 3 image (IMG4), with a beam size of 15$''$, is shown. The previously discovered diffuse emissions are also well recovered in our uGMRT images. The tail of the relic R1 (extended faint emission) is labelled in yellow with a naming convention of ``R1-TL''. Similarly, the newly discovered relic, named R2 (and better observable with this resolution), labelled in yellow. Contours level are drawn at [1,2,4,8...]$\times$ 3$\sigma_{\rm rms}$, where $\sigma_{\rm rms}$ = 34.0 $\mu$Jybeam$^{-1}$. \textit{Top right:}Similar to the precedent panel but at 650MHz. The spacing of the contour levels is similar with $\sigma_{\rm rms}$ = 21.6 $\mu$Jybeam$^{-1}$. \textit{Bottom left:} The cluster central diffuse sources are shown in a low resolution (20$''$) image (IMG5) at 400 MHz. The interested sources are mentioned above. The RMS noise of the image is 38.6 $\mu$Jybeam$^{-1}$ and the contour spacing is similar to the above panel. \textit{Bottom right:} Here we have shown the 20$''$ image (IMG10) of the cluster field at 650 MHz. The RMS noise of the image is 25.6 $\mu$Jybeam$^{-1}$.} 
    \label{img:3a}
    \label{img:3b}
    \label{img:3c}
    \label{img:3d}
\end{figure*}

\section{Results: continuum images}  \label{sec:4}

In Figure~\ref{img:1} we have shown the uGMRT band 3 continuum image of the A521 cluster field. At the central region, we have detected the arc-shaped relic (R1),  southeast of the cluster center. The central regions are dominated by the point sources and the low surface brightness radio halo emission. Figure~\ref{img:2} shows a multi-wavelength view of the cluster. The central radio halo emission is morphologically similar to the \textit{Chandra} X-ray thermal emission (blue). Many point sources are seen in the central regions and all over the field at the optical wavelength (yellow). In the following sections, we will discuss the previously known and newly discovered diffuse radio emissions at the cluster central regions.

\subsection{Radio halo emission}\label{sec:4.1}

We have detected the radio halo emission both at 400 and 650 MHz, shown at different resolutions in Figure~\ref{img:3a}. The radio halo is not circular in shape rather it is very elongated. The elongation is along the southeast to the northwest, which was supposed to be the merger axis \citep{2003A&A...399..813F}. In our sensitive high-resolution images, we have detected 33 discrete sources (\textgreater 3.5 $\sigma_{\rm rms}$) in the halo region. The total flux density of the 32 sources (excluding J0454-1016a, squared boxed region in Figure~\ref{img:1}) is 22.5 $\pm$ 1.7, which is $\sim$ 33 \% of the total flux density (radio halo $+$ individual sources)at 400 MHz. In Figure~\ref{img:4a} we have shown the point source subtracted image at 400 MHz (left) and 650 MHz (right). Radio emissions from both the radio halo and relic were recovered well. Similar to \cite{2013A&A...551A.141M} we have also detected the ``spur'' region connected to the radio relic at the southeast position in our low-resolution images. This is still unexplained whether the ``spur'' region is a part of the radio relic or the centrally located radio halo. At 400 MHz, the radio halo has been detected up to the Largest Linear Size (LLS) of 1.3 $\times$ 1.0 Mpc$^{2}$, whereas at 650 MHz it is 1.03 $\times$ 0.85 Mpc$^{2}$. \citet{2013A&A...551A.141M} had reported the size of the radio halo to be $\sim$ 1.35 Mpc at 153 MHz. The radio halo is more extended at lower frequencies. The radio halo brightness shows a rapid decline in surface brightness at the outermost regions. The overall radio surface brightness is very smooth over the spatial scale of the radio halo at this resolution.

The radio relic emission penetrates the southern part of the radio halo emission, which we define as the "bridge" region. This component was also detected significantly in previous studies from 150 to 1400 MHz. It is extremely difficult to distinguish the contribution of radio halo and radio relic emissions at a given observed frequency. This bridge region's properties will be discussed in Section~\ref{sec:6}. Many examples of overlapping emissions from relics, bridges, and halos are discussed in the literature, including the radio halo in toothbrush clusters \citep[e.g.][]{2016ApJ...818..204V}, halo in CIZA J2242.8+5301 \citep{2017MNRAS.471.1107H} and radio halo in Abell 3667 \citep{2022A&A...659A.146D}.

\subsection{New discoveries}\label{sec:4.2}

 Our deep and sensitive uGMRT observations have revealed some new insights into the A521 cluster central field. In Figure~\ref{img:3a} we have presented the central region of this cluster, where we have labeled the different extended sources in yellow color. The R1 is the brightest relic which was discovered to the southeast of the cluster center and studied in the literature \citep{2000A&A...355..461A,2006A&A...446..417F,2006NewA...11..437G,2008A&A...486..347G}. ``R1-TL'' is the low surface brightness extension of the relic, which is uncovered at 400 MHz at a significant level. Similarly, our new deep radio images revealed another relic ``R2'' northwest of the cluster center, first reported by \citet{2022A&A...657A..56K} using the MeerKAT.

\begin{table}
  \centering
  \caption{Flux density estimates for the radio halo.}
\begin{tabular}{@{}ccc@{}}
    \hline
      Frequency (MHz) & Flux density (mJy) & Ref. \\
      \hline\hline
    74 & 1500 &  B.08  \\

    153 & 328$\pm$ 66 & M.13    \\

    240 & 152$\pm$15 & B.08  \\

    330 & 90$\pm$ 7&  B.08  \\
    402 & 45$\pm$5& This work \\
    
    610 & 15$\pm$ 3.5 & B.08 \\
    
    650 & 15.31 $\pm$ 3&  This work\\
    
    1400& 6.4$\pm$ 0.6&  D. 09 \\
    
    \hline
\end{tabular}         
      \tablecomments{The references are B.08: \citet{2008Natur.455..944B}, D.09: \citet{2009ApJ...699.1288D}, M.13: \citet{2013A&A...551A.141M}. In Figure~\ref{img:7a} we have annotated the region in the plot in the blue circle, which is used to estimate the flux density.}
  \label{table3}
\end{table}

\begin{figure*}[t]
    \includegraphics[width=9.5cm,height=10cm]{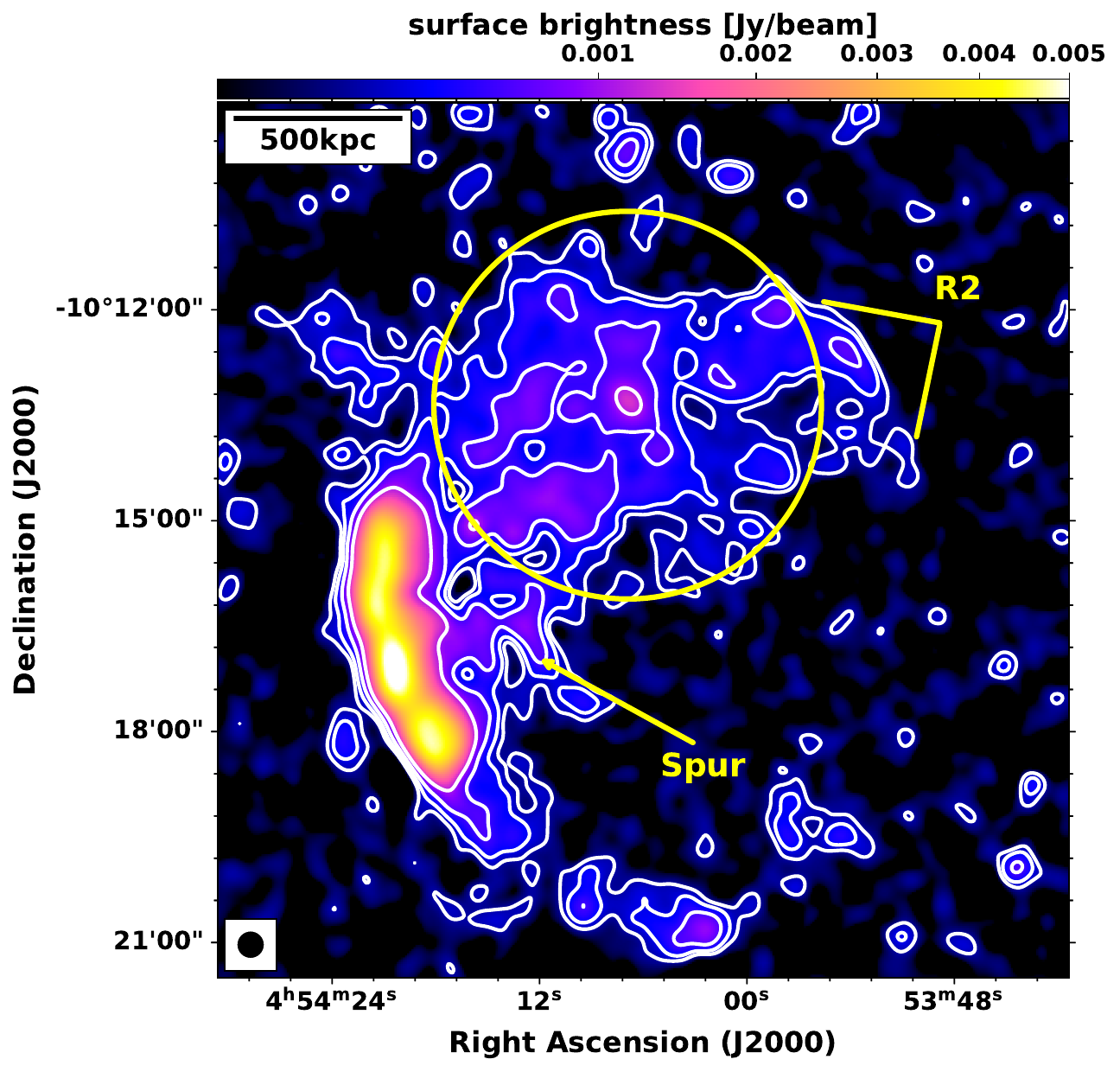}
    \includegraphics[width=8.3cm,height=10cm]{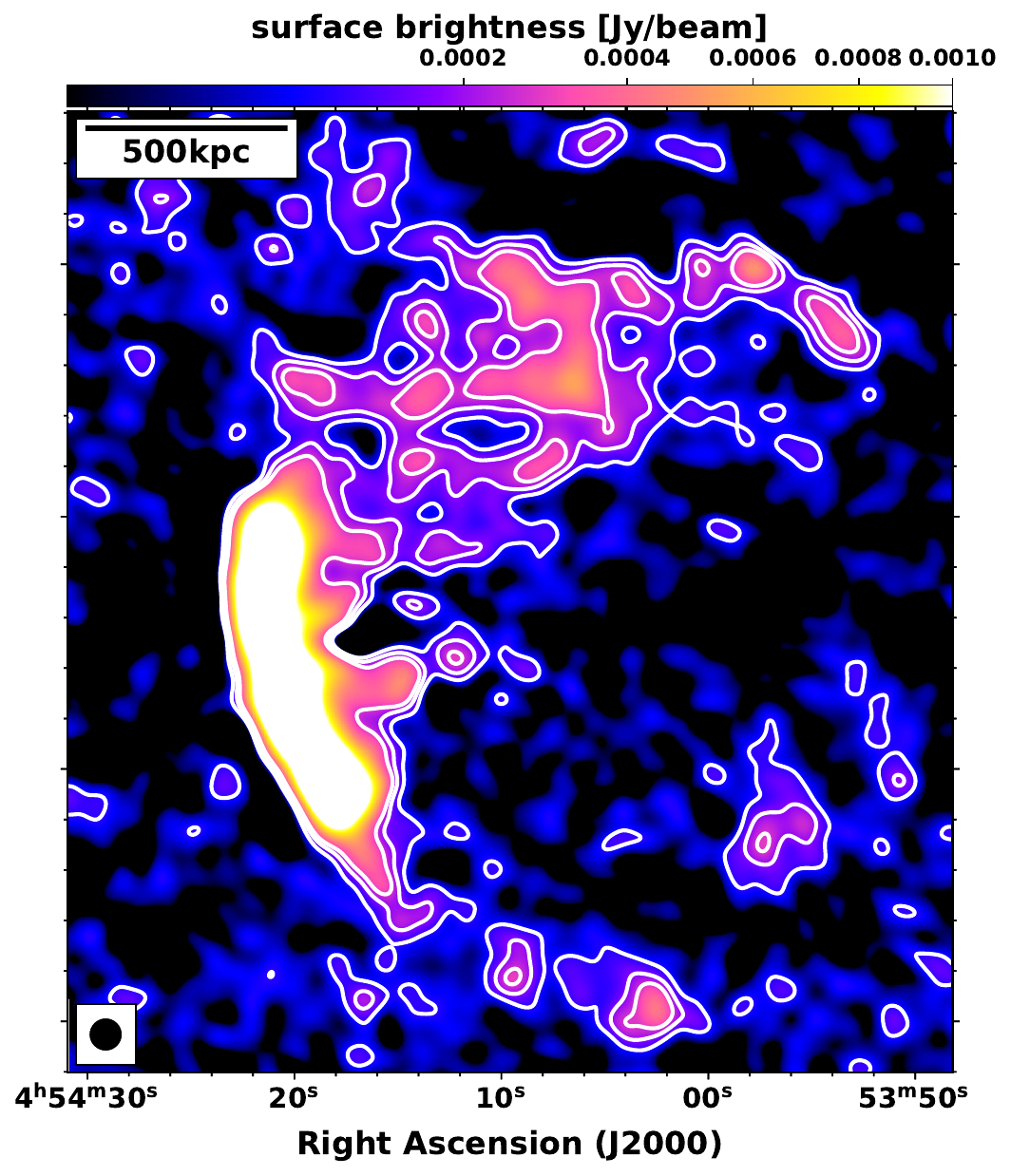}
    
    \caption{\textit{Left:} We have shown the point source subtracted image of the cluster central field at 400 MHz. The resolution of the image is 22$''$ $\times$ 22$''$. The yellow circle, having a size of $\sim$ 1 Mpc represents the regions of the radio halo emission, adopted from \citet{2008Natur.455..944B}. The spur emission, a bridge region between the radio halo and relic is labeled in yellow. Contours level are drawn at [1,2,4,8...]$\times$ 3$\sigma_{\rm rms}$, where $\sigma_{\rm rms}$ = 45 $\mu$Jybeam$^{-1}$ is the noise level. \textit{Right:} The point source subtracted image is shown at 650 MHz. The spur emission is also visible here. The RMS noise of the image is 30 $\mu$Jybeam$^{-1}$ and the spacing of the contour level is similar to the left image.}   
    \label{img:4a}
    \label{img:4b}
\end{figure*}

\begin{figure*}
    \includegraphics[width=9.5cm, height = 10cm]{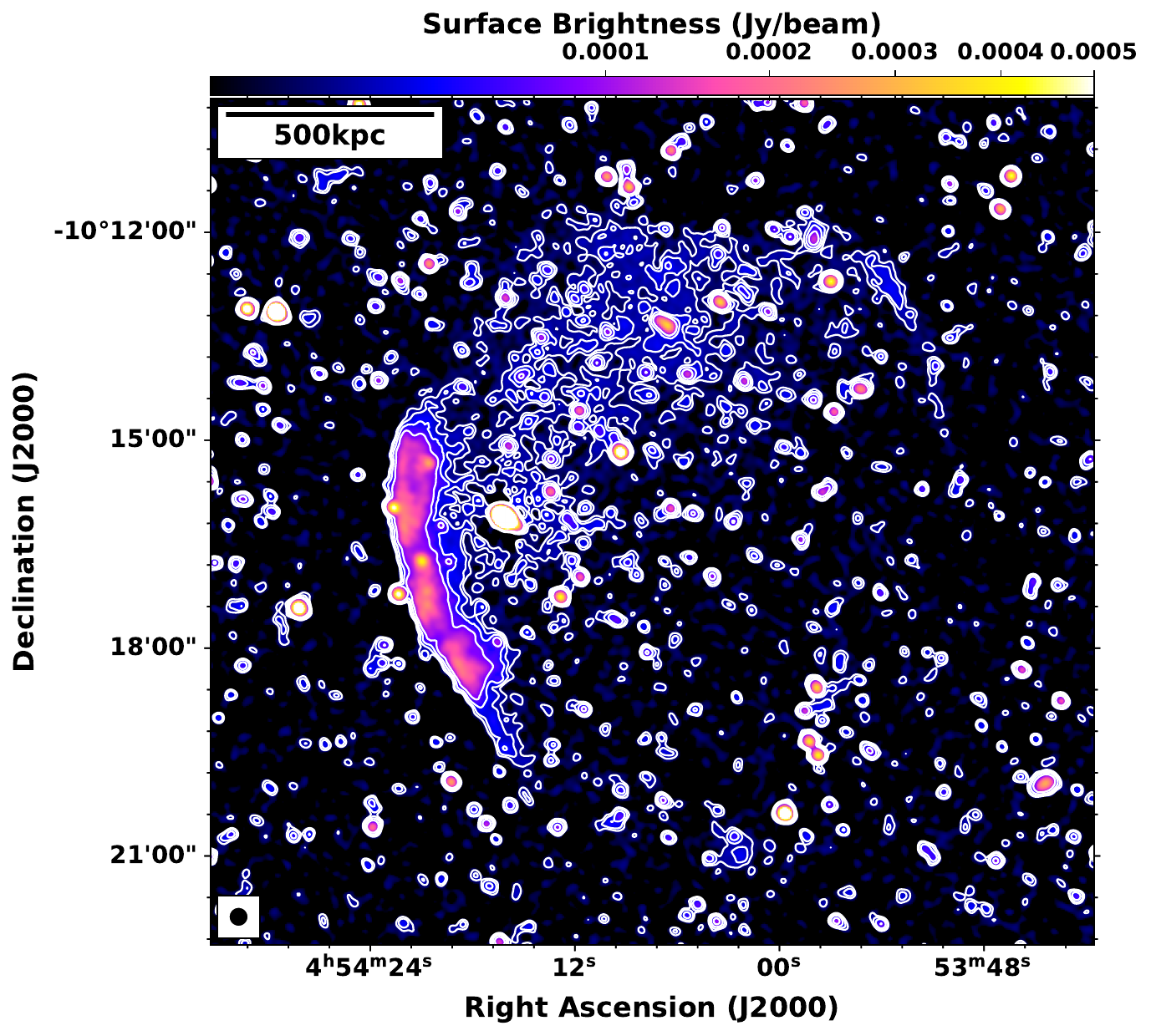}
    \includegraphics[width=8.5cm, height=10cm]{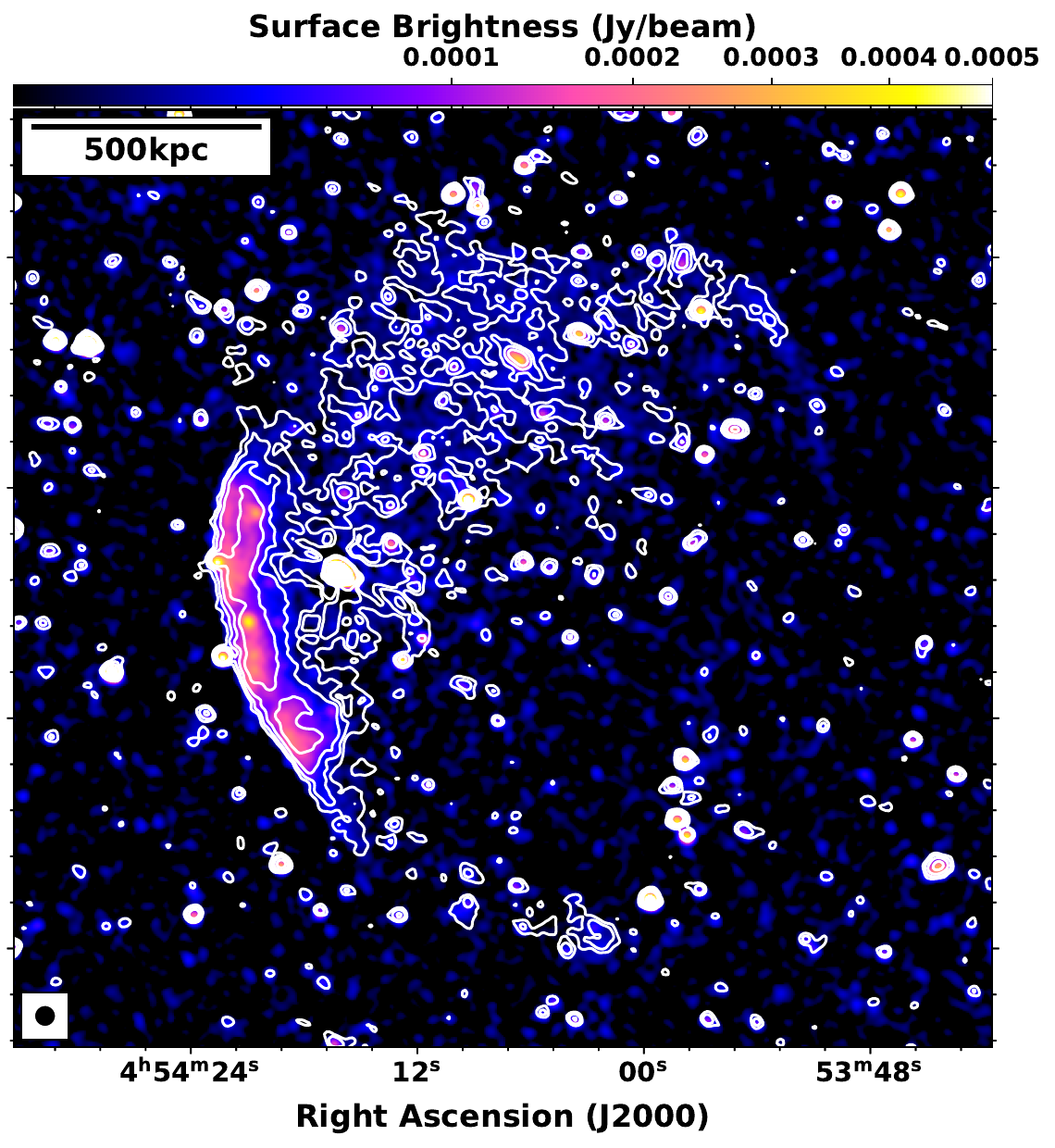}
    \caption{\textit{Left:} MeerKAT 1.28 GHz image of the A521 cluster is shown in colour and contours. The MeerKAT image has a beam size of 8.5$''$ $\times$ 7.5$''$. Contours levels are drawn at [1,2,4,8]$\times$ 3$\sigma_{\rm rms}$, where $\sigma_{\rm rms} = 3$ $\mu$Jybeam$^{-1}$ is the noise level, quoted from \citet{2022A&A...657A..56K}. \textit{Right:} uGMRT 400 MHz emission (full resolution, IMG1) is overlaid (white contour) on the MeerKAT image (color). Contour levels are at a similar level, with $\sigma_{\rm rms} = 24.9$ $\mu$Jybeam$^{-1}$. } 
    \label{img:5a}
    \label{img:5b}
\end{figure*}

\begin{figure}
	\includegraphics[width=\columnwidth]{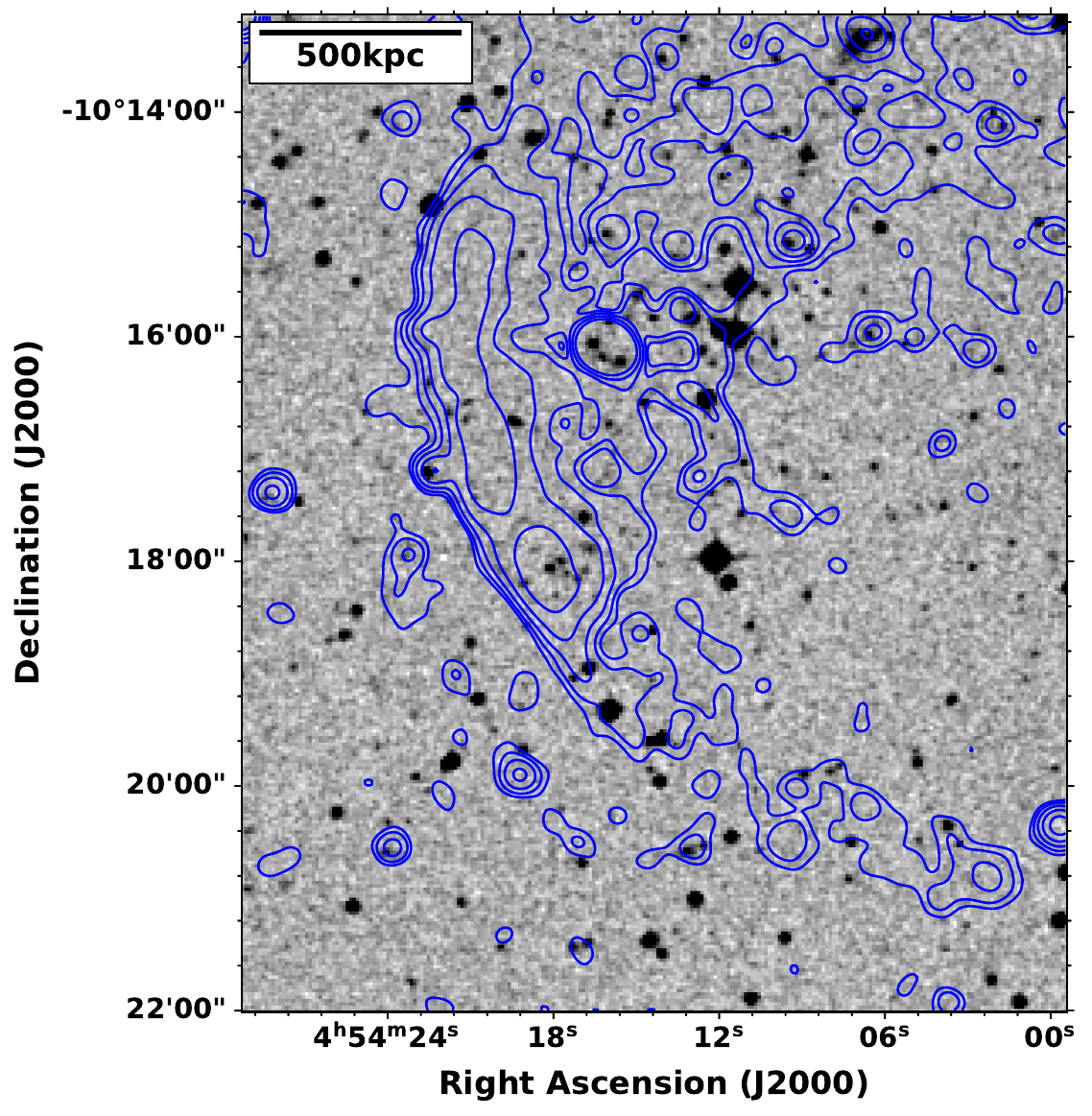}
        \caption{DSS2 image cutout  of the A521 SE relic field is shown (in grey), overlaid with the radio emission at 400 MHz, with a beam size of 10$''$ (IMG3). Contours level are drawn at [1,2,4,8...]$\times$ 3$\sigma_{\rm rms}$, where $\sigma_{\rm rms} = 31.7$ $\mu$Jybeam$^{-1}$ is the noise level. In the relic region, the point sources embedded in the radio emission are also seen here, similar to the \citep{2006NewA...11..437G}.}
    \label{img:6}
\end{figure}

\citet{2008Natur.455..944B} had also detected a very small portion of this ``R1-TL'' in the  240 MHz legacy GMRT image of the A521 cluster.
This low surface brightness extended region is visible in low-resolution images (15$''$, 25$''$) at 400 MHz (in Figure~\ref{img:3a}). At 650 MHz we have also detected the extension. In this region, some of the point sources are seen both in the optical and radio images. Therefore we have tried to see if this emission is seen after the point source subtraction. At 400 MHz, we have detected this ``R1-TL'' at a significant level of 12$\sigma_{\rm rms}$, shown in Figure~\ref{img:4a}. In Figure~\ref{img:5a} we have shown the publicly available MeerKAT image (at 1.28 GHz) from MGCLS survey \citep{2022A&A...657A..56K}. The faint extension is also seen in the MeerKAT image. In Figure~\ref{img:6}, the uGMRT 400 MHz image (beam size: 10$''$ $\times$ 10$''$) is overlaid on the DSS2 optical image (grey scale). Several point sources are present in the ``R1-TL'' region. At 400 MHz the linear size of the extension is $\sim$ 1 Mpc, with a width of $\sim$ 300 kpcs. The continuous decrement of the surface brightness along the arc of the relic is still unexplained, especially after a certain length. Considering this ``R1-TL'' extension, the total length of the relic arc is $\sim$ 2.2 Mpc. Till now only very few relics (sausage \citealt{2017MNRAS.471.1107H}, Toothbrush \citealt{2016ApJ...818..204V}, ZwCl 0008.8+5215 \citealt{2011A&A...528A..38V}) have been detected with such a large extent.

Our uGMRT band 3 and band 4 images revealed the presence of another relic at the northwest position of the cluster. In Figure~\ref{img:3a} (top panel), we have labeled the position of the newly discovered relic, R2. In Figure~\ref{img:5a} (right panel) we have overlaid the 400 MHz (in white contours) on the MeerKAT image (in colors), where the relic R2 is seen clearly up a larger extent compared to our analyzed images. Thanks to the superb sensitivity of the image, MeerKAT has been able to detect the R2 relic up to such a low surface brightness extent. In both our 400 and 650 MHz this relic is detected up to the Largest linear size of $\sim$ 650 kpc, and 450 kpc respectively, due to the sensitivity limit of our data quality. In Figure~\ref{img:4a} the R2 relic is seen at a significant level for both bands 3 and 4 in the low-resolution image (beam size: 22$''$ $\times$ 22$''$)

\begin{table}
  \centering
  \caption{Flux density estimates for the two radio relics.}
\begin{tabular}{@{}cccc@{}}
    \hline
     Relic & Freq. (MHz) & Flux density (mJy) & Ref. \\
      \hline\hline
    R1&74 & 660 &  G.08  \\
      & 153 & 299.9 $\pm$ 59 & M.13\\
    &235 & 181.8$\pm$ 10 & G.08    \\

    &327 & 115.14$\pm$6 & G.08  \\

    &402 & 87.8$\pm$10& This work \\
    
    &610 & 42$\pm$ 2 & G.08 \\
    
    &650 & 41.6 $\pm$ 3&  This work\\
    
    &1410& 14.1$\pm$ 1&  G. 08 \\

    &4890 & 2.02 $\pm$ 0.2 & G.08 \\
    \hline

    R2 & 400 & 3.45 $\pm$ 0.34& This work \\
    & 650 & 2.01 $\pm$ 0.2& This work \\

    & 1280 & 0.800$\pm$ 0.09& K.22 \\
    \hline
\end{tabular}
     
     \tablecomments{The flux density reported for the R1 relic for our work, only takes emission contained within 3$\sigma_{\rm rms}$. The references are G.08: \citet{2008A&A...486..347G}, K.22: \citet{2022A&A...657A..56K}. The flux density reported above for our work does not include the ``R1-TL'' during the calculation. Including the ``R1-TL'', the average flux density of the relic is 100.1 $\pm$ 12.0 mJy and 48.3 $\pm$ 7.0 mJy at 400 and 650 MHz, respectively. }
     
  \label{table4}
\end{table}

\begin{figure*}
    \includegraphics[width=9.5cm, height = 8.15cm]{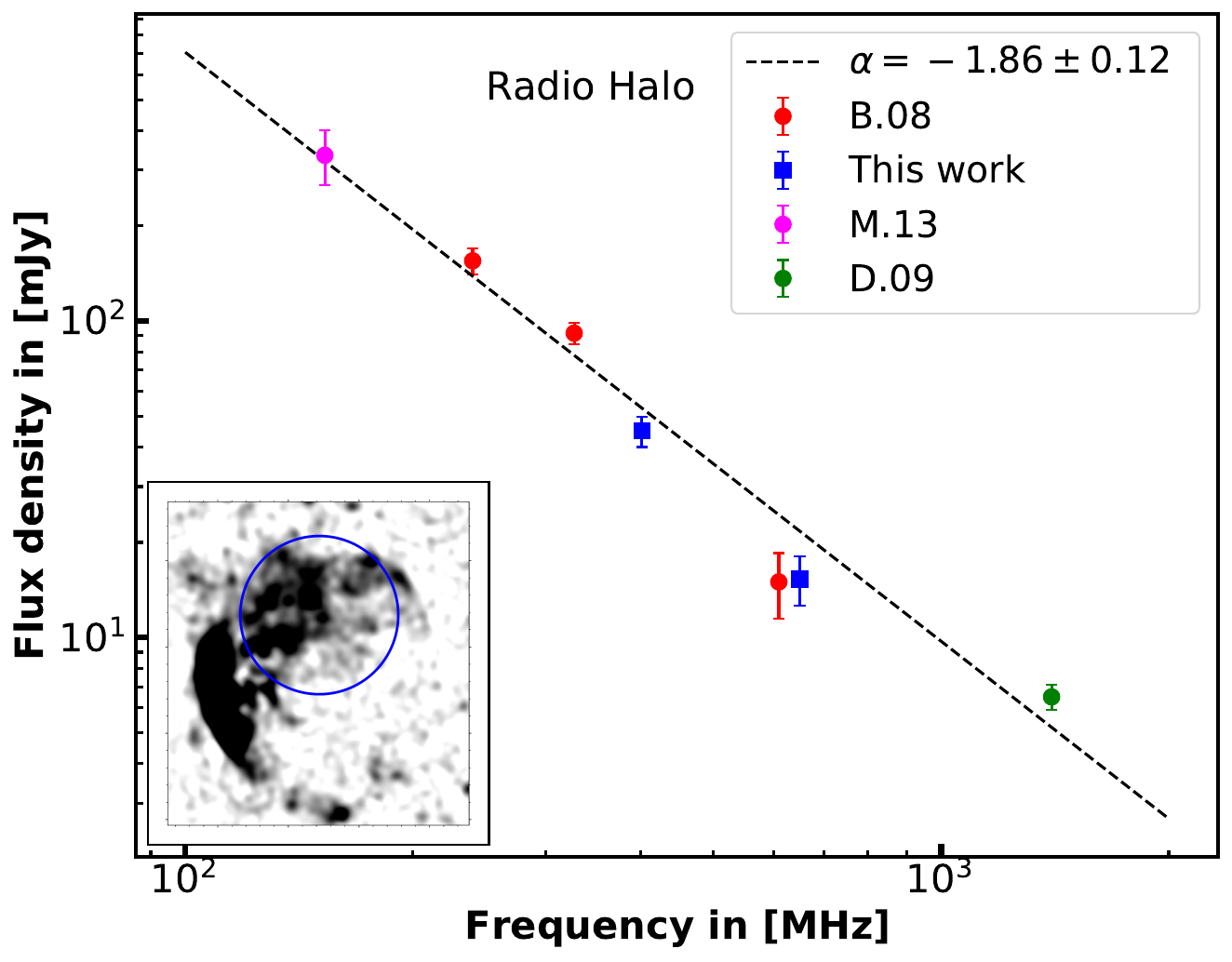}
    \includegraphics[width=8.55cm, height=8.15cm]{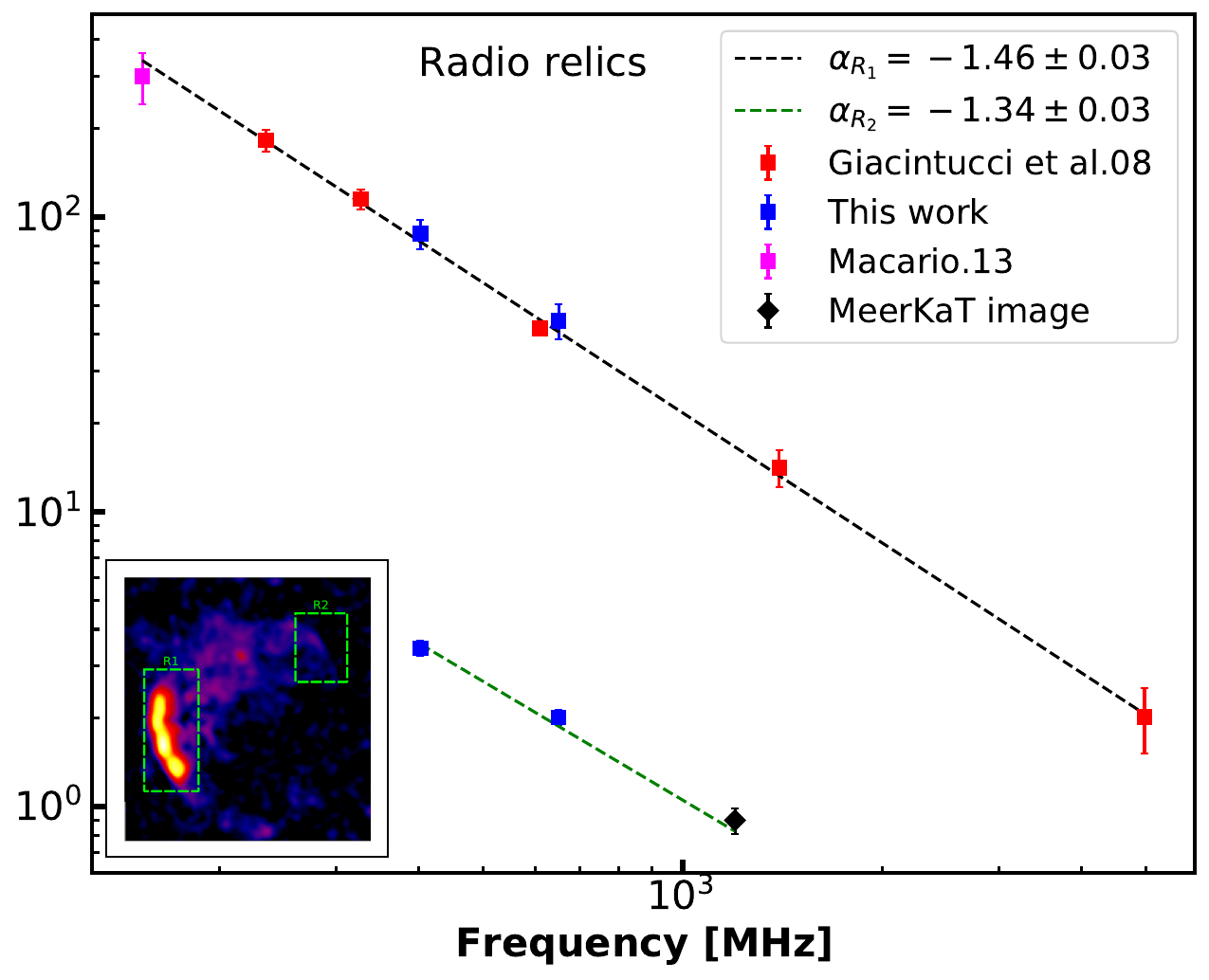}
    \caption{\textit{Left:} Integrated radio spectral index is shown for the radio halo between 150-1400 MHz. The black dashed line shows the overall spectrum of the halo follows a single power law. The blue points denote the results from our work, the other values are taken from the literature. The annotated figure shows the 400 MHz radio emission (in grey). \textit{Right:} The integrated spectrum of both the radio relics, R1 and R2 are shown. The black and green dashed lines represent the overall spectrum of the relics R1 and R2. The annotated figure (color) shows the emission of the two relics (green box) at 400 MHz. }
    \label{img:7a}
    \label{img:7b}
\end{figure*}

\section{Spectral analysis} \label{sec:5}
To provide new insights into the origin of the diffuse radio sources, in this section, we present the integrated spectrum, radial profile, and resolved spectral index map, using our uGMRT observations at bands 3 and 4.

\subsection{Integrated spectrum}\label{sec:5.1}

 Using the point source subtracted image, we have calculated the flux density of the halo and estimated the integrated spectral index, combined with the available literature values. We have tabulated the flux density values in Table~\ref{table3}. The integrated flux density values of the radio halo are calculated over the same circular region, used in \citet{2008Natur.455..944B}. 

In Figure~\ref{img:7a} the resulting integrated spectrum is shown for the radio halo in A521. The overall spectrum of the radio halo is well fitted with the single power law between the frequency range of 150 - 1400 MHz, having an integrated radio spectral index of -1.86 $\pm$ 0.12. Therefore, our results are in line with the previous results, which showed that the radio halo in the A521 is an ultra-steep spectrum radio halo. The value of the spectral index quoted here has not used the 610 MHz flux density value during the estimation of the spectral index due to the uncertainty and poor recovery of the extended radio halo at the cluster central position. The inclusion of the 610 MHz flux density value will lead to a spectral index of -1.92 $\pm$ 0.13, which is also very steep for a radio halo in A521. The flux density value at 74 MHz is reported from \citet{2008Natur.455..944B}, but due to the uncertainty in the measurement of the flux density, we have not accounted for this value during the fitting. We have combined the flux densities from the previous studies \citep[Flux density values are taken from;][]{2008Natur.455..944B,2009ApJ...699.1288D,2013A&A...551A.141M} with ours. So, before doing the estimation, all the flux density values were converted into a common flux density scale, \texttt{Perley - Butler 2017} \citep{2017ApJS..230....7P}. There are only a handful of clusters, where sensitive wide-band radio observations have been done and the radio halos are found to be a USSRH. Some of them (MACS J0717.5+3745; \citealt{2021A&A...646A.135R}) also show high-frequency steepening in the integrated spectrum, while the rest show a single power law spectrum. 

We have also estimated the integrated spectral index of the R1 and R2 relics by combining the previously available flux density values and our analysis. The integrated spectrum of the R1 relic is well fitted with a single power-law of spectral index -1.46 $\pm$ 0.03. For the stationary shock, the integrated spectral index, $\alpha_{\rm int}$, is steeper by 0.5 from the injection spectral index, $\alpha_{\rm inj}$, 
\begin{equation}
    \alpha_{\rm int} = \alpha_{\rm inj} - 0.5
\end{equation}

Assuming the Diffusive Shock Acceleration to be the origin of the particle acceleration in the relic region, the Mach number (M$_{\rm S}$) is related to the injection spectral index $\alpha_{\rm inj}$ as 

\begin{equation}\label{eq.4}
    \rm M_{\rm S} =\sqrt{\frac{2\alpha_{\rm inj} -3}{2\alpha_{\rm inj} +1}} = 2.30\pm0.02 
\end{equation}

 \citet{2008A&A...486..347G} had reported the integrated spectral index of the relic to be -1.48 $\pm$ 0.01, with a Mach number of shock is 2.27 $\pm$ 0.02. Using the equation~\ref{eq.4}, we have estimated the Mach number to be 2.30 $\pm$ 0.02, consistent with the previous work. We have calculated the integrated spectral index value of the R2 relic using the uGMRT (point source subtracted image) and MeerKAT images. The spectrum is well represented by a single power law with a spectral index of -1.34 $\pm$ 0.03 over a frequency range of 400-1280 MHz. Assuming DSA to be the origin of the R2 relic also, the Mach number estimated from the integrated radio spectral index is 2.67$\pm$ 0.09. Although constraining the acceleration efficiency at these shocks is beyond the aim of our paper, we stress that weak shocks with Mach number $\sim$ 2.2$-$2.7 sit exactly in the transition region where direct DSA from thermal pool becomes energetically difficult \citep[e.g.][]{2020A&A...634A..64B}

\begin{figure}
	
	\includegraphics[width=\columnwidth]{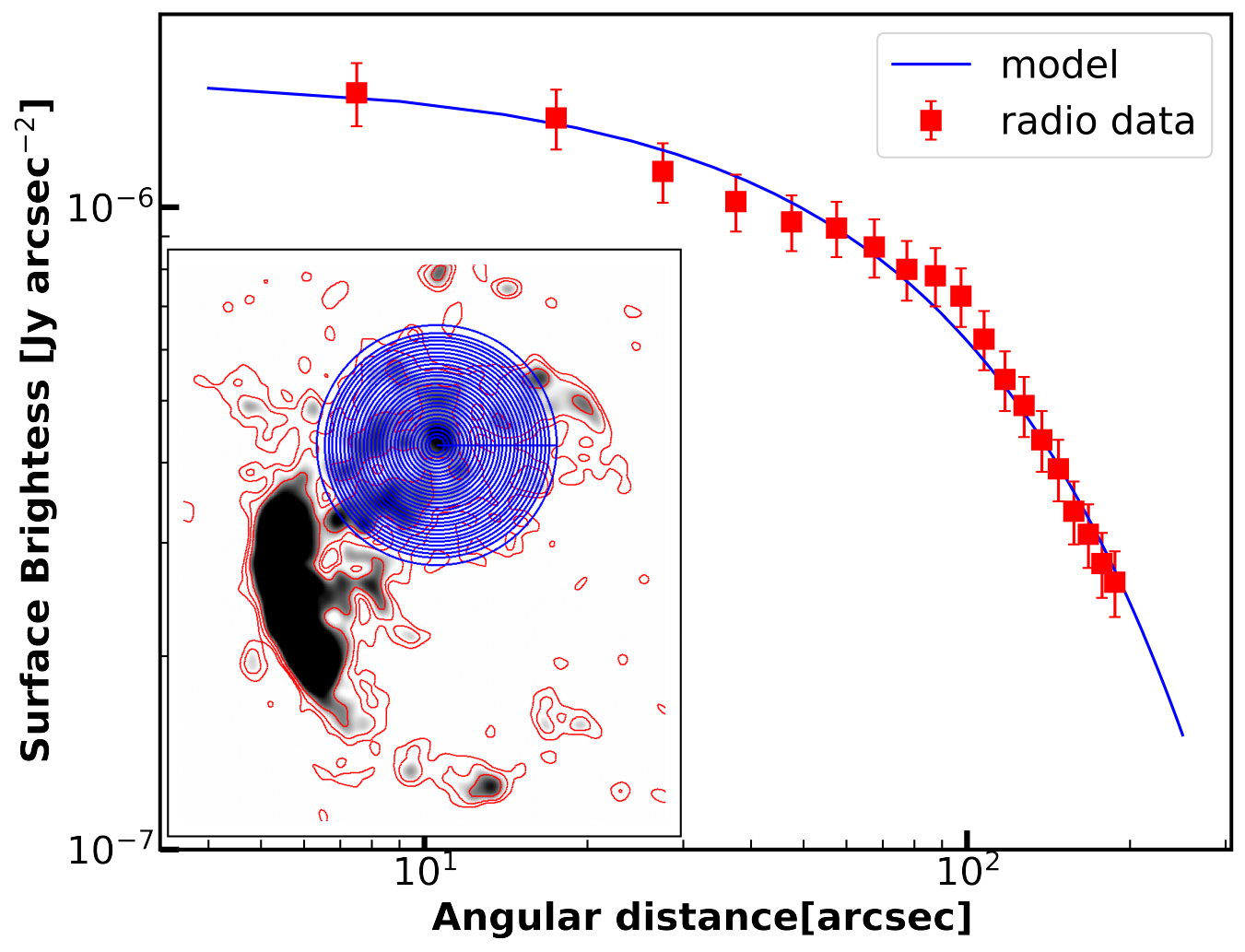}
         \caption{The radial profile of the radio surface brightness is shown. The annotated image shows the 400 MHz radio emission (in grey) and the contours (in red), starting with 3$\sigma_{\rm rms}$ $\times$ [1,2,4,8....]. Each of the red squares denotes the azimuthally averaged radio surface brightness values, within the annular regions. The error bars represent the error on the mean. The blue line shows the two-dimensional elliptical model fitted to the radio surface brightness profile.}
         \label{img:8}
\end{figure}

\begin{figure*}[t!]
    \includegraphics[width=9.5cm, height = 9.5cm]{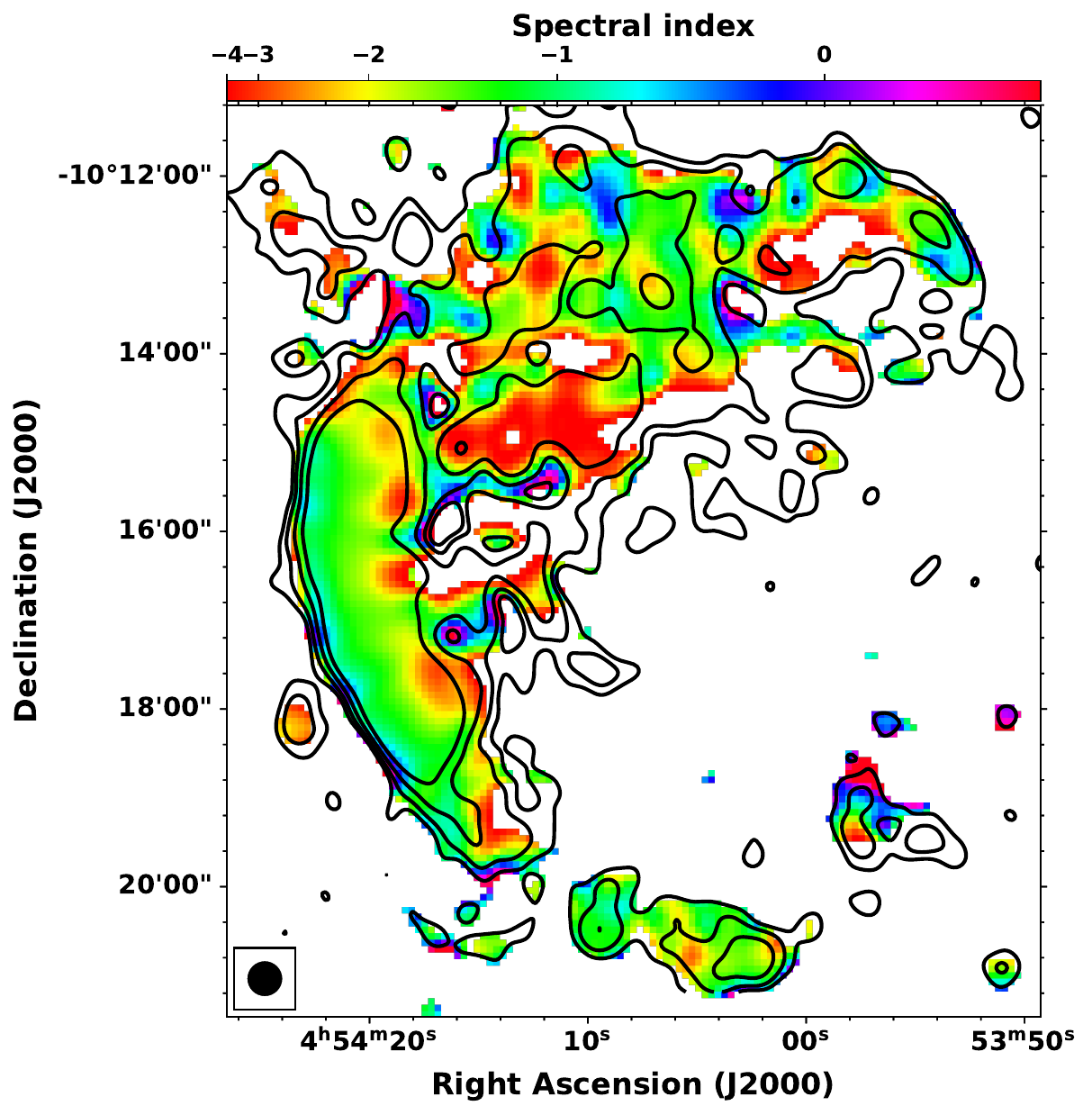}
    \includegraphics[width=8.3cm, height=9.5cm]{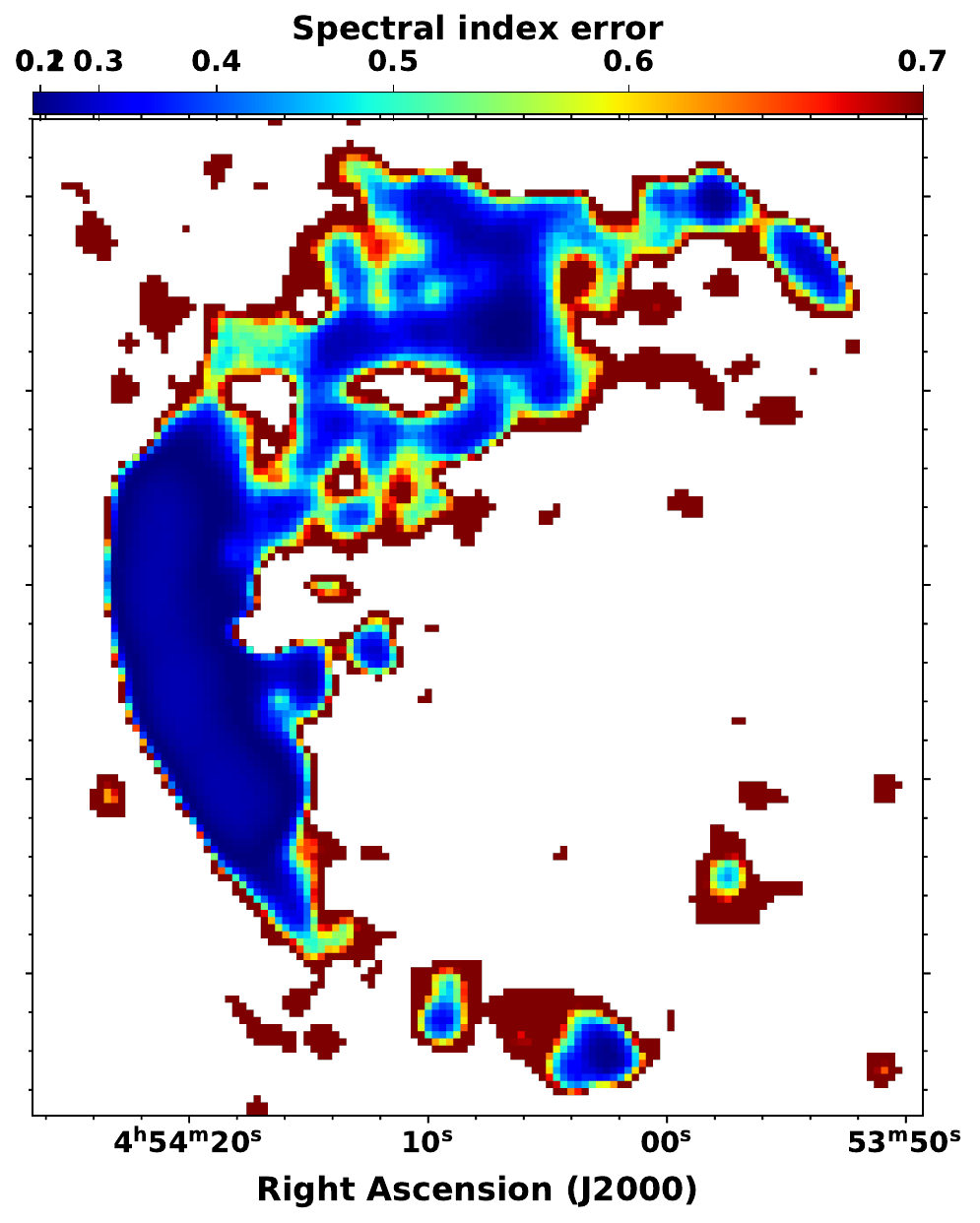}
     
    \caption{\textit{Left:} Spectral index map for the radio halo in A521 at 22$''$ is shown between 400 MHz and 650 MHz. The black contours are drawn at 3$\sigma_{\rm rms}$ $\times$ [1,2,4,8..] from the 400 MHz image. The relic is showing a spectral steepening towards the cluster center. While the radio halo region shows fluctuations in the spectral indices. \textit{Right:} The corresponding error map is shown here. The procedures for the estimation of the error map are mentioned in Section~\ref{sec:5.3}.}
    \label{img:9a}
    \label{img:9b}
    
\end{figure*}

\subsection{Radial profile of the radio surface brightness}\label{sec:5.2}

To characterize the radio halo properties, we have fitted the surface brightness model and checked how the averaged radio surface brightness is varying with distance. As suggested by \cite{2009A&A...499..679M} and very recently by \cite{2021A&C....3500464B}, we modeled the surface brightness variation by taking into account the asymmetric and elongated shape of the radio halo. The fitting procedures mentioned in \citet{2021A&C....3500464B} allow one to fit the elliptical and skewed (asymmetrical) models. The surface brightness profile is given by:

\begin{equation}
    I(r) = I_{0} \exp^{-G(r)}
\end{equation}

where I$_{0}$ is the surface brightness in the central region, and the G(r) is the radial function given by: G(r) = $\left(\frac{|r|^{2}}{r_{e}^{2}}\right)^{0.5}$ for the circular shape, G(r) = $\left(\frac{x^{2}}{r_{1}^{2}} + \frac{y^{2}}{r_{2}^{2}}\right)^{0.5}$ for elliptical models. Here r$_{e}$ denotes the e-folding radius and r$^{2}$ = x$^{2}$ + y$^{2}$.

\begin{table}
  \centering
  \caption{Radio halo surface brightness profile fitting parameters}
  \begin{tabular}{@{}ccccc@{}}
    \hline
      Model & $\chi_{r}^{2}$ & I$_{0}$ ($\mu$Jyarcsec$^{-2}$)& r$_{1}$ (kpc) & r$_{2}$ (kpc) \\
    \hline\hline
    Circular & 1.76&1.28 $\pm$ 0.04 & 380 $\pm$ 40  \\

    Elliptical & 1.15&1.42$\pm$ 0.03 & 420$\pm$ 22 & 360 $\pm$ 20   \\

    \hline
  \end{tabular}
  \label{table6}
\end{table}

In Figure~\ref{img:8}, we have shown the radial profile of the radio surface brightness at 400 MHz and the model fitted to that. The total radio-emitting regions are divided into concentric elliptical annular regions to estimate the azimuthally averaged surface brightness of the radio halo. The central annular region is drawn at 22$''$, which is the beam size of the image. The cluster center is chosen to be the center of the annuli. The red data points in Figure~\ref{img:8} are the estimations of the averaged surface brightness from each of these annular regions. Then we fitted the elliptical and circular surface brightness model on the estimated averaged radio surface brightness and estimated the parameters, using the Bayesian analysis. In Table~\ref{table6} we have reported the parameters obtained from the fitting. The central radio brightness is $\sim$ 1.35 $\mu$Jyarcsec$^{-2}$. The extension of the radio halo is detected up to $\sim$ 3r$_{1}$ and 2.5r$_{2}$. In following this, we refer to the halo ``core" region as the emission contained within the ellipse, having major and minor axis as 0.5r$_{1}$ and 0.5r$_{2}$ respectively, while the outer halo region is the emissions contained within the region beyond the core.

\begin{figure}
	
	\includegraphics[width=\columnwidth]{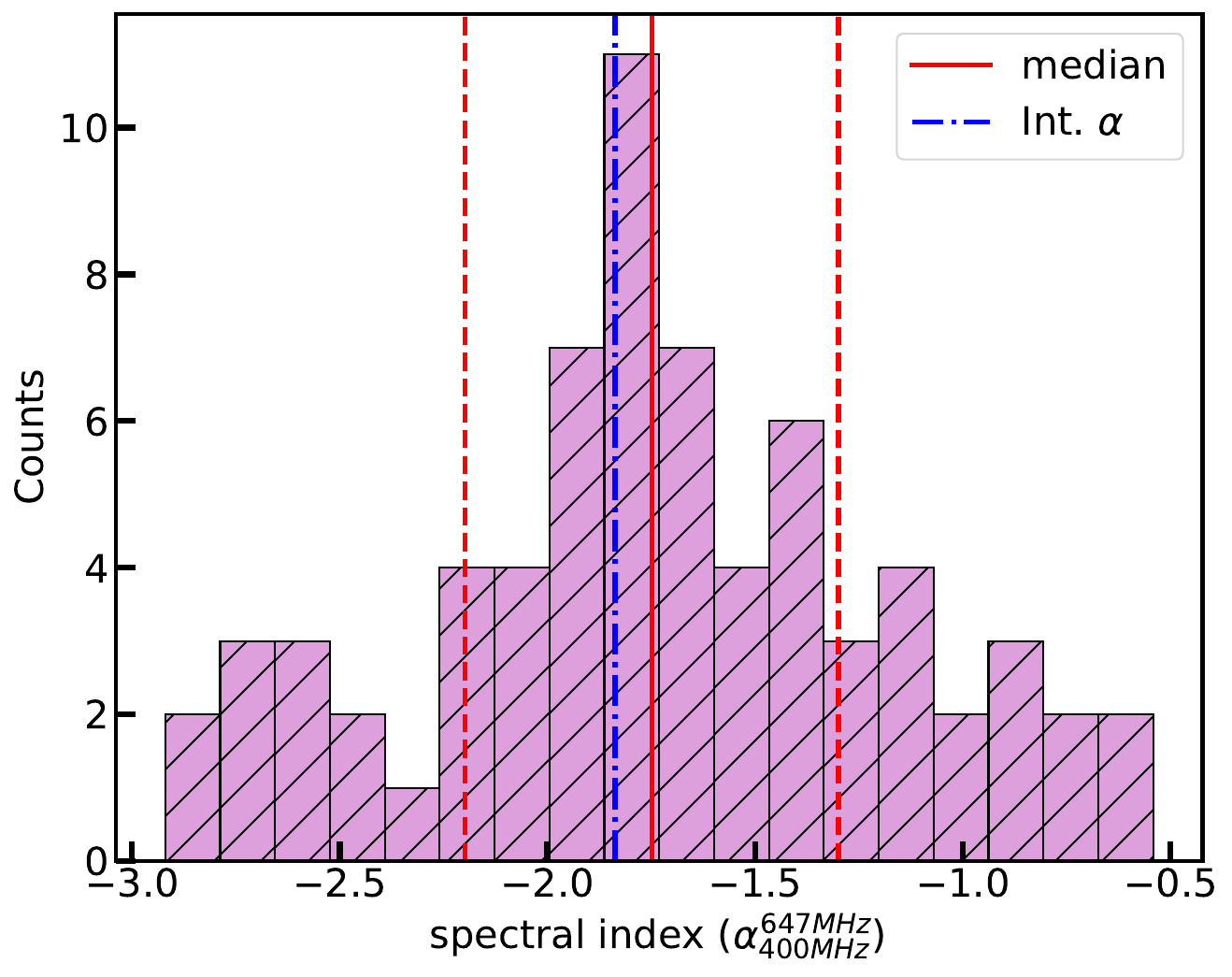}
    \caption{Histogram of the spectral index distribution for the halo in A521 between 400 MHz and 650 MHz is shown. The values for the spectral indices were obtained by sampling the halo into square boxes of size 22$''$. The red solid line is the median value of the spectral index is $\alpha_{\rm med}$ = -1.75. The two dashed line shows the standard deviation around the median value, with $\sigma = 0.45$. The blue dashed-dotted line shows the integrated spectral index value.}
    \label{img:10}
\end{figure}

\subsection{Spectral index map of the halo} \label{sec:5.3}

Spatially-resolved spectral index maps are a very useful tool in understanding the different physical processes related to mergers happening at the local scale and the origin behind the radio halo emission. Our uGMRT band 3 and band 4 images have allowed us to study the resolved spectral map for the radio halo and radio relic region. Despite some problems with the band 4 point source subtracted image, the resolved spectral index map for the central region is obtained with fair accuracy.   

The spectral index maps are created using the point source subtracted visibilities in band 3 and band 4. The images were made with a similar inner \textit{uv} cut-off (\textgreater 0.2 k$\lambda$) and uniform weighting to match the spatial scale of the radio emission at both 400 and 650 MHz. Then both the images are convolved to a common resolution of 22$''$, such that the radio halo emission can be seen properly in both frequencies. The resolution of the image is chosen such that there is an optimization between probing the spectral variation in the radio-emitting region and having a good Signal to Noise Ratio. For each image, the pixel value of less than 3$\sigma_{\rm rms}$ is blanked. Flux density values are obtained 1000 times from a Gaussian distribution that has the mean as the estimated flux density value and the standard deviation as the background rms of the images. The errors regarding the spectral indices are obtained from the standard deviation of the extracted 1000 values.

Figure~\ref{img:9a} shows the  resolved spectral index map (left panel) and corresponding error map (right panel) for the radio halo. A local variation of the spectral index is visible.  In the northern part, some flat spectral regions are seen. The central part of the radio halo shows some uniform spectral index regions. The southern part of the radio halo is seen to have a very patchy distribution of the spectral index. Fluctuations in the spectral index over the radio halo can occur due to the non-uniformity of the magnetic field distribution and different acceleration efficiencies \citep[e.g.][]{2001MNRAS.320..365B,petrosian2001nonthermal,2007MNRAS.378..245B}. However, artificial patches in a resolved spectral index map can also be originated due to unmatched \textit{uv}-coverage in the different frequencies.

In Figure~\ref{img:10} we have shown the distribution for the spectral index over the radio halo extent. The median spectral index is $-1.75$ and the standard deviation is $\sigma =0.45$. According to \citet{2014A&A...561A..52V, 2017ApJ...845...81P}, if the variations in the spectral index are the result of measurement errors, the median error value is expected to be comparable or similar to the standard deviation. The estimated median error is 0.35, which is lower than the dispersion of our distribution, which indicates small-scale fluctuations over the extent of the radio halo. Also, this is a projected spectral index. It is averaged along the Line Of Sight (LOS) and the ``intrinsic'' scatter in the volume is larger ($\sim$ $\sqrt N$), where N is the number of beams/cells along the LOS. Fluctuations in the resolved spectral map are seen in some recent studies for the halos in Abell 2255 \citep{2020ApJ...897...93B}, Abell 520 \citep{2021A&A...656A.154H}, MACSJ017.5+3745 \citep{2021A&A...646A.135R}, and A2256 \citep{2022arXiv220903288R}. The median spectral index of -1.75 is slightly flatter compared to the integrated spectral index. However, the integrated spectrum is calculated based on the flux density estimation, which is brightness weighted, and the resolved spectral maps are created by giving each pixel the same relevance.

\begin{figure}
	\includegraphics[width=\columnwidth]{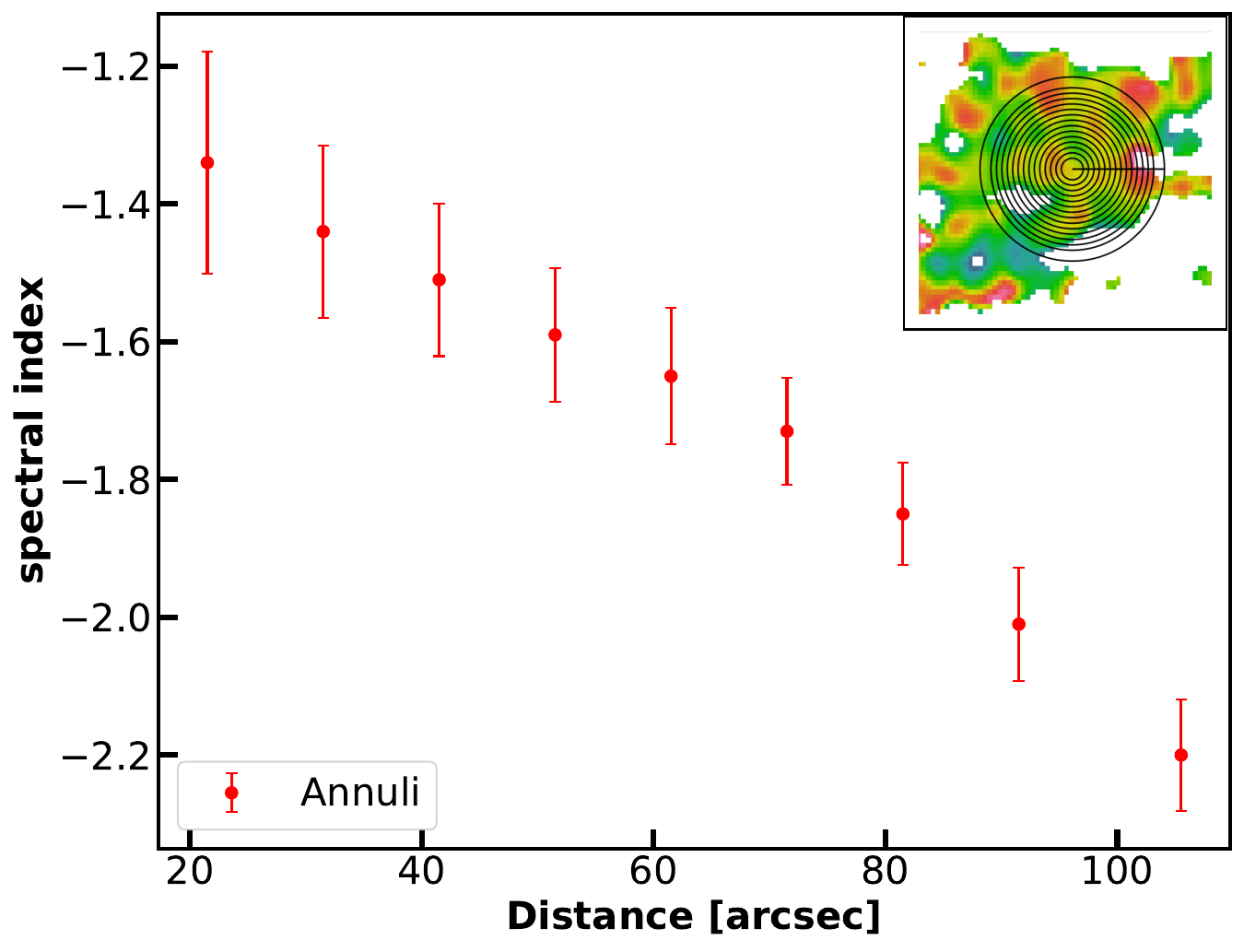}
    \caption{Here the spectral index profile across the radio halo region is shown. The annotated figure shows the annular regions, which sampled the radio halo. The average spectral indices estimated from each annulus region are fitted in red data points. There is an indeed clear radial steepening in the spectral index value of the radio halo with the increasing radius.}
    \label{img:11}
\end{figure}

We show the spectral index as a function of distance from the cluster center in Figure~\ref{img:11}. By sampling the total radio halo into different annular regions, the average spectral index values were estimated. The cluster center is chosen to be the center of the annulus, and we have checked the radial profile of average spectral index by shifting the center 20 $-$ 30 $''$ from its original position to check the robustness of the profile. In all cases, the steepening is present along the radially outward direction. Only the measurement errors are shown in the error bars. The radio halo in A521 has a distinct outward radial spectral steepening. It begins at -1.4 and rises to -2.2 as we move towards the outskirts. The steepening of the spectral index along outwards is expected either due to the decline of the magnetic field (for a constant acceleration efficiency) or due to the decline of the acceleration efficiency (i.e. less acceleration at the larger distance from the center) \citep{2001MNRAS.320..365B}. Insufficient \textit{uv}-coverage at the upper frequency can also lead to a radial steepening. However, we have a sufficient \textit{uv}-range at both frequencies to map the extended emission of the A521.

Observations have shown a variety regarding the radial variation of the spectral index. An outward radial spectral steepening is seen for Abell 2744, \citet{2017ApJ...845...81P} and MACS J0717.5+3745 \citet{2021A&A...646A.135R};  uniformity in the spectral index is seen for the Toothbrush cluster \citet{2016ApJ...818..204V, 2018ApJ...852...65R, 2020A&A...642A..85D}, the Sausage cluster \citet{2017MNRAS.471.1107H}. No firm detection of the radial spectral steepening has been made in A2163 \citep[][]{2004A&A...423..111F}, A2219 \citet{2006AN....327..565O}. However, the studies of A2163 and A2219 are very old and done at different (very poor) resolutions.

\begin{figure*}
    \includegraphics[width=10cm, height = 10cm]{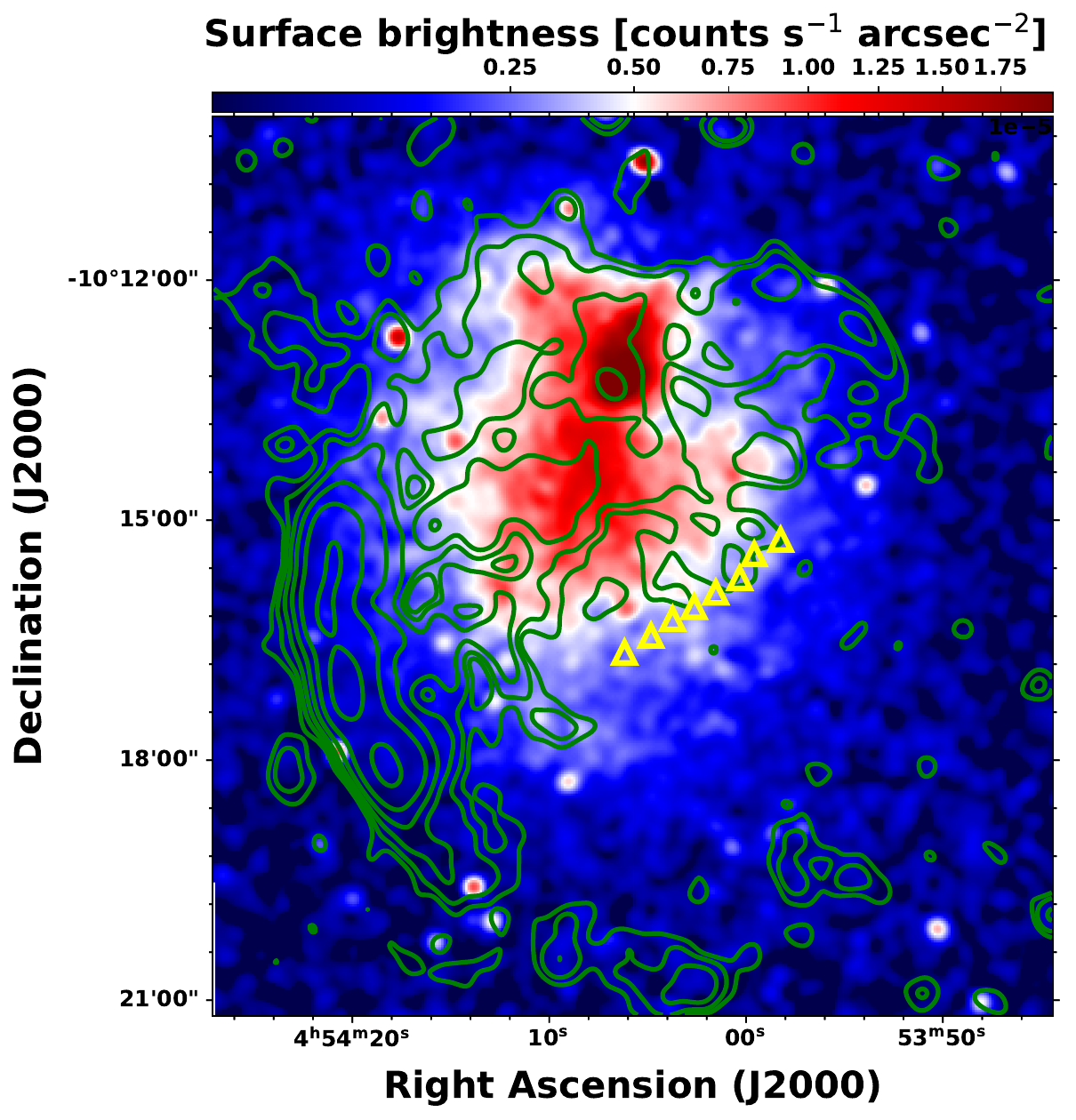}
    \includegraphics[width=8.2cm, height=10cm]{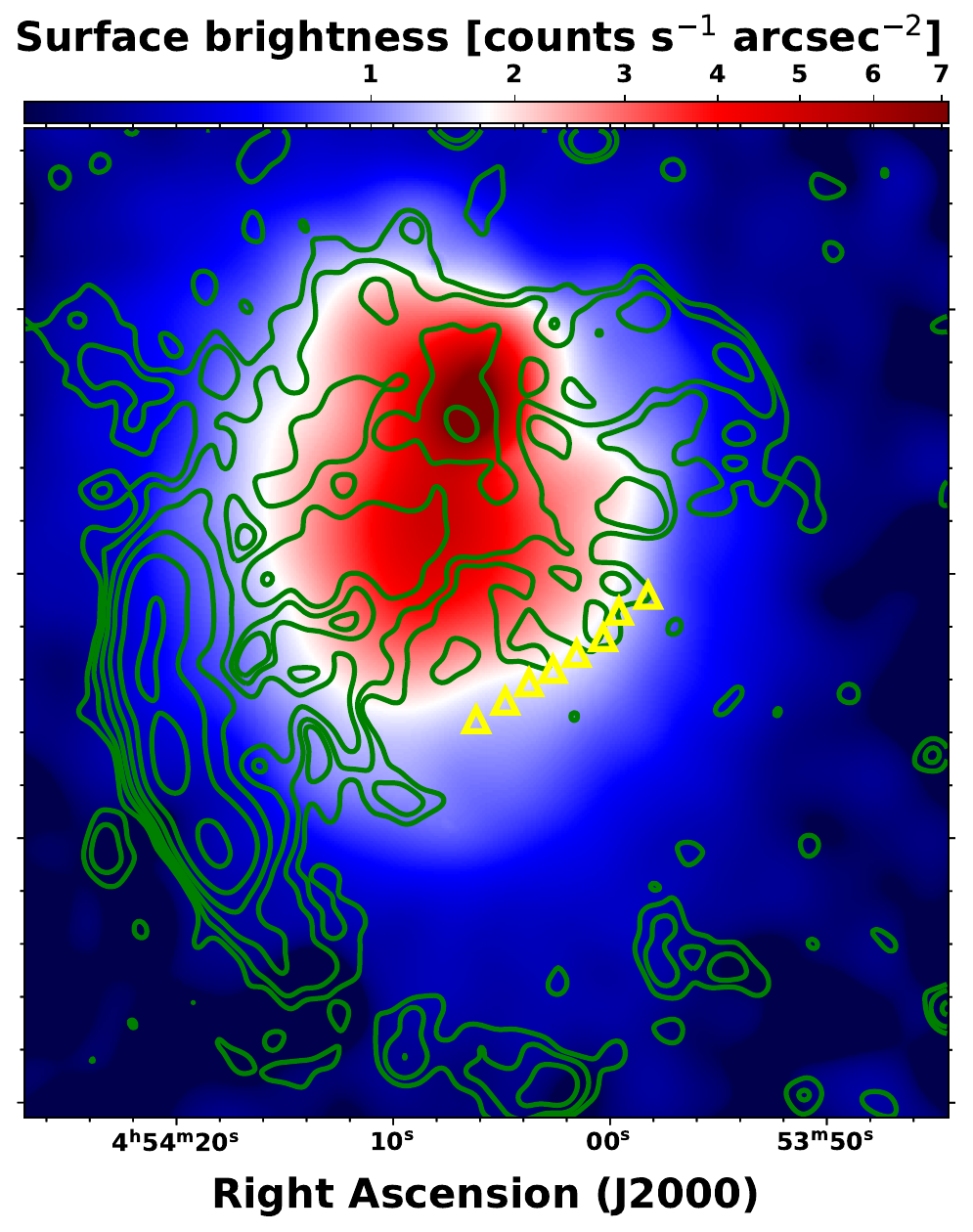}
    \includegraphics[width=10cm, height = 10cm]{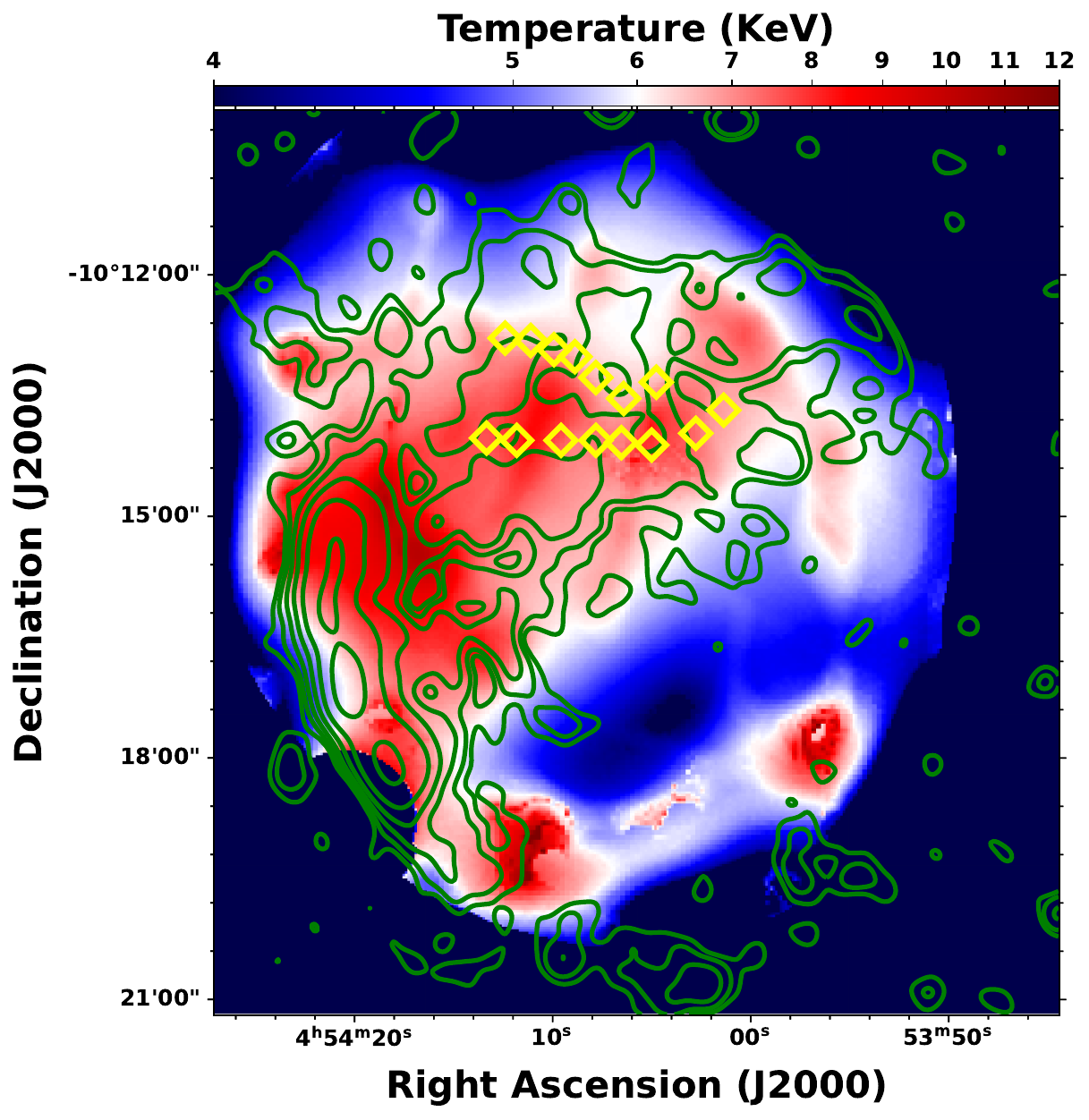}
    \includegraphics[width=8.2cm, height=10cm]{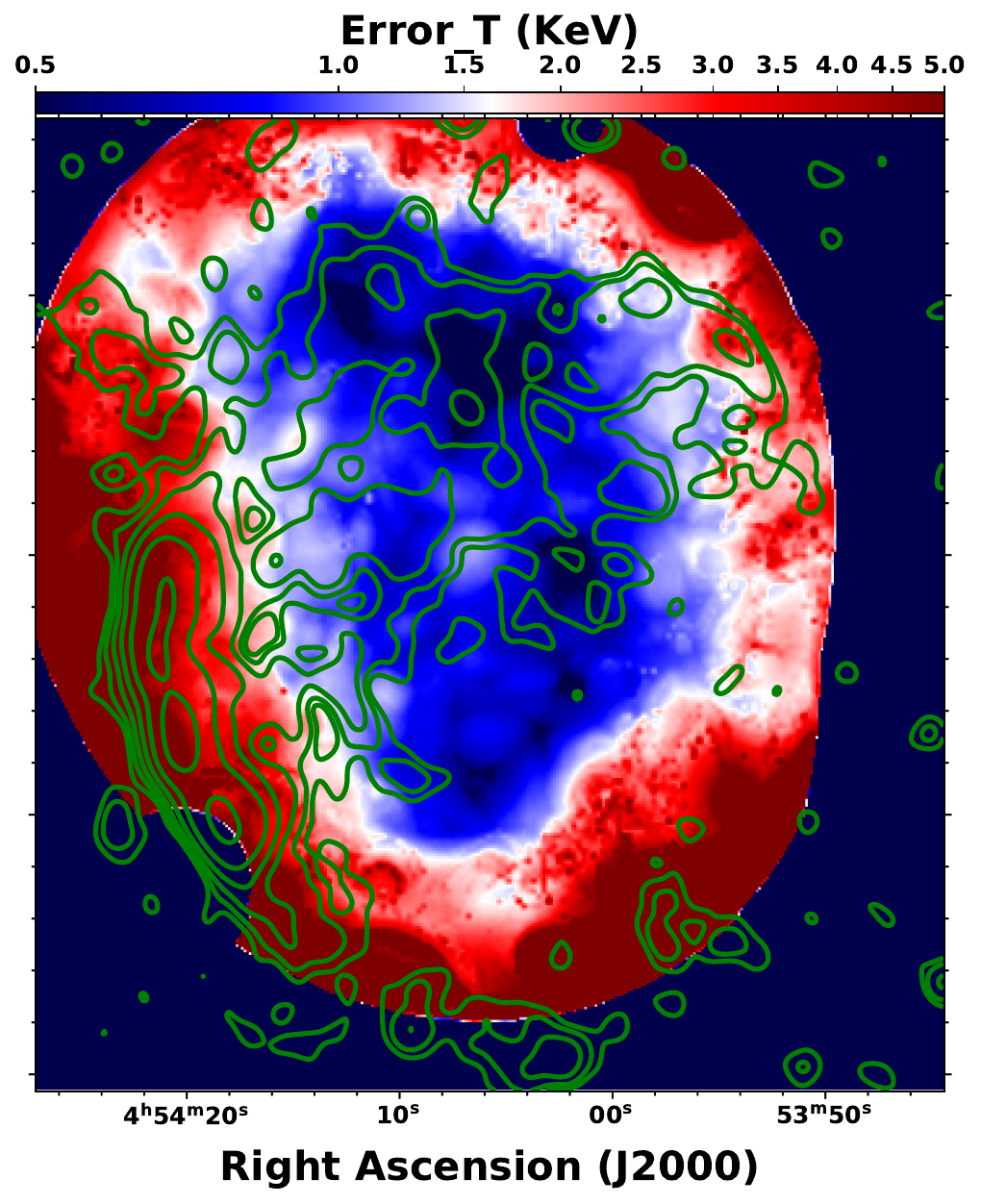}
     
    \caption{\textit{Upper left:} \textit{Chandra} image of A521 at 0.5-4 keV band is shown, smoothed with a Gaussian Full-width half maximum of 5$''$, also overlaid (green contours) with radio emission at 400 MHz. The contours level starts from 3$\sigma_{\rm rms}$ $\times$ [1,2,4..], where $\sigma_{\rm rms}$ = 45 $\mu$Jy/bm. The X-ray image is not point-source subtracted. The surface brightness gradient is shown as a color scale. The yellow triangle indicates the position, where a shock font has been detected by \citet{2013ApJ...764...82B}. \textit{Upper right:} Here the XMM-\textit{Newton} surface brightness map (0.5$-$2.5 keV) is shown overlaid with the same radio map. The radio emission is very much correlated with the thermal emission from the ICM. \textit{Lower left:} temperature map extracted from the XMM-\textit{Newton} surface brightness map is shown, overlaid with the radio emission at 400 MHz. Most of the radio emission occurs at the comparable hotter part of the cluster. The yellow diamond regions show the interaction region. \textit{Lower right:} Here the error map for the temperature is shown. At the edge region, the errors are high compared to the central regions.}
    \label{img:12a}
    \label{img:12b}
    \label{img:12c}
    \label{img:12d}
\end{figure*}

\section{X-ray and Radio correlation}\label{sec:6}

All models for the origin of the radio halos predict similarity between the X-ray and radio component of the ICM \citep[e.g.][]{2014IJMPD..2330007B}. Emission from the radio halos typically follows the X-ray distribution of the ICM \citep[][]{2001A&A...369..441G}. However, there are a few clusters whose radio emission does not trace the thermal emission with the same morphology \citep{2005A&A...440..867G}. Morphological similarity between the X-ray and radio can probe the interactions and different physical processes between the thermal and non-thermal plasma of the ICM.

In Figure~\ref{img:12a} we show the thermal X-ray emission (both from the \textit{Chandra} and XMM-\textit{Newton}) of the A521, overlaid with radio emission at 400 MHz. Overall, the radio emission from the halo follows the X-ray emission, despite the elongation of the radio halo along the merger axis. The position of the second relic (R2) is also seen in the peripheral region of the cluster and it is associated with low surface brightness ICM thermal emission, similar to relic R1. A significant part of the radio halo seems to be strongly correlated to thermal gas. However, after a certain distance ($\sim$ 600 kpc) from the cluster center (marked yellow triangle), the radio emission rapidly drops, even if the X-ray emissions are still detected in that region. \cite{2013ApJ...764...82B} have detected a shock front at that location. We have shown the XMM-\textit{Newton} temperature map and overlay our 400 MHz radio emission on that in Figure~\ref{img:12c}. The northern part of the cluster is cooler compared to the southern one. The former shows  morphological similarity with the radio halo emission. At the interacting region between the northern and southern clumps, patches of the high-temperature region are seen (yellow diamond), as also reported in \citet{2013ApJ...764...82B}. This hot gas may be a result of the compression between the two cold fonts. The bridge region between the R$_{1}$ and halo shows a high temperature ($\sim$ 10 keV) region. A temperature drop across the shock front (situated in SE to NW) is also seen in Figure~\ref{img:12c}.

\begin{figure*}
    \includegraphics[width=5.8cm, height=5cm]{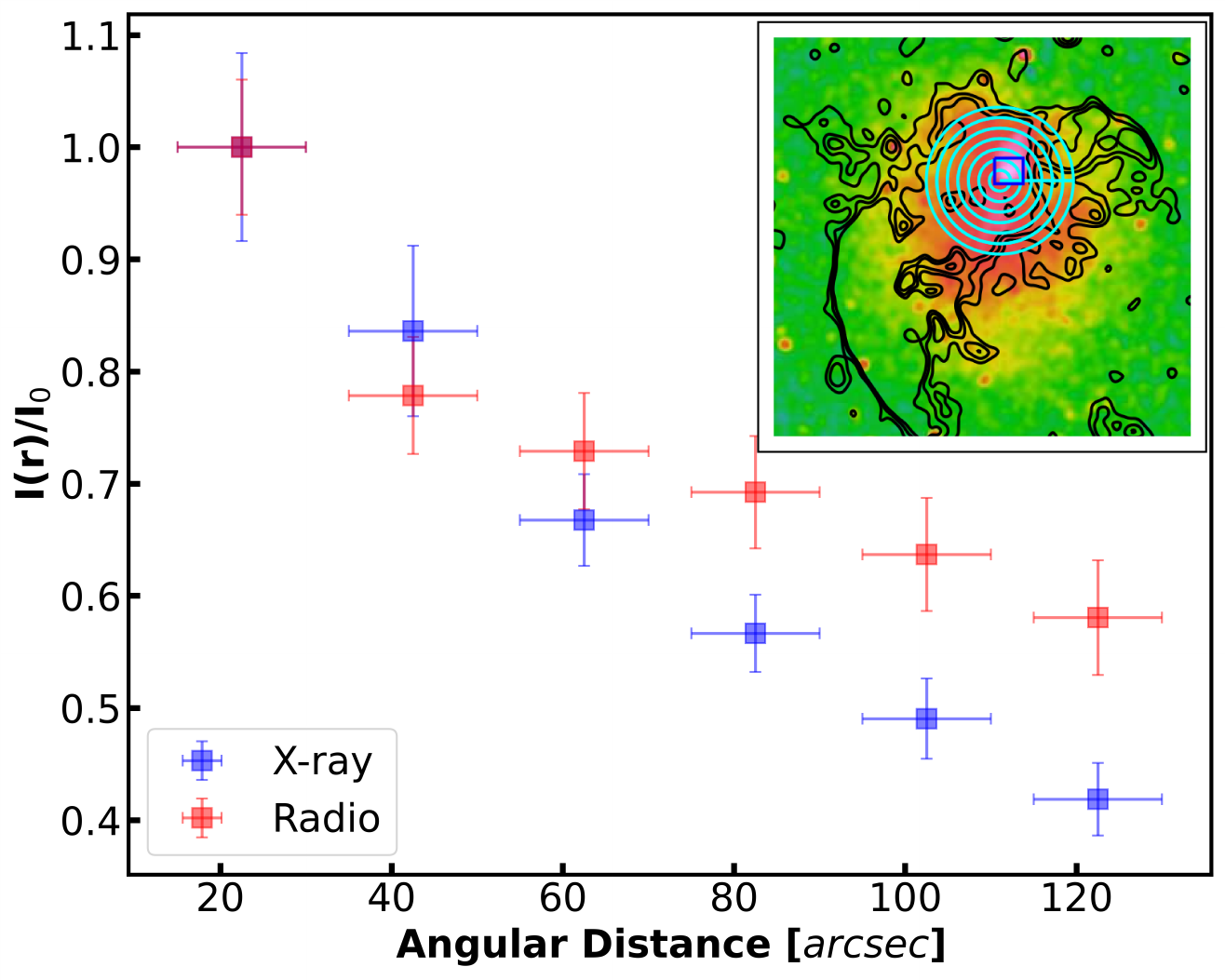}
    \includegraphics[width=5.8cm, height=5cm]{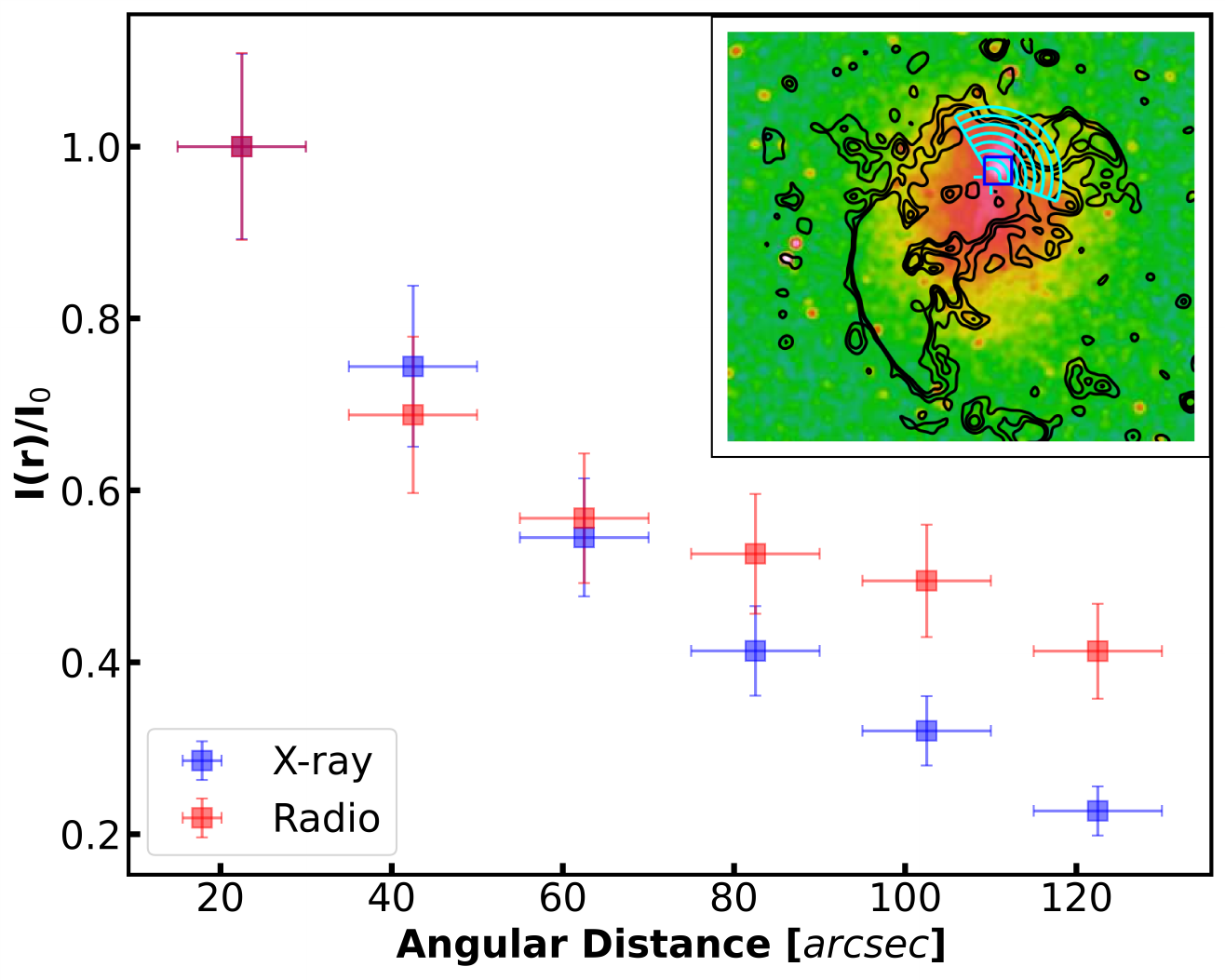}
     \includegraphics[width=5.8cm, height=5cm]{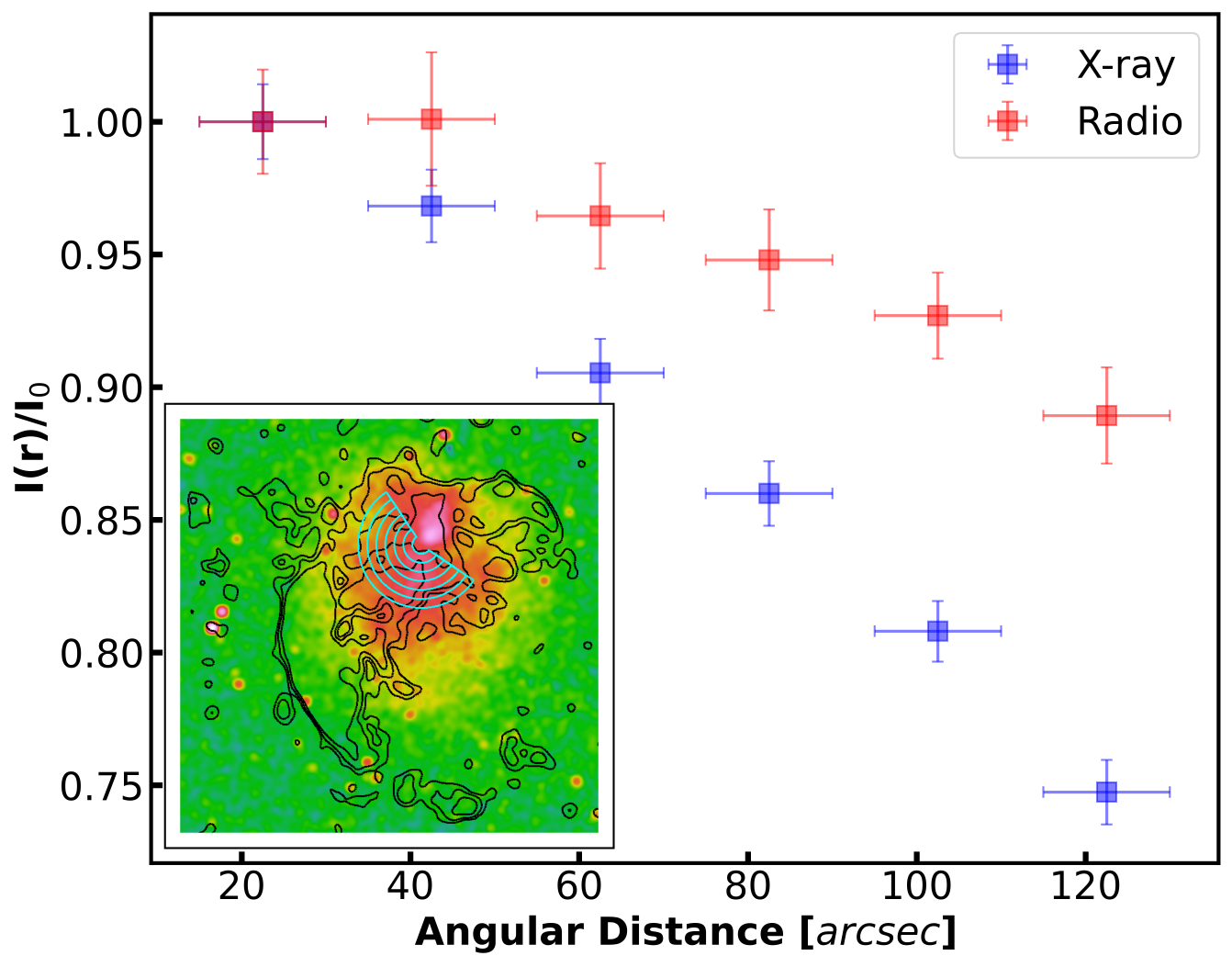}
    \caption{\textit{Left:} Radio (red squares) and X-ray (blue squares) surface brightness radial profiles within the halo region, normalized at their maximum are shown. Inset at the top right corner is the X-ray \textit{Chandra} image, overlaid with the radio contours (black contours) at 400 MHz. The annular regions in cyan are used to estimate the averaged radio and X-ray surface brightness profiles. The blue square indicates the position of the BCG, which is masked during the calculation. The error bars represented here are purely statistical.  
    \textit{Centre:} The similar radial profile is shown here for the northern part of the halo. The semi-circle (cyan) annular regions are used to estimate the averaged surface brightness values.
    \textit{Right:} The radial profiles for the southern part of the radio halo is shown with the corresponding regions in the inset.}
    \label{img:13a}
    \label{img:13b}
    \label{img:13c}
\end{figure*}

We also study the radial variation of radio and X-ray surface brightness profiles. To investigate the surface brightness variation, we divided the radio halo into concentric annular regions (centered at the cluster center). In each annular region, the mean surface brightness and standard deviations are calculated. The relic R1, R2, and discrete unresolved sources (from X-ray images) were masked and excluded from the calculations. The resulting radial profiles are shown in Figure~\ref{img:13a}a. The figure shows that the surface brightness of radio (red squares) and X-ray (blue squares) correlates well, close to the central region. Non-thermal components decay at a slower rate than thermal gas. Similarly, we investigated the radial profiles of surface brightness in the halo's northern (middle panel) and southern (right panel). Both sub-clusters exhibit a similar pattern, with the non-thermal components experiencing a less steep decline than the thermal components.

\subsection{Spatial correlation between the X-ray and radio surface brightness}\label{sec:6.1}

To date, point-to-point analysis between the radio surface brightness and the X-ray surface brightness has been done for many radio halos, mostly at 1.4 GHz. For giant radio halos the correlation slope predominantly shows sub-linear to linear values \citep[e.g.][]{2001A&A...369..441G, 2005A&A...440..867G, 2018ApJ...852...65R, 2019A&A...622A..20H,2019A&A...628A..83C,2020A&A...636A...3X,2021A&A...654A..41R,2021A&A...650A..44B,2021A&A...656A.154H,2022ApJ...933..218B,2022arXiv220903288R}. This relationship between the X-ray (I$_{\rm X}$) and radio surface brightness (I$_{\rm R}$) is generally described in the form of a power law:
\begin{equation}
    \log(\rm I_{\rm R}) = a + b\log(\rm I_{\rm X})
\end{equation}
where b is the slope of the correlation. The correlation slope b=1 is for a perfect linear correlation. The value of the b determines whether the thermal components (gas density, temperature) of the ICM decay faster (b \textless 1) than the non-thermal components (relativistic electron populations, magnetic field) or vice-versa (b \textgreater 1). 

\begin{figure}
	
	\includegraphics[width=\columnwidth]{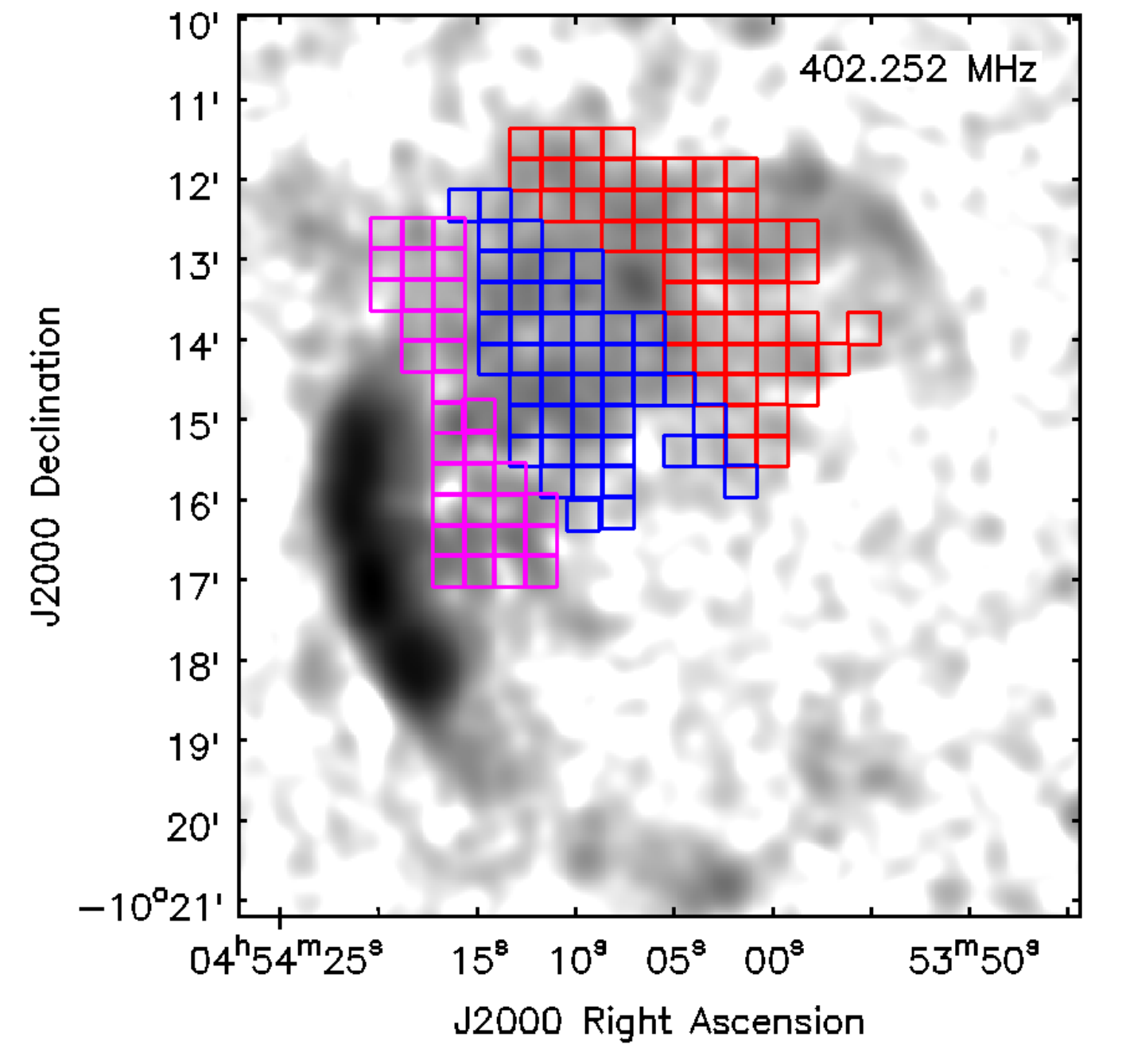}
    \caption{The grey-scale image shows the uGMRT 400 MHz image. Each of the square boxes has a size of 22$''$. The red boxes sample the northern part of the radio halo, whereas the blue ones are for the southern part. The magenta square boxes are used to estimate the surface brightness values at the bridge region between the relic R1 and the radio halo.}
    \label{img:14}
\end{figure}

High-resolution, sensitive uGMRT, \textit{Chandra} X-ray observations allow us to do a detailed investigation of the interplay between the thermal and non-thermal components of the ICM. For the analysis, we have used the 400 MHz, and 650 MHz uGMRT point-source subtracted images. \textit{Chandra} X-ray image is smoothed to a resolution of 5$''$. We have made use of the publicly available software \texttt{PT-REX\footnote{\url{https://github.com/AIgnesti/PT-REX}}} (\texttt{Point-to-point TRend EXtractor}) \citep{2020A&A...640A..37I} to sample the radio and X-ray emissions in radio halo regions and exclude the relics and the discrete unresolved sources. The properties of the bridge region between the relic R1 and the halo are unclear, therefore we have done a correlation study separately for this region. For estimation of the radio and X-ray surface brightness, we have chosen a grid size of 22$''$ (88 kpc). To retain a good signal-to-noise ratio, we have included only regions with above 3$\sigma_{\rm rms}$, below that, are included as the upper limit. The radio brightness is expressed in the units of Jy arcsec$^{-2}$ and X-ray in units of Counts s$^{-1}$arcsec$^{-2}$.

Due to the threshold and large intrinsic scatter of the data, a sophisticated linear regression method should be taken into account the Malmquist bias for both variables. Here we have used the \texttt{LinMix\footnote{For more information on LinMix check \url{https://linmix.readthedocs.io/en/latest/src/linmix.html}}} \citep{2007ApJ...665.1489K} to determine the best-fitting parameters from the observed data set. \texttt{LinMix} is an MCMC hierarchical approach, that considers the uncertainties of both variables and incorporates the upper limits (non-detection) of the y-axis, allowing the estimation of intrinsic scatter ($\sigma_{\rm int}$). The strength of the correlation is measured using the \texttt{Pearson} (r$_{\rm p}$) and \texttt{Spearman} (r$_{\rm s}$) correlation coefficients.   

In Table~\ref{table7} we have summarised all the best-fit parameters and correlation coefficients for each frequency of our analysis. The radio (I$_{\rm R}$) and X-ray (I$_{\rm X}$) surface brightness show a significant positive correlation both at 400 and 650 MHz. A sub-linear slope of $b_{\rm 400 MHz}$ = 0.61 $\pm$ 0.06, $b_{\rm 650 MHz}$ = 0.55 $\pm$ 0.08, is obtained at 400 and 650 MHz respectively. Point-to-point correlation studies for the ultra-steep spectrum radio halo (Abell 2256; \citealt{2022arXiv220903288R} and MACSJ1149+222.3; \citealt{2021A&A...650A..44B}) shows a similar range of values. 

\begin{figure*}
    \includegraphics[width=9.1cm, height=7.85cm]{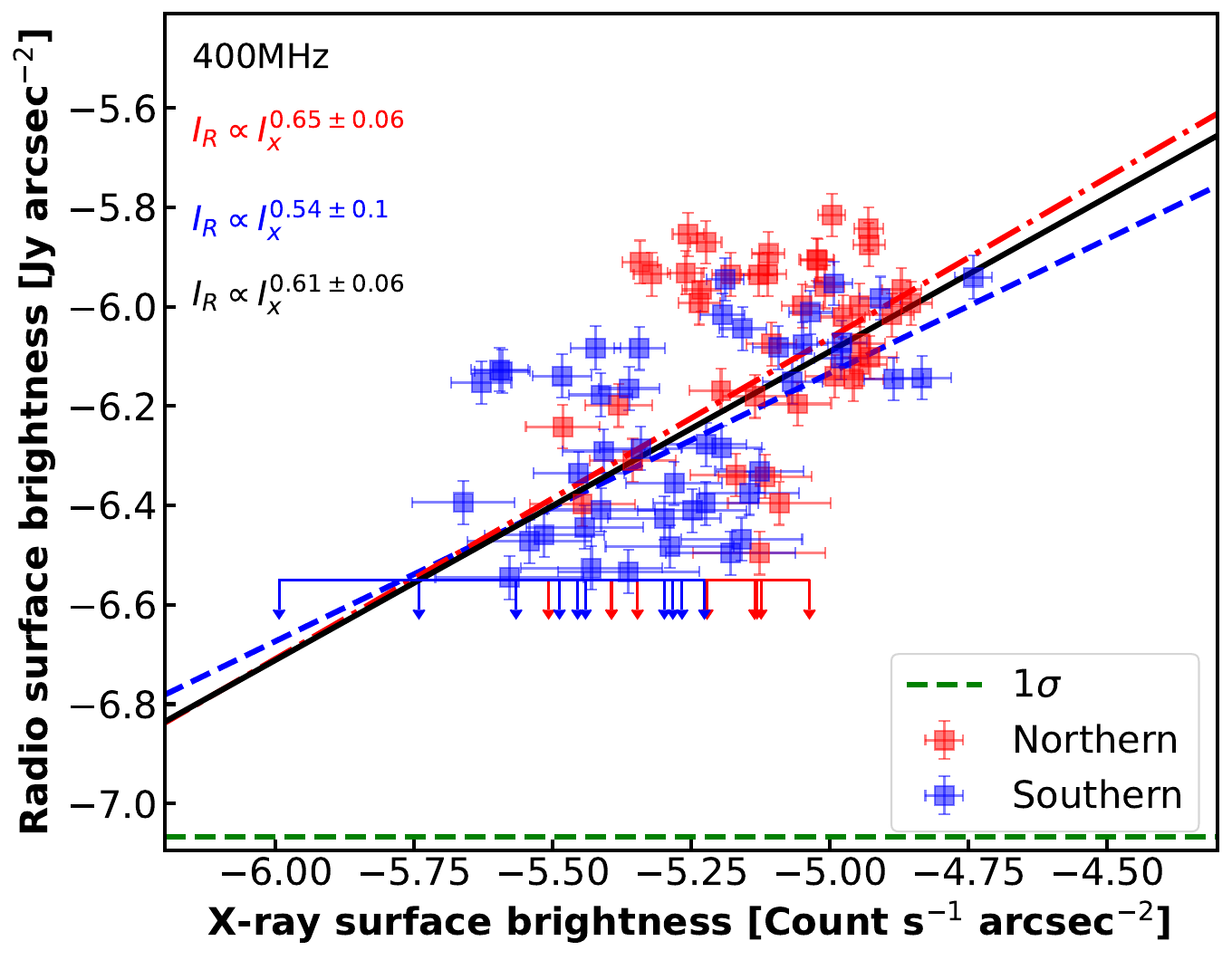}
    \includegraphics[width=9.1cm, height=7.75cm]{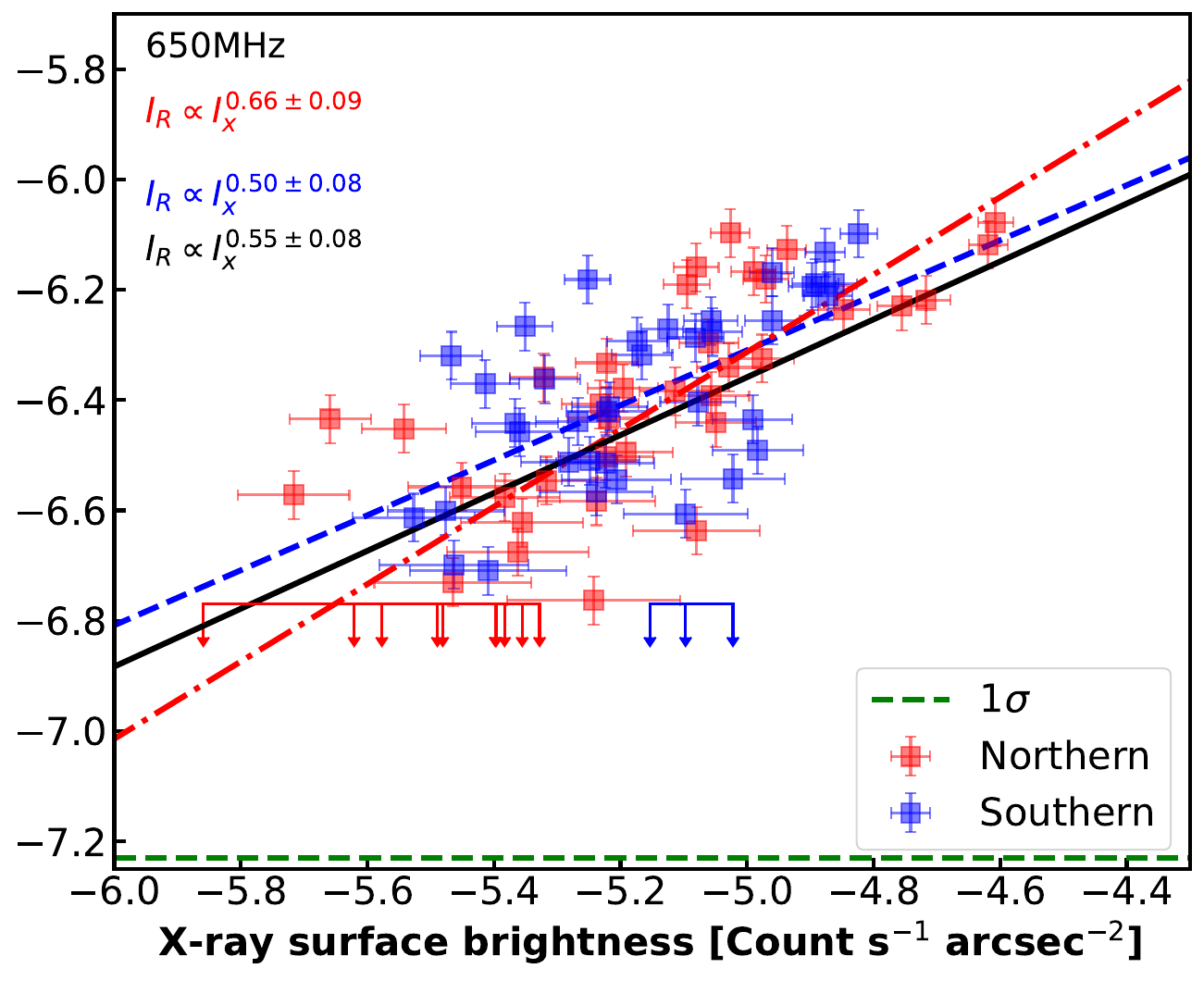}
    \caption{ Radio Vs X-ray surface brightness (on a logarithmic scale) plots for the halo in A521 at 400 (left panel) and 650 MHz (right panel) are shown. The LinMix best-fit line is shown in the solid black (entire halo), blue dashed (southern), and red dashed-dotted (northern) lines. The red and blue square boxes show the detection of the radio halo above 3$\sigma_{\rm rms}$. The 2$\sigma_{\rm rms}$ are considered as the upper limits and indicated by arrows. The green dashed line shows the 1$\sigma_{\rm rms}$ line at both frequencies.}
    \label{img:15a}
    \label{img:15b}
   
\end{figure*}

As discussed in Section~\ref{sec:6}, the cluster is composed of two main  gas clumps (northern and southern). In Figure~\ref{img:15a} we show our test to search for a correlation between the northern and southern clumps. Our best-fit parameters are consistent with a sub-linear slope of the radio-X-ray surface brightness, regardless of the radio frequency we consider. The correlation is stronger in the northern clump, and is also steeper than the southern clump, being in all cases sub-linear. In the bridge region (shown in the magenta boxes in Figure~\ref{img:14}) the linear regression shows that the radio and X-ray surface brightness is very weakly correlated, with a linear correlation coefficient of 0.30. Due to the uncertain boundary of the bridge region, it is still very hard to draw any strong conclusion on the properties of the bridge region. 
   
In past, the frequency evolution of the correlation slope has been studied for many radio halos, exceptional case: Abell 520 \citep{2019A&A...622A..20H}, MACSJ017.5+3745 \citep{2021A&A...646A.135R}, where the investigation for the I$_{\rm R}$ - I$_{\rm X}$ correlation have been done for multiple frequencies. They report a significant variation of correlation slope over frequency, indicating the presence of a high-frequency cut-off. However, there are cases where the slope remains constant as a function of frequency: Abell 2744 \citep{2021A&A...654A..41R}, Abell 2256 \citep{2022arXiv220903288R}. Similarly, in A521 we discovered that the correlation slope is nearly  constant across the frequency range of our study. However, the distance between the frequencies in our study is not very big.

It is worth checking the effect of the box size on the correlation slope by using the lower-resolution images. We thus repeated the fitting procedures using the LinMix for 28$''$ and 35$''$ resolution images. We have created two new grids, sampling the total radio halo emission. We have obtained a slope of $b_{28''}$ = 0.58 $\pm$ 0.09, $b_{35''}$ = 0.56 $\pm$ 0.1. The results indicate that the correlation slope does not change very much with changing the box size. \citet{2021A&A...654A..41R} have shown the effect of the grid size on the correlation slope in Abell 2744, where the slope becomes flatter at higher resolutions and steeper at lower resolutions.

\begin{table*}
\centering
\caption{Best-fit slopes and Spearman (r$_{s}$) and Pearson (r$_{p}$) correlation coefficients of the radio and X-ray surface brightness data are summarized.}
 \begin{tabular}{@{}cccccccccc@{}}
 \hline\hline
 \multicolumn{0}{c}{}&\multicolumn{0}{c}{Frequency(MHz)}& \multicolumn{4}{c|}{3$\sigma$}& \multicolumn{4}{c|}{2$\sigma$} \\

\cline{3-10}
 & &slope (b) &$\sigma_{\rm int}$ & r$_{s}$ & r$_{p}$  & slope (b) &$\sigma_{\rm int}$ & r$_{s}$ & r$_{p}$ \\
\hline
 Entire halo &400 & 0.55 $\pm$ 0.07 & 0.02 & 0.72 & 0.76 & 0.61 $\pm$ 0.06 &0.04 & 0.68 & 0.71  \\

  &650 & 0.50 $\pm$ 0.08 &0.02 & 0.65& 0.72& 0.55 $\pm$ 0.08 & 0.02 & 0.65 & 0.70  \\
\hline

Northern & 400 & 0.55 $\pm$ 0.06 & 0.02 & 0.80 & 0.83 & 0.65 $\pm$ 0.06 & 0.03& 0.72 & 0.74 \\

& 650 &0.53$\pm$0.08 & 0.01 & 0.77& 0.73& 0.66 $\pm$ 0.09 & 0.02 & 0.76 & 0.81  \\

\hline

Southern & 400 & 0.50 $\pm$ 0.07 & 0.02 & 0.80 & 0.81 & 0.54 $\pm$ 0.1 & 0.07 & 0.65 & 0.70   \\

& 650 & 0.50 $\pm$ 0.09 & 0.2 &0.73 &0.75 & 0.50 $\pm$ 0.08 & 0.02& 0.69 & 0.68 \\

\hline
\end{tabular}
     
\label{table7}
\end{table*}

\subsubsection{Radial variation of the correlation strength}\label{sec:6.1.1}

\begin{figure}
	
\includegraphics[width=\columnwidth]{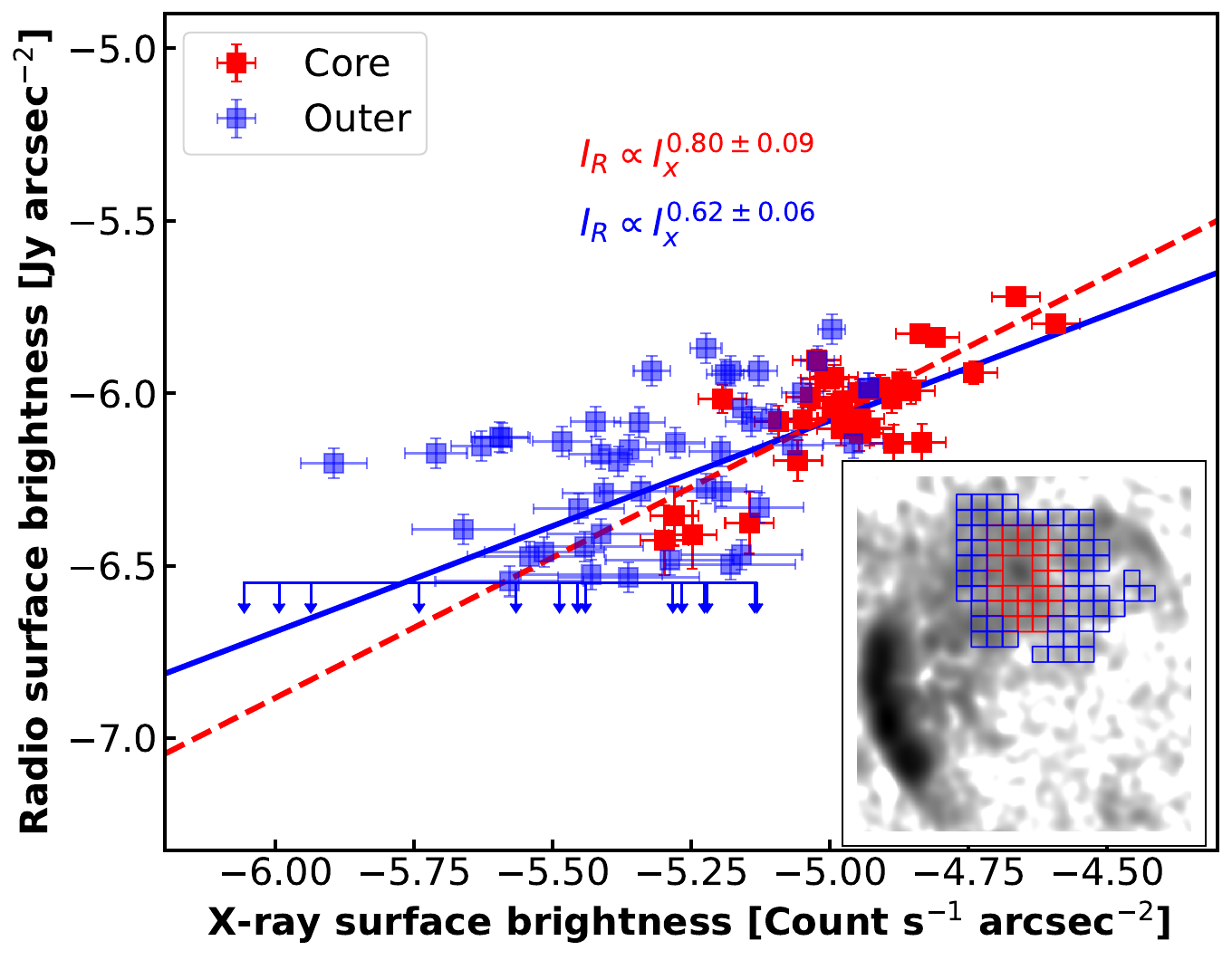}
    \caption{We have shown the X-ray vs radio surface brightness at 400 MHz for the core and outer region of the halo. The annotated figure shows the uGMRT 400 MHz emission (in grey). Different regions are shown in square boxes with different color-coded. The blue solid line is the best-fit line for the outer regions and the red dotted line is the best-fit line for the core region.}
    \label{img:16}
\end{figure}

We investigated the relationship between radio and X-ray surface brightness at various sub-regions (core and the outer region) shown in Figure~\ref{img:16}. The boundaries of the core and outer regions were defined in Section~\ref{sec:5.2}. Because the radio halo is much more extended at 400 MHz, we used that image as well as the same \textit{Chandra} X-ray image. The slope is 0.80$\pm$ 0.09 in the core regions and 0.62$\pm$0.06 in the outer regions. Although we saw a sub-linear correlation strength in both regions, a gradient in the slope can still be seen. \citet{2022ApJ...933..218B} discovered an increase in correlation slope for the Coma cluster. While \citet{2022arXiv220903288R} has also reported variation in the correlation slope across sub-regions. In our case, we discovered different slopes in the two sub-regions at increasing radii. For some cool core clusters that host a hybrid halo, that is, a mini halo and halo-type component, such as RXC J1720.1+2638 \citep{2021MNRAS.508.3995B}, MS 1455.0+2232 \citep{2022MNRAS.515.1871R}, a steeper slope is found in the core region followed by flattening in the outermost region. The above-mentioned behavior can be possible due to the complex interplay between the thermal and non-thermal components of the ICM in a different way for the central and outer regions.

\begin{figure*}
    \includegraphics[width=9.5cm, height=7.9cm]{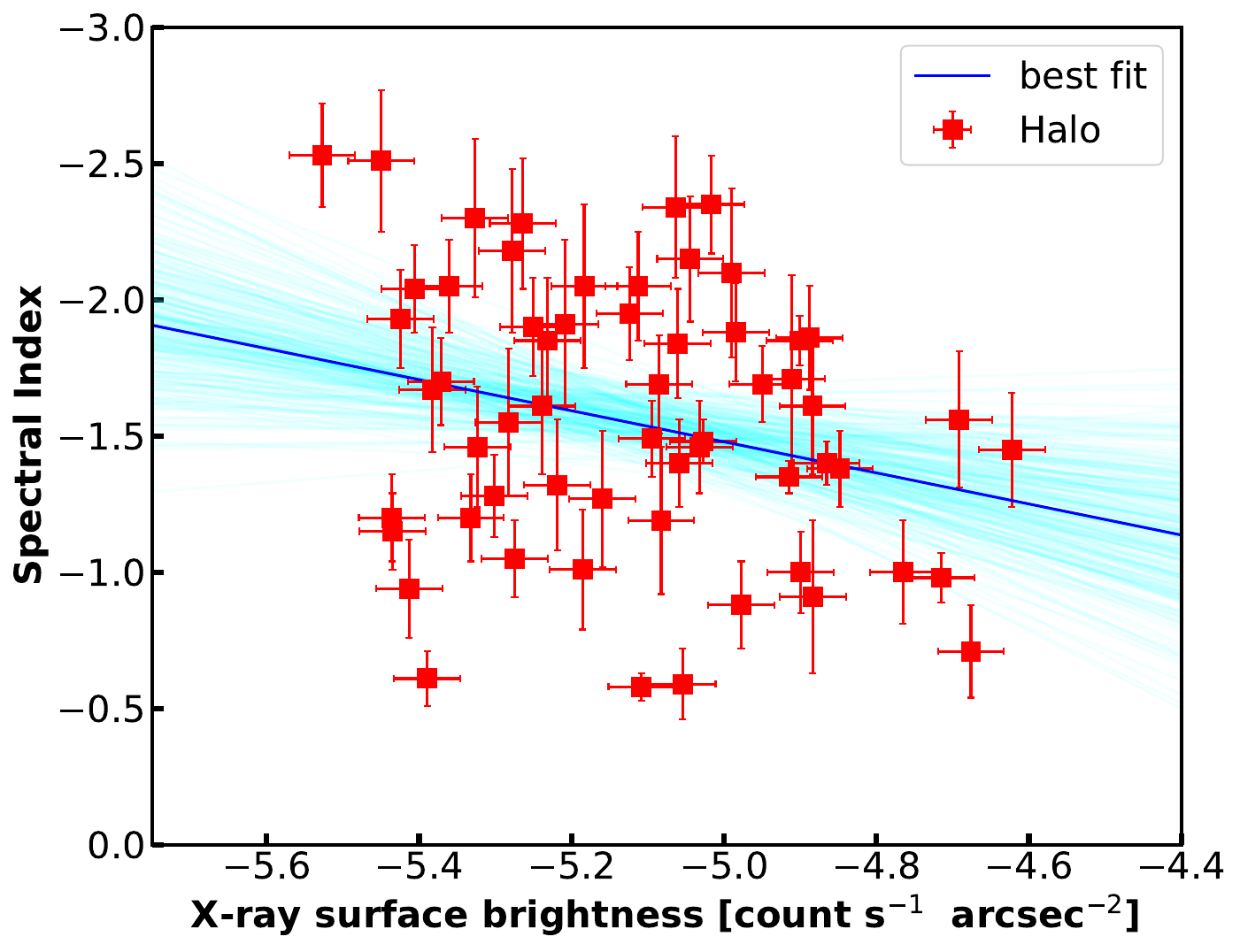}
    \includegraphics[width=8.7cm, height=7.85cm]{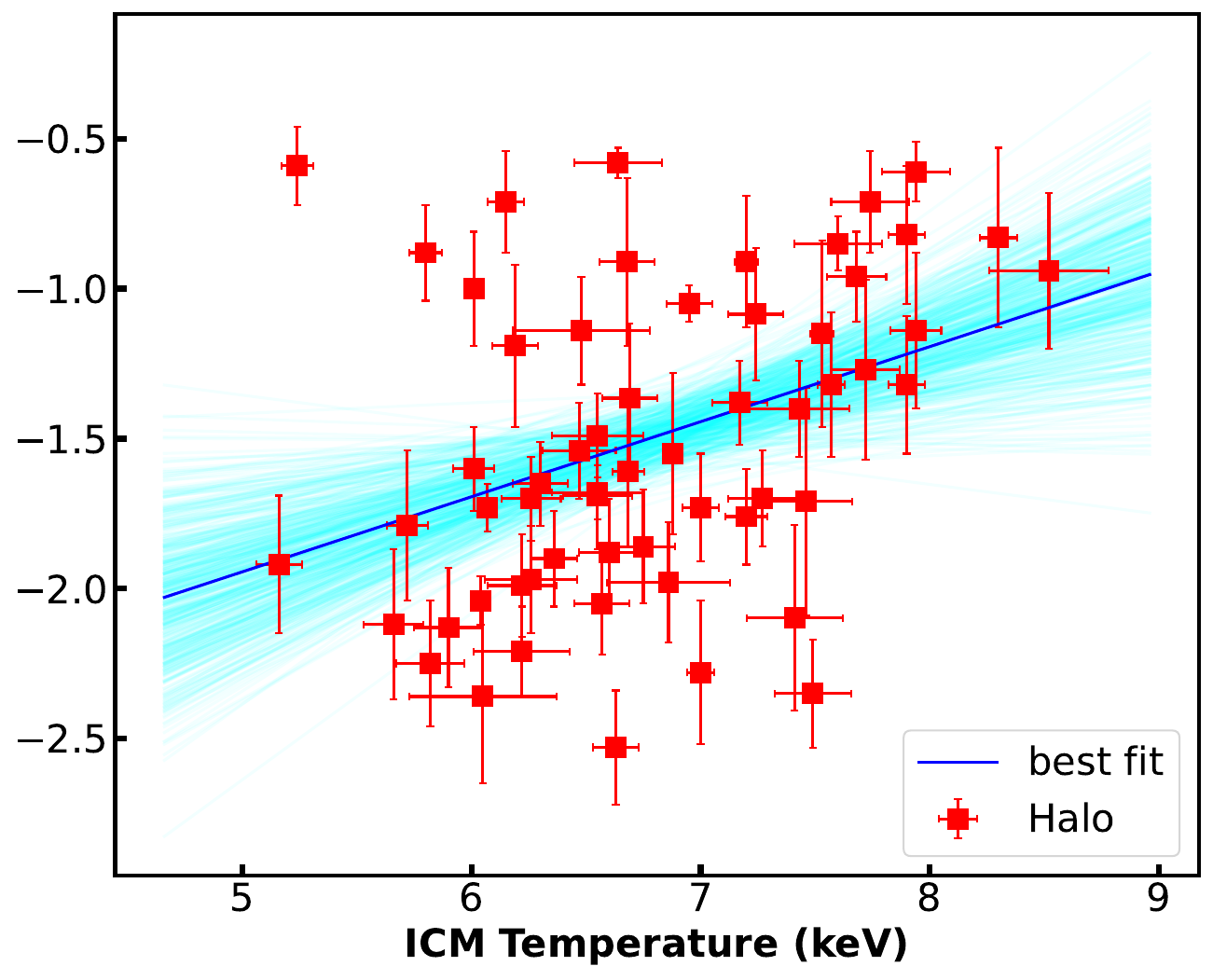}
     
    \caption{\textit{Left:} X-ray surface brightness versus spectral index relation for the A521 radio halo is shown at 400 MHz. The cyan lines represent the posterior of the MCMC chain and the blue line indicates the best fit for the regression. \textit{Right:} $\alpha$$-$T relation for the A521 radio halo. The temperature extracted from the XMM-\textit{Newton} temperature map is shown in Figure~\ref{img:12c}, (bottom left). }
    \label{img:17a}
    \label{img:17b}
    
\end{figure*}

\subsection{Spatial correlation in the spectral index vs X-ray surface brightness}\label{sec:6.2}
We also checked the point-to-point relationship between the spectral index and the X-ray surface brightness in A521. We used the same sampling method described in the previous section to obtain the spectral index and X-ray surface brightness values. We extracted the surface brightness using the same \textit{Chandra} image as in Section~\ref{sec:6}. We fitted the data with \texttt{LinMix} to do linear regression between the spectral index and X-ray surface brightness and to test the significance of the correlations. We have assumed the relation between the spectral index and X-ray surface brightness in the form of:

\begin{equation}
       \alpha = \rm a + \rm b\log (\rm I_{\rm X})
\end{equation}

The spectral index and the X-ray surface brightness appear to have a mild anti-correlation (Figure~\ref{img:17a}). To test the significance of the correlation between these two quantities, we estimated the \texttt{Spearman} and \texttt{Pearson} correlation coefficients for the given data set. \texttt{Spearman} coefficient is $\approx$ -0.37, with a slope of 0.46$\pm$0.10. The weak anti-correlation indicates that the brighter X-ray-emitting regions correspond to flatter spectral indices. This is consistent with the radial spectral steepening in the radio halo region.

To our best knowledge, the studies for the correlation between the spectral index and the X-ray surface brightness have been done for a very small number of halos like A2744 \citep[][]{2021A&A...654A..41R}, A2255 \citep{2020ApJ...897...93B} and CIG0217+70 \citep{2019A&A...622A..20H}, A2256 \citep{2022arXiv220903288R}. A2255 has a mildly positive correlation between the two quantities. However, in A2744, both positive and negative correlations were discovered, indicating the presence of multi-component radio halo and complex merger systems. While there is no significant correlation between the spectral index and radio surface brightness for CIG0217+70. A recent study by \cite{2022arXiv220903288R} found a moderate anti-correlation between these two quantities for the A2256, which contains an ultra-steep spectrum radio halo. Based on the correlation coefficients, a weak negative correlation between the spectral index and X-ray surface brightness can be seen. However, further investigations are needed to establish any strong trend.

\begin{table}
  \centering
  \caption{Linmix best fitting slopes and Spearman and Pearson coefficient for the data plotted in Figure~\ref{img:17a} }
  \begin{tabular}{@{}ccc@{}}
    \hline\hline
      &  $\alpha$-I$_{X}$ & $\alpha$-T  \\
    \hline
     Correlation slope &0.46 $\pm$ 0.10   & 0.37$\pm$ 0.09  \\
     Spearman Coff. & -0.37  & 0.43 \\
     Pearson Coff. & -0.35  & 0.39\\
   \hline
  \end{tabular}

  \label{table8}
\end{table}

\subsection{Correlation between the spectral index and ICM temperature}\label{sec:6.3}

 The XMM-\textit{Newton} map was used for the analysis because it covers a larger area than \textit{Chandra}. The resulting plot is in Figure~\ref{img:17b} (right panel). We used \texttt{LinMix} to look for any possible correlations. The Spearman correlation coefficient is 0.43. This indicates a very weak positive correlation between the spectral index and the ICM temperature. The positive correlation between these two quantities strengthens the fact that a fraction of the gravitational energy dissipated during cluster mergers heats up the thermal plasma as well as accelerates relativistic particles in the ICM. 

A comparison between the resolved spectral index and ICM temperature values has been studied in \cite{2006AN....327..565O} for A2744, where they reported that flatter spectrum emission would trace higher temperature regions. However, the results were questioned by the \cite{2017ApJ...845...81P} using deeper radio and X-ray observation and found no strong correlation between these two. A mild anti-correlation has been found for radio halo in A2255 \cite{2020ApJ...897...93B}. While we have found a weak positive correlation for the radio halo in A521. High-temperature regions are expected to trace the heated regions, excited by turbulence, hence, the gas dynamics and the turbulent energy flux, that is dumped into the particles will correlate well with flat spectral index regions. However, \cite{2010ApJ...718..939K} argued about the validity of any possible correlation and anti-correlation between the spectral index and ICM temperature due to the different cooling time scales for the thermal gas and non-thermal plasma. The other caveat about these correlations is the effect of plasma instability and micro-physics, which is not well understood in theoretical models. The plasma collisionality may play a crucial role in the acceleration efficiency \citep[e.g.][]{2011MNRAS.412..817B}. An increase in temperature may affect collisionality and mean free path of (thermal) particles which potentially could drive effects on the effective acceleration rate. It is also possible to imagine situations where decreasing collisionality (hot regions) might decrease the acceleration rate.

\subsection{X-ray shock at R2?}

\begin{figure*}
    \includegraphics[width=9cm, height=9cm]{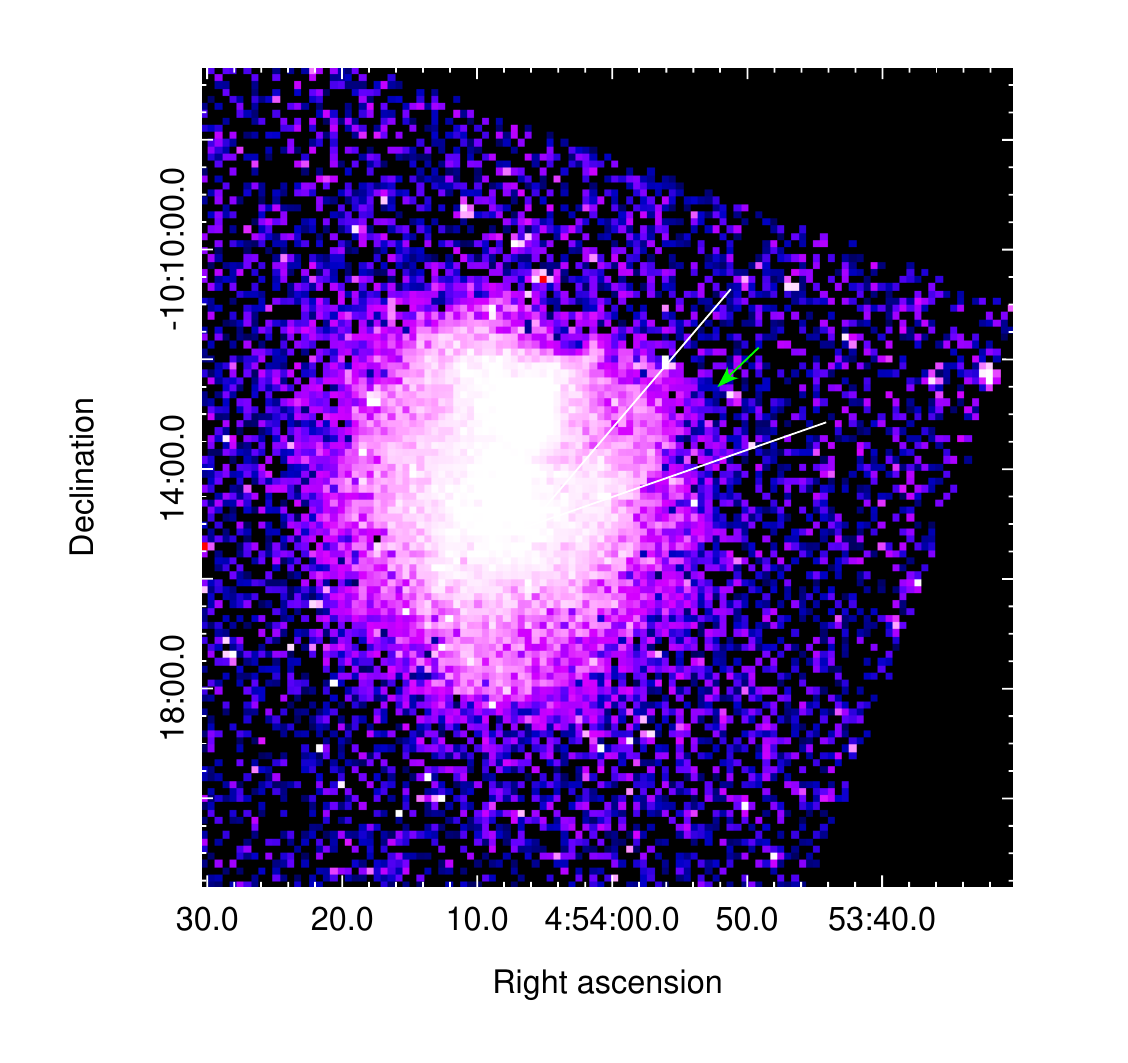}
    \includegraphics[width=9.2cm, height=9.5cm]{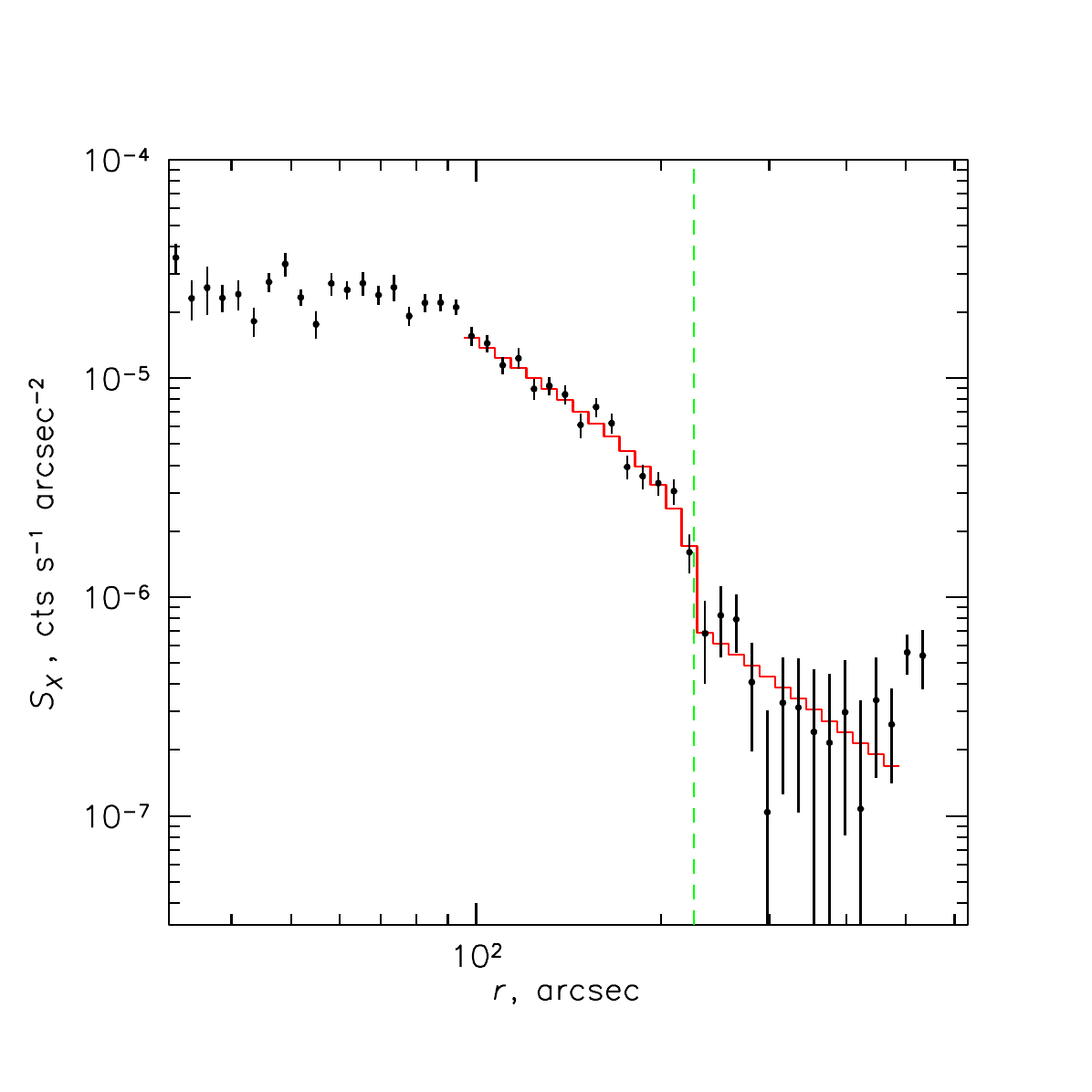}
    
    \caption{\textit{Left:} Chandra image of A521 in the 0.5$-$4.0 keV energy band 
    binned to 8$''$ pixels. The image is background subtracted and divided by the 
    exposure map. A subtle edge (marked by a green arrow) is coincident with the brightest and sharpest segment of the relic R2 (Figure~\ref{img:3a}, upper-left panel). The white sector is used for the radial profile shown in the right panel. \textit{Right:} X-ray 0.5$-$4.0 keV surface brightness profile across the edge in the region of the relic R2, extracted in the white sector indicated in the right panel. Error bars are 1$\sigma$. The red line is a fit with a shock discontinuity. The dashed green vertical line marks the position of the outer edge of the relic.}
    \label{img:shocka}
    \label{img:shockb}
\end{figure*}

Using the \textit{Chandra} image, we searched for an X-ray brightness edge at the position of the newly-detected radio relic R2, which would indicate a possible shock front. The \textit{Chandra} image binned to $8''$ pixels to emphasize low surface brightness features in the cluster outskirts is shown in the left panel of Fig.~\ref{img:shocka}. It indeed shows a subtle edge (green arrow) coincident with the brightest and sharpest segment of the relic (Figure~\ref{img:3a}, upper-left panel). To quantify this edge and see if it is consistent with a shock front, we extracted an X-ray radial brightness profile in the white sector shown in Figure~\ref{img:shocka} (left). The sector is centered on the curvature center of the relic and covers the most prominent northern half of the relic. We used an image binned to $1^{\prime\prime}$ pixels and masked out the X-ray point sources in the sector.
The X-ray profile is shown in the right panel of Figure~\ref{img:shockb}. A brightness edge is clearly seen at $r=226^{\prime\prime}$, corresponding to the outer edge of the relic, which is shown by a dashed green vertical line. The brightness edge has the characteristic shape of a spherical density discontinuity projected on the sky plane \citep[e.g.][]{2007PhR...443....1M}. To quantify this discontinuity, we modeled the brightness profile close to the edge ($\pm$ factor 2 from the radius of the edge) using a spherical gas density distribution with the center of symmetry at the center of the sector and the gas density described by two power laws on the two sides of the edge and a sharp jump at the edge, with the power laws and the position and amplitude of the jump being free parameters \citep{2000HEAD....5.1701M}. The best-fit model is shown in red in the right panel of Figure~\ref{img:shockb}. A continuous density profile is inconsistent with the data at a $\sim3\sigma$ confidence; the density jump amplitude is $2.4^{+2.3}_{-0.8}$ at 90\% (where we did not apply the precise conversion factor between the emission measure and X-ray brightness, which is uncertain without the temperature measurements but sufficiently close to 1 for the \textit{Chandra} instrument response in this energy band and our large error bars). If this density jump is a shock front --- which is plausible but cannot be determined with certainty without the temperature measurements on both sides of the edge --- it would correspond to $M=2.1$ or $M>1.4$ at 90\%. The X-ray Mach number (M$_{\rm X}$ is lower as compared to the radio Mach number (M = 2.67$\pm$0.09) at R2. The discrepancy between the radio and X-ray Mach numbers may indicate a non-uniform distribution of the Mach number over the shock surface \citep{2019ApJ...883...60R,2021MNRAS.506..396W}. The radio observations trace the higher Mach number regions, where the particles get more accelerated. Also, the radio Mach number is estimated assuming the standard DSA to be the possible origin of the R2. The estimated Mach numbers of the shock are low ($\sim$ 2.5) to generate the observed emission of the relic in a standard DSA scenario \citep{2020A&A...634A..64B}. Therefore a shock re-acceleration of supra-thermal electrons might be needed to understand the relic emission.

\section{Discussion} \label{sec:7}

\subsection{Spectral properties and re-acceleration models }

A strong indication of spectral fluctuations and radial spectral steepening may be reconciled with the absence of a clear spectral curvature in the integrated spectrum assuming in-homogeneous turbulent re-acceleration. Homogeneous turbulent re-acceleration models predict the spectral steepening to occur at frequencies ($\nu_{s}$) larger than a few times the critical frequency ($\nu_{c}$) of the high energy electrons \citep{2006AN....327..557C}, $\nu_{s}$ $\sim$ $\xi$$\nu_{c}$, where $\xi$ was 6-8 estimated by \citet{2012A&A...548A.100C}. However, if the magnetic field intensity, acceleration time scale vary in the emitting volume (in-homogeneous case) and along the line of sight, the spectrum may get stretched in frequency and the curvature becomes less evident \citep{2021A&A...646A.135R, 2013MNRAS.429.3564D}. The similar in-homogeneous behavior of the magnetic field and acceleration time may lead to the observed spectral fluctuations in the radio halo regions. Radial spectral steepening demonstrates the presence of a high-frequency break in the spectrum, however, the in-homogeneous conditions might smooth the effect of the break in the integrated spectrum. This combined with other results \citep[e.g.][]{2021A&A...646A.135R}, demands further theoretical steps to understand the spectral shape of radio halos.

The turbulent re-acceleration model has predicted that a large number of radio halos (at GHz frequencies) should host the ultra-steep spectrum radio halos \citep{2006AN....327..557C,2008Natur.455..944B}. \citet{2013A&A...551A.141M} have taken a homogeneous (homogeneous in the emitting volume) re-acceleration model to explain the integrated spectrum of A521. The curved spectra had shown a gradual steepening of the spectrum at high frequencies, although the steepening can not be very drastic since the observations for A521 are limited up to 1.4 GHz \citep{2009ApJ...699.1288D}. However, they have used the 74 MHz upper limit, which we decided to avoid. The steepening frequency ($\nu_{s}$) of the synchrotron emission is given by:
\begin{equation}
    \nu_{s} \propto \tau_{\rm acc}^{-2} \frac{\rm B}{(\rm B^{2} + \rm B_{\rm \rm IC}^{2})^{2}},
\end{equation}
where B is the magnetic field and \textbf{$\tau_{\rm acc}$} is the re-acceleration time \citep{2007MNRAS.378..245B}. If $\tau_{\rm acc}$ is constant, the steepening frequency depends on the magnetic field. The highest frequency will be emitted at a field $\rm B_{\rm cr} = \frac{B_{\rm IC}}{\sqrt{3}}$, where the  CRe has a maximum lifetime at a given frequency. With B \textless B$_{\rm cr}$, a radial steepening of the spectral index can be seen at distances larger from the cluster center, at a lower observing frequency. Assuming a spectral steepening above 1.4 GHz (due to the stretch of curvature in frequency domain by in-homogeneous ICM properties) implies the minimum acceleration time (Eq.8 in \citealt{2021A&A...646A.135R}) in the emitting volume is $\sim$ 157 Myr, providing that the magnetic field fluctuations in that region are B $\sim$ B$_{cr}$.

The limits obtained for the turbulent acceleration timescale can be used to constrain the energy density and scale of the turbulence. This will depend on the adopted re-acceleration model, as different models consider the nonlinear interactions between the turbulence and the particles. Stochastic re-acceleration model \citep{2016MNRAS.458.2584B}, based on the scattering of particles in super-Alfvenic in-compressible turbulence, connects the steepening frequency and the turbulent Mach number via:

\begin{equation} \label{eq:16}
    M_{\rm t}\sim0.2 \frac{2000 \rm km \rm s^{-1}}{\rm c_{s}} \left(\frac{\nu_{s}}{1 \rm GHz}\right)^{1/6} \left(\frac{L}{10^{2} \rm kpc}\right)^{1/2} (1 + z)^{\frac{11}{6}},
\end{equation}

where c$_{\rm s}$ is the sound speed in that medium and $L$ is the injection length scale. Using eq~\ref{eq:16}, the value of $M_{\rm t}^{2}$ to be 0.3 and the ratio of the turbulent energy to the thermal energy $\epsilon_{\rm t}/ \epsilon_{\rm ICM}$ $\sim$ 1/3$\Gamma M_{\rm t}^{2}$ $\sim$ 0.19, considering the adiabatic index $\Gamma = 5/3$ and injection scale $\sim$ 100 kpc. The above calculation assumes that turbulence and particle acceleration occur everywhere (filling factor of 1). However, there can be situations with a filling factor \textless 1, leading to a smaller energy budget for the turbulence. The obtained value is consistent with results found in the high-resolution cosmological simulations of the ICM \citep[e.g.][]{2015ApJ...800...60M, 2017MNRAS.464..210V}.

\begin{figure}
	
	\includegraphics[width=\columnwidth]{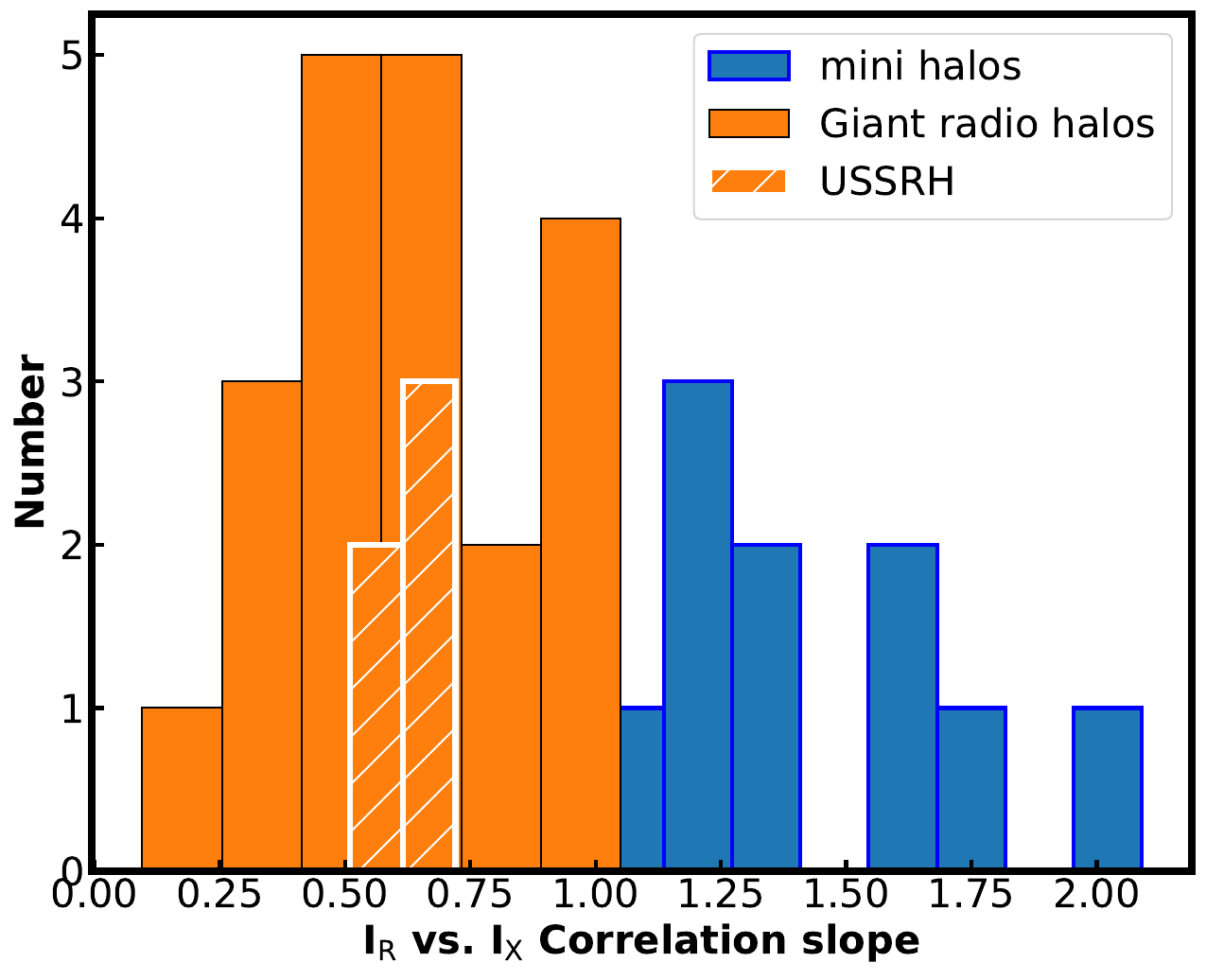}
    \caption{Distribution of the I$_{\rm R}$ vs I$_{\rm X}$ correlation slope for the different radio halos. The cyan color shows the statistics for the giant radio halos, whereas the value line is for the mini halos. We have collected the data for 22 giant radio halos and 9 mini halos. The USSRH populations are shown in the white hatch line. The giant radio halos mostly fall in \textless 1 region, having a sub-linear slope.}
    \label{img:18}
\end{figure}

\subsection{Thermal and non-thermal correlation}

Despite the complex morphology of the radio halo in A521, the radio and X-ray surface brightness show a positive correlation at both frequencies. In Figure~\ref{img:18}, we have shown the distribution of the radio vs X-ray surface brightness correlation slope. We emphasize that the sample space is very low, but a significant number of giant radio halos ($\sim$ 22) have shown a sub-linear correlation slope. Whereas the USSRH spans over a very narrow range of correlation slopes.

The correlation slope provides the information for the different acceleration processes and the magnetic field distribution \citep[e.g.][]{2001A&A...369..441G,2014IJMPD..2330007B,2015MNRAS.448.2495S}. Most of the thermal energy content of the ICM results from the dissipation of the kinetic energy of the gas, falling to the cluster, and a part of the thermal energy channeled into the non-thermal components. In the turbulent re-acceleration model, the synchrotron emissivity can be expressed as 

\begin{equation}\label{eq:9}
    \epsilon_{\rm R} \propto \eta_{\rm acc} F \frac{\rm B^{2}}{\rm B^{2} + \rm B_{\rm IC}^{2}}
\end{equation}
    
Where $\eta_{\rm acc}$ is the acceleration efficiency (fraction of turbulent energy flux converted into relativistic electrons) and $F$ is the turbulent energy flux in a unit volume:

\begin{equation}
    F = \frac{1}{2} \rho \frac{dv^{3}}{L}
\end{equation}

If we assume a constant temperature and constant Mach number, from equation ~\ref{eq:9}, the synchrotron emissivity will be:

\begin{equation}
    \epsilon_{\rm R}= \epsilon_{\rm R}(0) \frac{\rm X(r)}{\rm X(0)} \left(\frac{\epsilon_{\rm X} (r)}{\epsilon_{\rm X} (0)}\right)^{\frac{1}{2}} \frac{D_{\rm pp} (r)}{D_{\rm pp} (0)} \frac{1 + \left(\frac{\rm B_{\rm IC}}{\rm B(0)}\right)^{2}}{1+\left(\frac{\rm B_{\rm IC}}{\rm B(0)}\right)^{2}  \left(\frac{\epsilon_{\rm X} (0)}{\epsilon_{\rm X}(r)}\right)^{\frac{1}{2}}}
\end{equation}

where $\epsilon_{\rm R}, \epsilon_{\rm X}$ are the radio and X-ray emissivities ($\epsilon_{\rm X} \propto \rm n_{\rm ICM}^{2}$), B$_{\rm IC}$ is the equivalent CMB magnetic field, and X is the ratio of the energy density of the cosmic rays to the thermal gas. In terms of turbulent Mach number, the above equations can be written as:

\begin{equation}\label{eq:12}
    \epsilon_{\rm R} (r) \propto \rm X(r) \frac{\rm M_{\rm t}^{\alpha} (r)}{L(r)} \epsilon_{\rm X} (r) ^{\frac{1}{2}} \frac{1}{1+\left(\frac{\rm B_{\rm IC}}{\rm B(0)}\right)^{2} \left(\frac{\epsilon_{\rm X} (0)}{\epsilon_{\rm X} (r)}\right)^{\frac{1}{2}}}
\end{equation}

where M$_{\rm t}$ is the turbulent Mach number and $\alpha = 4$ for the Transit-time-damping \citep{2007MNRAS.378..245B} and $\alpha = 3$ for the acceleration by in-compressible turbulence \citep{2016MNRAS.458.2584B}. Therefore, different re-acceleration models would lead to sublinear slopes depending on the ICM conditions. Assuming X(r) and M$_{\rm t}$ constant with radius, equation ~\ref{eq:12} would lead to a sub-linear ($\epsilon_{\rm R} \propto \epsilon_{\rm X}^{0.5}$) and linear correlations for B(0)$^{2}$/B$_{\rm IC}^{2}$ \textgreater \textgreater 1 and  B(0)$^{2}$/B$_{\rm IC}^{2}$ \textless \textless 1, respectively.

A change in the correlation slope (flattens outside and steeper in the core region) has been obtained in the radio halo in A521. According to equation ~\ref{eq:12}, a possible scenario to explain the observed behavior is to assume the increase in the ratio of the CR energy to the thermal energy (X(r)) with the increasing radial distance from the cluster center. This is expected due to the lowering in the electron energy losses. Also, an increase of the Mach number with radius would make the radio-X ray scaling flatter in the external regions (equation ~\ref{eq:12}). However, a Mach number  increasing significantly with radius would make the synchrotron spectrum flatter in the outskirts and this situation is ruled out by the observed radial spectral steepening.

\subsection{Merging scenario}

\begin{figure}
	\includegraphics[width=\columnwidth]{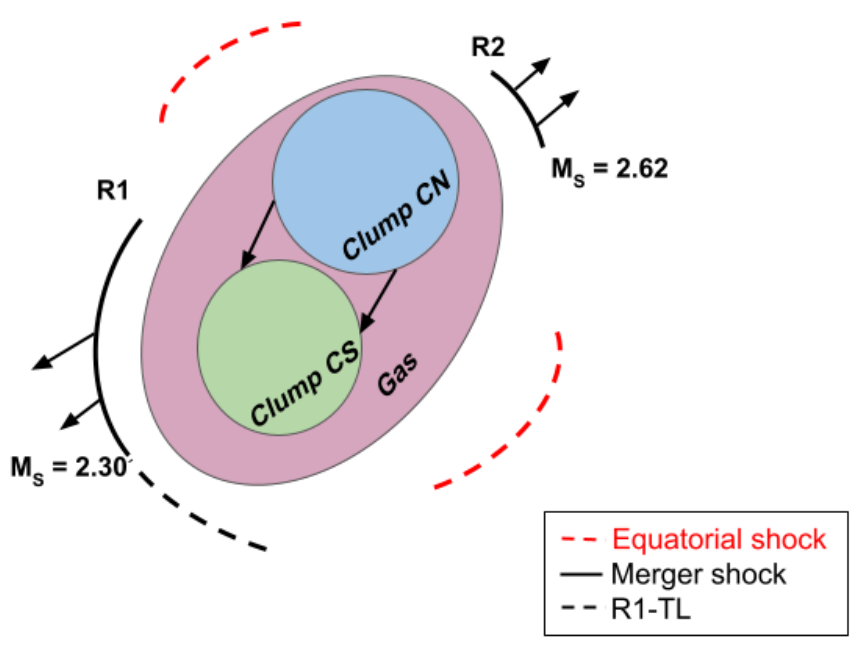}
    \caption{The schematic diagram for the merging scenario of cluster A521 is shown. The position of the R1 and R2 and the Mach number of the shock are labeled. The two clumps CN and CS are colliding with each other along the southeast and northwest direction. The red dashed lines are the equatorial shock wave outwards in the equatorial plane.}
    \label{img:19}
\end{figure}

A521 is dynamically active at the central region \citep{2020ApJ...903..151Y}. \cite{2006A&A...446..417F} had interpreted the \textit{Chandra} X-ray observation, by suggesting two-step merging events; pre-merger scenario and a post-merger scenario. HST Weak Lensing studies of A521 have shown that the cluster consists of three main clumps C, NW, SE \citep{2020ApJ...903..151Y}. The C clump consists of two more substructures, CN and CS. In Figure \ref{img:19} we have shown the schematic diagram of the Merging process.

The double relic in A521 confirmed a binary merger of two clumps at the cluster central region. In a binary merger, two ``equatorial'' shocks will move outward in the equatorial plane, perpendicular to the merger axis. Following the passage of the dark matter core, two ``merger'' shocks will travel in opposite directions along the merger axis. \citet{2011A&A...528A..38V} demonstrated that the collision of two similar mass clumps results in diametrically opposite relics with similar size and transverse width and luminosity. However, for a different mass ratio, the size and luminosity of the relics changes; the brighter relic will be situated ahead of the lower mass clump and the fainter will be in front of the higher mass clump. \cite{2020ApJ...903..151Y} has quoted of the masses of two sub clumps; M$_{\rm CN} = 3.5\times 10^{14}$M$_{\odot}$, M$_{\rm CS} = 1.5\times 10^{14}$M$_{\odot}$. The ratio of the LLS for the R1 and R2 relic is $\sim$ 3, whereas the mass ratio between the mass clumps (CN and CS) is $\sim$ 2. A similar value of mass ratio explains the observed emissions for the Sausage relic \citep{2011MNRAS.418..230V}. Because of the lack of X-ray observations of the NW and SE clumps, the merging scenario favors an off-axis collision between the CS and CN subclumps. Later in the phase, the ICM gains angular momentum and, as a result, lags behind. In the X-ray, it appears as a binary distribution, while the two DM halos begin their second infall. The observed features, such as double relics and hot intermediate regions, support a binary merging picture between two mass clumps.

Some clusters have a radio halo and double relics, while others do not. \citet{2017MNRAS.470.3465B} argued that clusters with double relics and a radio halo had a different time-since-merger ( t$_{\rm merger}$) than clusters without a radio halo. It should be noted that the t$_{\rm merger}$ derived here is not the time after the core passage. Taking the distance between the two relics as a proxy of merger time and estimating the shock velocity of $\sim$ 1500 km.s$^{-1}$, the time-since-merger is $\sim$ 0.45 Gyr.  \cite{2008Natur.455..944B} had shown that for the observed radio halo emissions, the electrons should be accelerated on a time scale of $\sim$ 0.1 Gyr. Therefore the turbulence generated during the merger will get the time to develop. Steep spectrum radio halos are predicted as an outcome of the less energetic merger or high energy losses \citep[e.g.][]{2010ApJ...721L..82C}. A binary merger between the two small clumps (CN and CS), in a bigger potential driven by the rest of the cluster, is well suited for the origin of the less energetic turbulence.

\section{Summary and Conclusions} \label{sec:8}

We present here the first deep uGMRT observations (300 - 850) MHz of the galaxy cluster A521. These sensitive uGMRT observations enable us to obtain detailed spectral characteristics of the radio halo. Previous studies of this radio halo at frequencies below 1 GHz were limited by the telescopes' poor resolution and insufficient \textit{uv}-coverage. Our deep observations, combined with available X-ray observations (\textit{Chandra}, XMM-\textit{Newton}), provide profound physical insights into the thermal and non-thermal connections in the ICM, as well as the origin of the radio halo. The overall findings are summarised as follows: 
\begin{enumerate}
    \item We present the first deep and high resolution (7$''$ - 22$''$) uGMRT radio images of the galaxy cluster A521. The extended radio halo emission at the central region and radio relic emission at the southeast outskirt regions recovered well, with an LLS of the radio halo is 1.3 Mpc at 400 MHz and 1 Mpc at 650 MHz respectively. 
\item Our deep and sensitive uGMRT images have revealed the presence of another relic, R2, at the northwest position of the cluster. The LLS of the R2 is 650 and 450 kpc at 400 and 650 MHz respectively. This new relic appears to be associated with a candidate shock front seen in the Chandra X-ray image. We have also detected a low surface brightness extension ($\sim$ 1 Mpc) of the R1 relic, R1-TL, which makes the total length of the relic to be $\sim$ 2.2 Mpc.

\item  The radio halo emission follows a single power law between 153 to 1400 MHz with an integrated spectral index of -1.86 $\pm$ 0.12, consistent with the previously obtained values. Using our uGMRT analysis and previously published literature values, we have obtained the integrated spectral index of the R1 relic to be -1.46 $\pm$ 0.03. We have also reported the integrated spectral index of the R2 relic to be -1.34 $\pm$ 0.03. 

\item The spatially-resolved spectral index map may suggest fluctuation over the extent of the radio halo. The median spectral index estimated from small regions is within the error bars of the integrated spectral index, indicating the self-similarity of the global spectral index to the local kpc scales. Resolved spectral maps also reveal a spectral steepening of the radio halo along the outward directions. This behavior incorporates the in-homogeneous conditions (variation in B, acceleration time) in the emitting volume.

\item The non-thermal emission from the radio halo has been found to have a strong morphological correlation with the thermal emission from the ICM, indicating that the hot gas and non-thermal plasma have a close relationship. The radio emission is mostly concentrated in the hotter regions of the cluster, as seen in the XMM-\textit{Newton} temperature map. Furthermore, a comparison of the radial profiles of radio (I$_{\rm R}$) and X-ray (I$_{\rm X}$) surface brightness has revealed a slower decline of non-thermal components than the thermal ones, a possible indication of the radial declination of the magnetic field.

\item A tight sub-linear correlation between the radio and X-ray surface brightness has been discovered through a point-to-point analysis across the entire extent of the radio halo. This correlation was investigated using different fitting methods and thresholds, and the slope was found to be relatively constant across the frequencies studied. There are slight changes in the correlation slope at the core (I$_{\rm R}$ $\propto$ I$_{\rm X}^{0.80 \pm 0.09}$) and outer (I$_{\rm R}$ $\propto$ I$_{\rm X}^{0.62 \pm 0.06}$) regions, which may be attributed a balance between the increase in energy density of the CRe and decrease of the magnetic field strength.

\item A weak anti-correlation (r$_{\rm s}$ = -0.37) has been observed between the spectral index and X-ray surface brightness over the spatial extent of the radio halo. This finding is in agreement with the radial spectral steepening and the dissipation of gravitational energy into the non-thermal component of the ICM. 

\item A weak positive correlation (r$_{\rm s}$ = 0.43) was found between the average temperature of the ICM and the spectral index. This observation is consistent with the idea that regions with higher temperatures will have flatter spectral indices: the energy dissipated by gravitational forces in the ICM will accelerate the seed electrons. However, accounting for the plasma collisionality and the microphysics (poorly understood) may affect the correlation.      
\end{enumerate}

Several radio halo observations support the turbulent re-acceleration model, including spectral index fluctuations across its spatial extent, a radial steepening in the spectral index, and a strong sub-linear correlation between radio and X-ray surface brightness. Because of the high energy budget of thermal protons in the ICM, as demonstrated by \citet{2008Natur.455..944B}, the pure hadronic model faces challenges. The R1 radio relic shows significant substructures in the uGMRT images, making it a candidate for future low and high-frequency polarization studies to understand magnetic field orientation in the shock plane better. Also characterizing the properties at the shock downstream region is one of the frontier goals for the future study of the relics. The detection of another relic at the northwest position has placed this cluster in a poorly understood class of objects featuring radio halo and double relic systems. Investigating the complexity of radio emission through high-resolution MHD simulations is necessary for the future.

\begin{acknowledgments}
We thank the anonymous referee for their comments that have improved the
clarity of the paper. RS acknowledges Alessandro Ignesti for his help regarding the technical issues of PT-REX. RS also thanks Luca Bruno and Francesco de Gasperin for their quick replies to many useful queries. RS and RK acknowledge the support of the Department of Atomic Energy, Government of India, under project no. 12-R\&D-TFR-5.02-0700. SG acknowledges that the basic research in radio astronomy at the Naval Research Laboratory is supported by 6.1 Base funding. F.D. and H.B. acknowledge the financial contribution from the European Union’s Horizon 2020 Programme under the AHEAD2020 project (grant agreement n.871158). We thank the staff of the GMRT that made these observations possible. The GMRT is run by the National Centre for Radio Astrophysics (NCRA) of the Tata Institute of Fundamental Research (TIFR). The scientific results reported in this article are based in part on data obtained from the \textit{Chandra} Data Archive. This work made use of observations obtained with XMM-Newton, an ESA science mission funded by ESA Member States and the USA (NASA). This research had made use of the NASA/IPAC Extragalactic Database (NED), which is operated by the Jet Propulsion Laboratory, California Institute of Technology, under contract with the National Aeronautics and Space Administration. 

\end{acknowledgments}

%

\vspace{5mm}
\facilities{upgraded Giant Metrewave Radio Telescope (uGMRT), \textit{Chandra}, XMM-\textit{Newton}}


\software{CASA, numpy, aplpy, matplotlib, ds9,   }



\begin{figure*}[t!]
    \includegraphics[width=5.78cm, height=5.9cm]{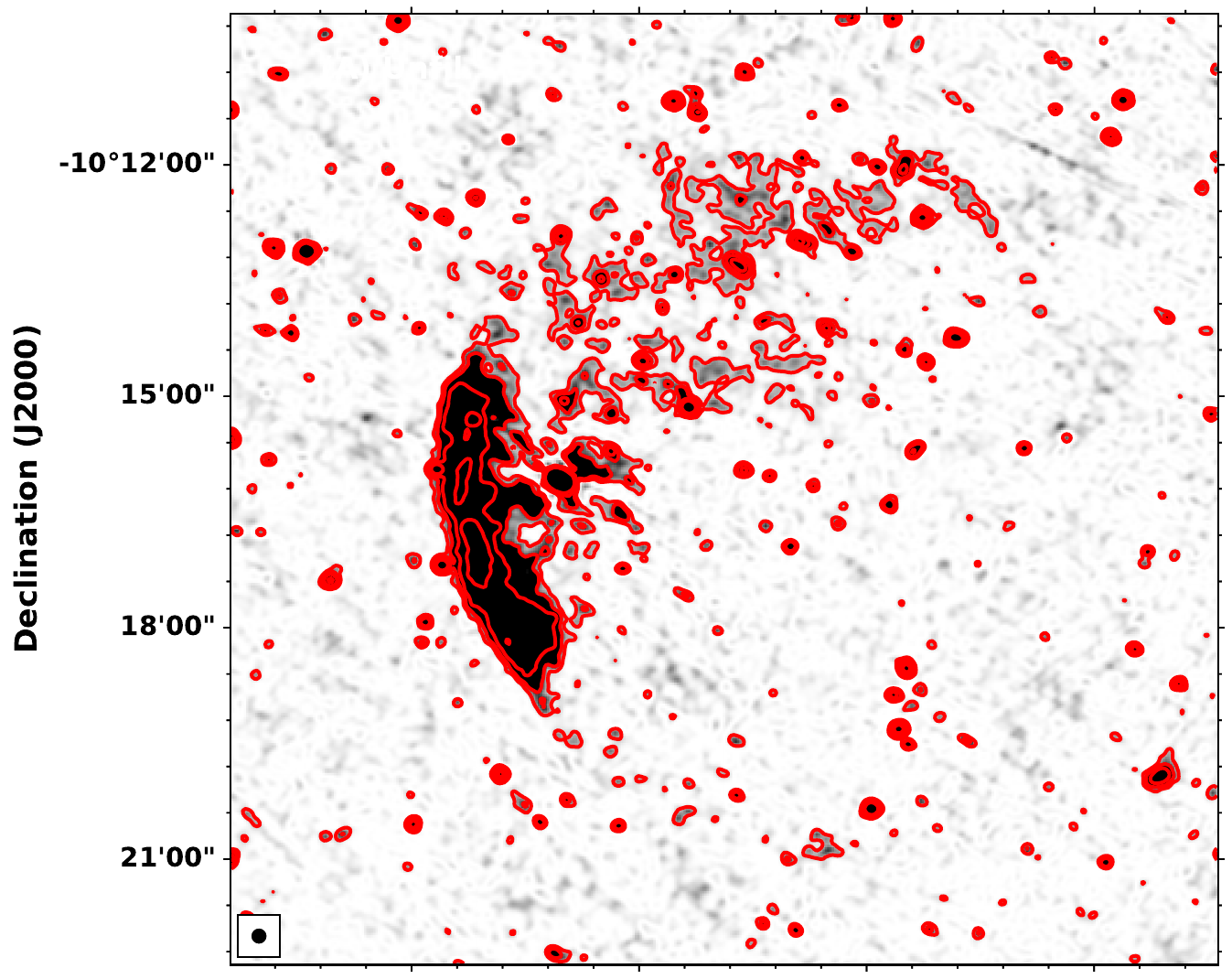}
    \includegraphics[width=5.9cm, height=5.9cm]{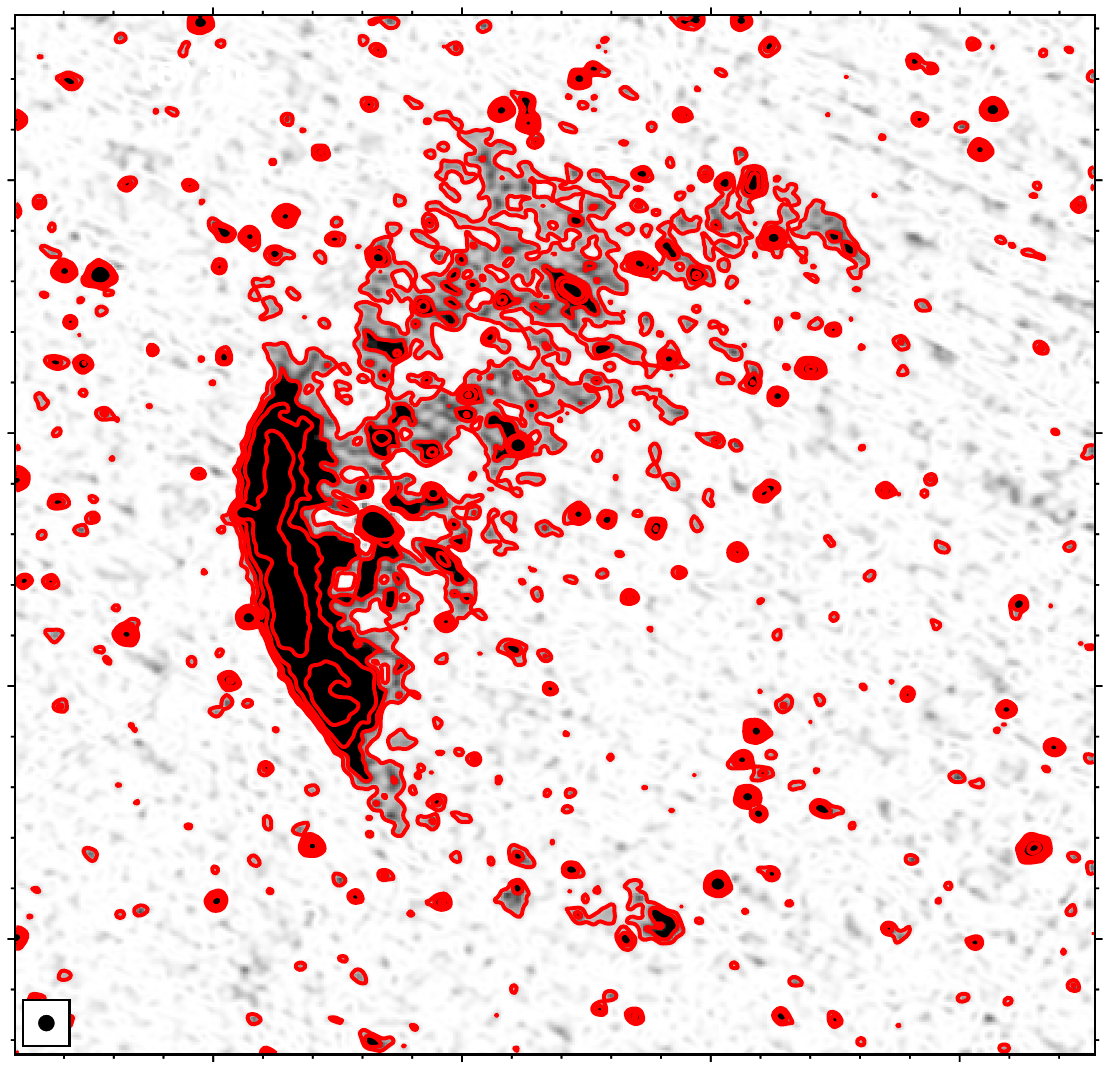}
    \includegraphics[width=5.99cm, height=5.98cm]{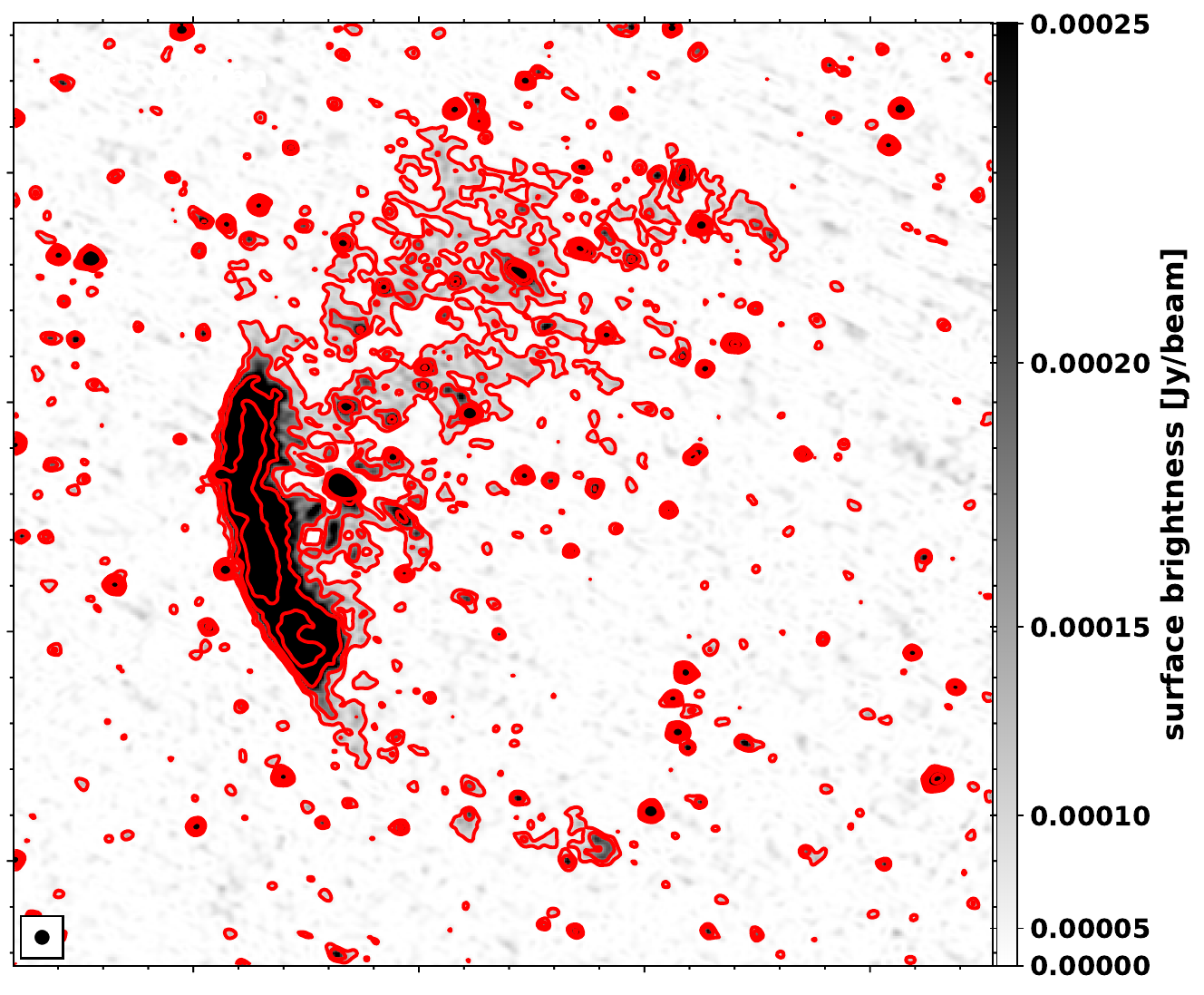}
     \includegraphics[width=5.78cm, height=5.9cm]{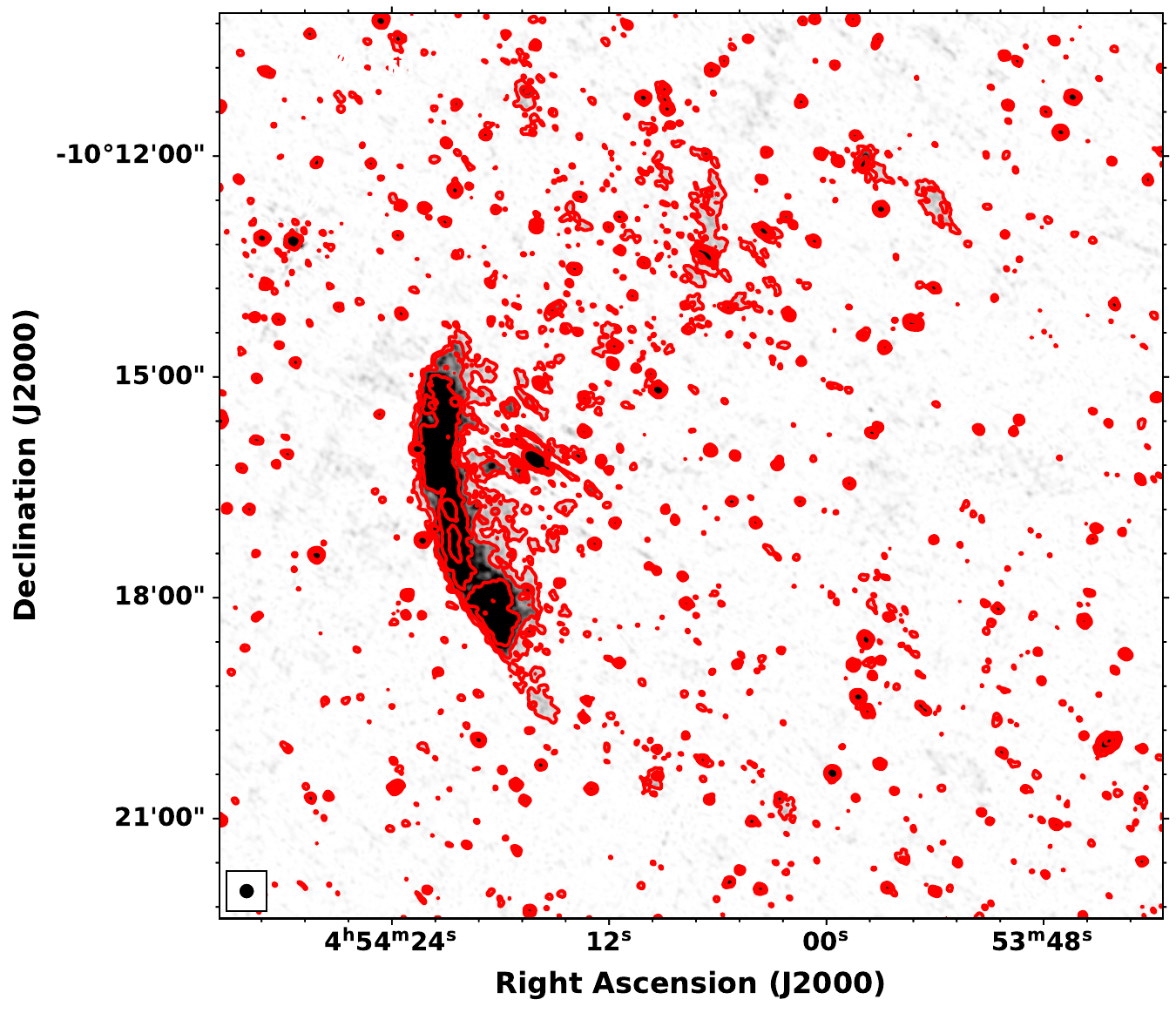}
    \includegraphics[width=5.9cm, height=5.9cm]{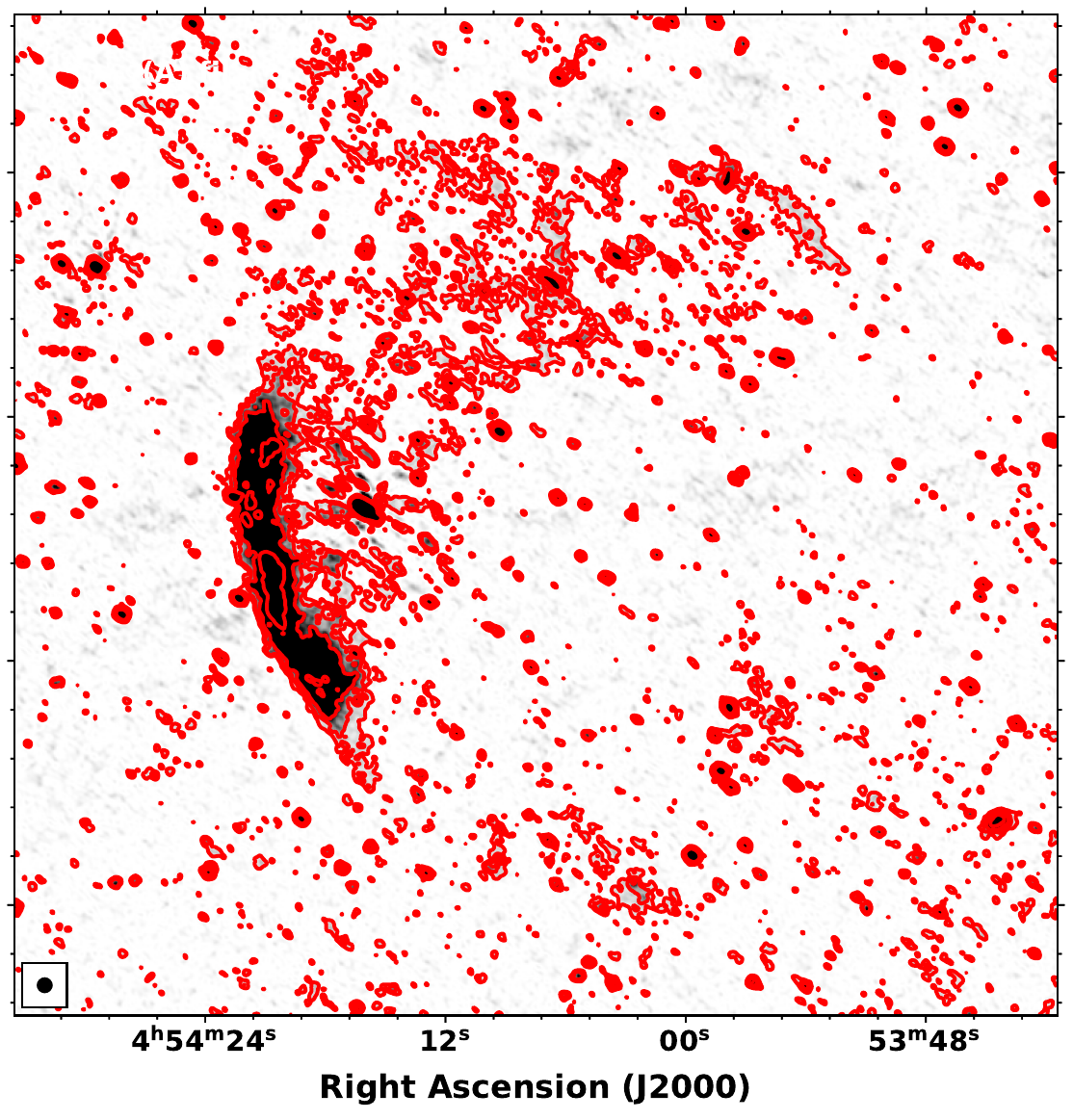}
     \includegraphics[width=6.1cm, height=5.9cm]{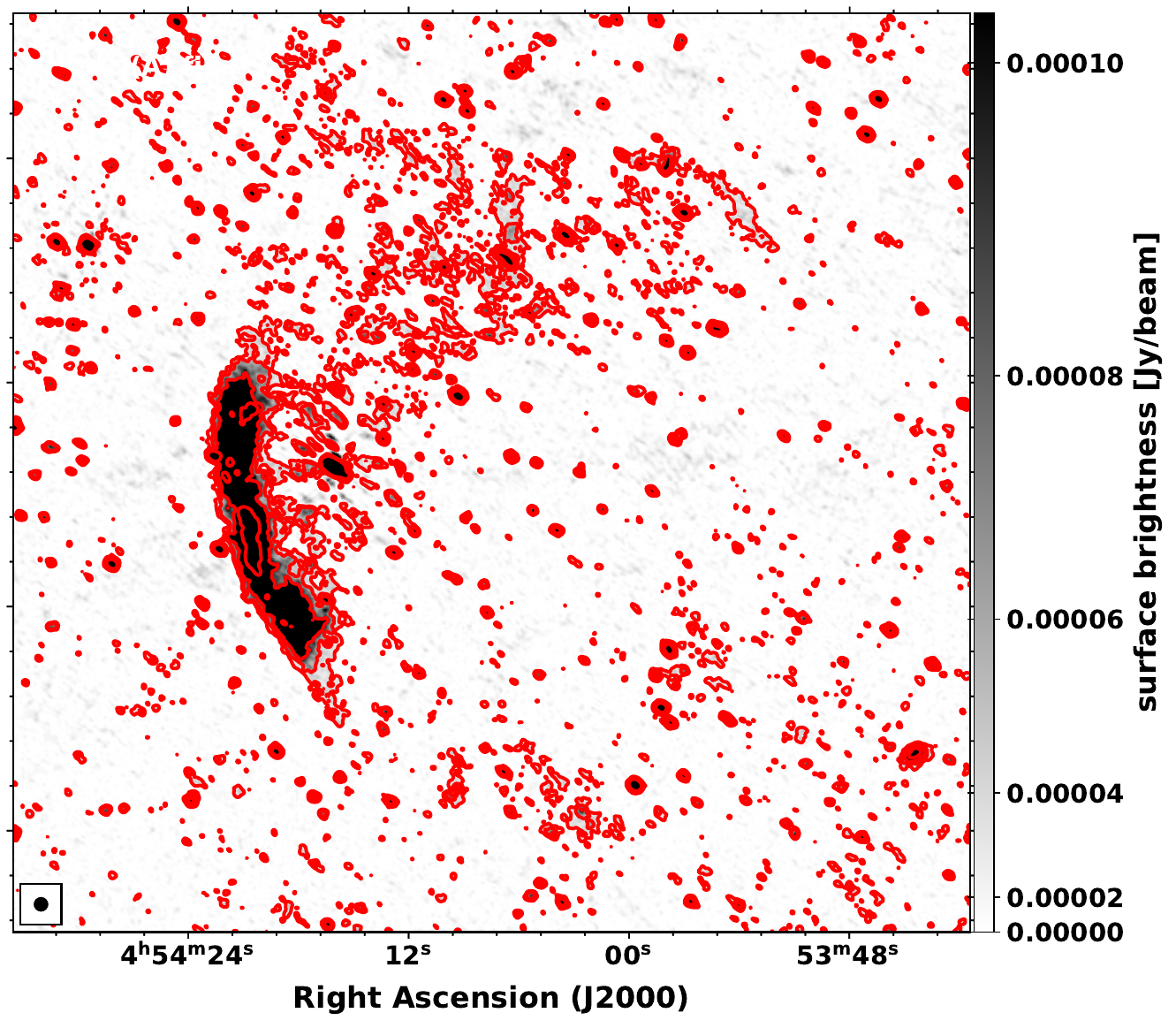}
    \caption{\textit{upper panel:} Here we have shown the full resolution uGMRT band3 images, for the unfiltered, filtered, and combined data sets. The beam size of each image (left: 8.19 " $\times$ 7.78$''$, middle: 8.54$''$ $\times$ 7.51$''$, right: 8.51$''$ $\times$ 7.5$''$) is given in the left bottom corner. The contours start from 3$\sigma_{\rm rms}$ $\times$ [1,2,3...], where $\sigma_{\rm rms}$ = 28, 24.6, 25 $\mu$Jybeam$^{-1}$ for the unfiltered, filtered and combined image respectively. \textit{Lower panel:} Here we have shown the same but at 650 MHz. The beam size of each image (left: 5.4 " $\times$ 4.7$''$, middle: 6.0$''$ $\times$ 4.6$''$, right: 5.8$''$ $\times$ 4.7$''$) is given in the left bottom corner. The contours start at a similar significance as the upper panel, where $\sigma_{\rm rms}$ = 10, 8, 8.5 $\mu$ Jybeam$^{-1}$. The RFI-filtered image shows a better recovery of extended emission. }
    \label{img:20a}
    \label{img:20b}
    \label{img:20c}
    \label{img:20d}
    \label{img:20e}
    \label{img:20f}
\end{figure*}

\begin{figure*}[t!]
        \includegraphics[width=.33\textwidth]{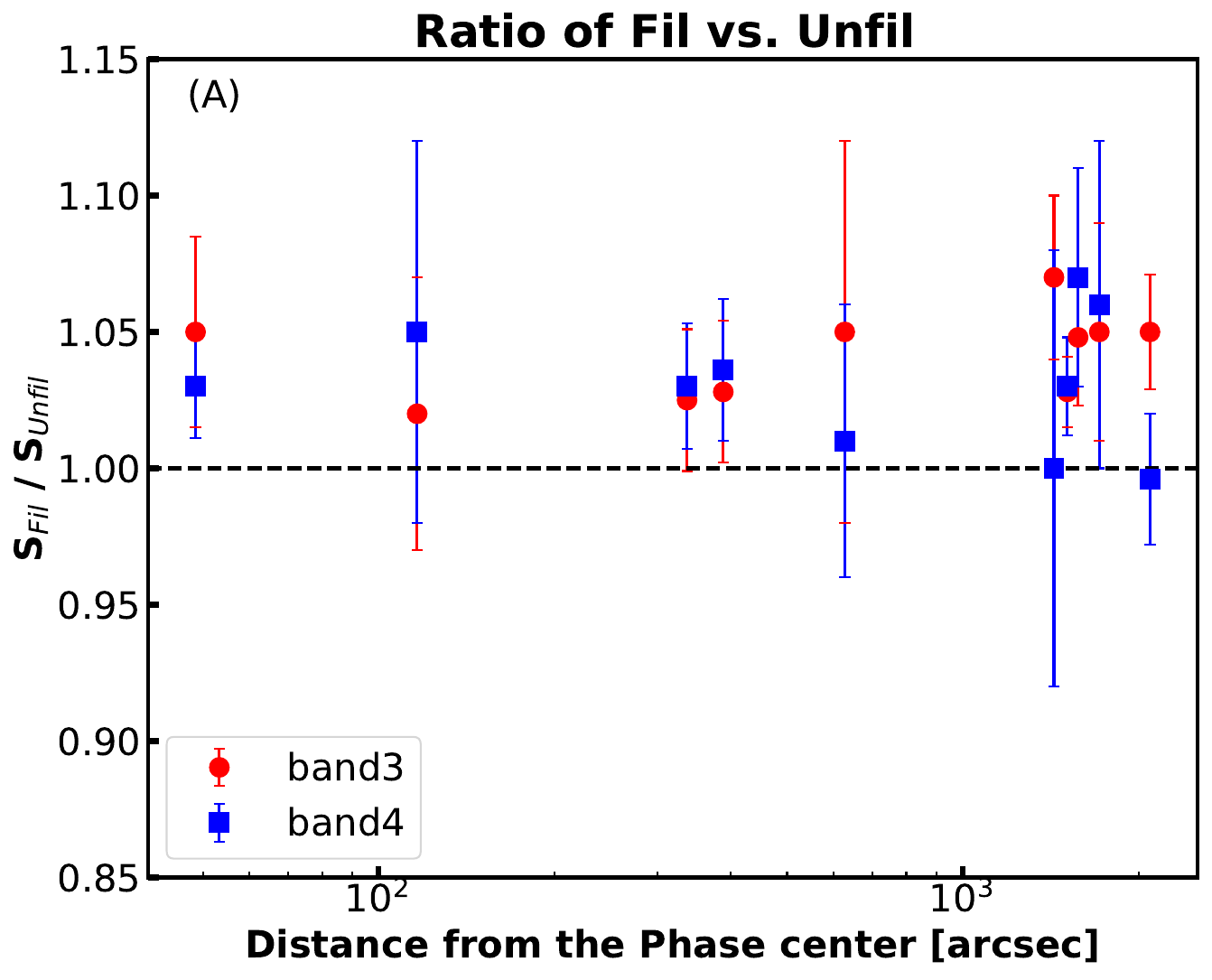}
        \includegraphics[width=.33\textwidth]{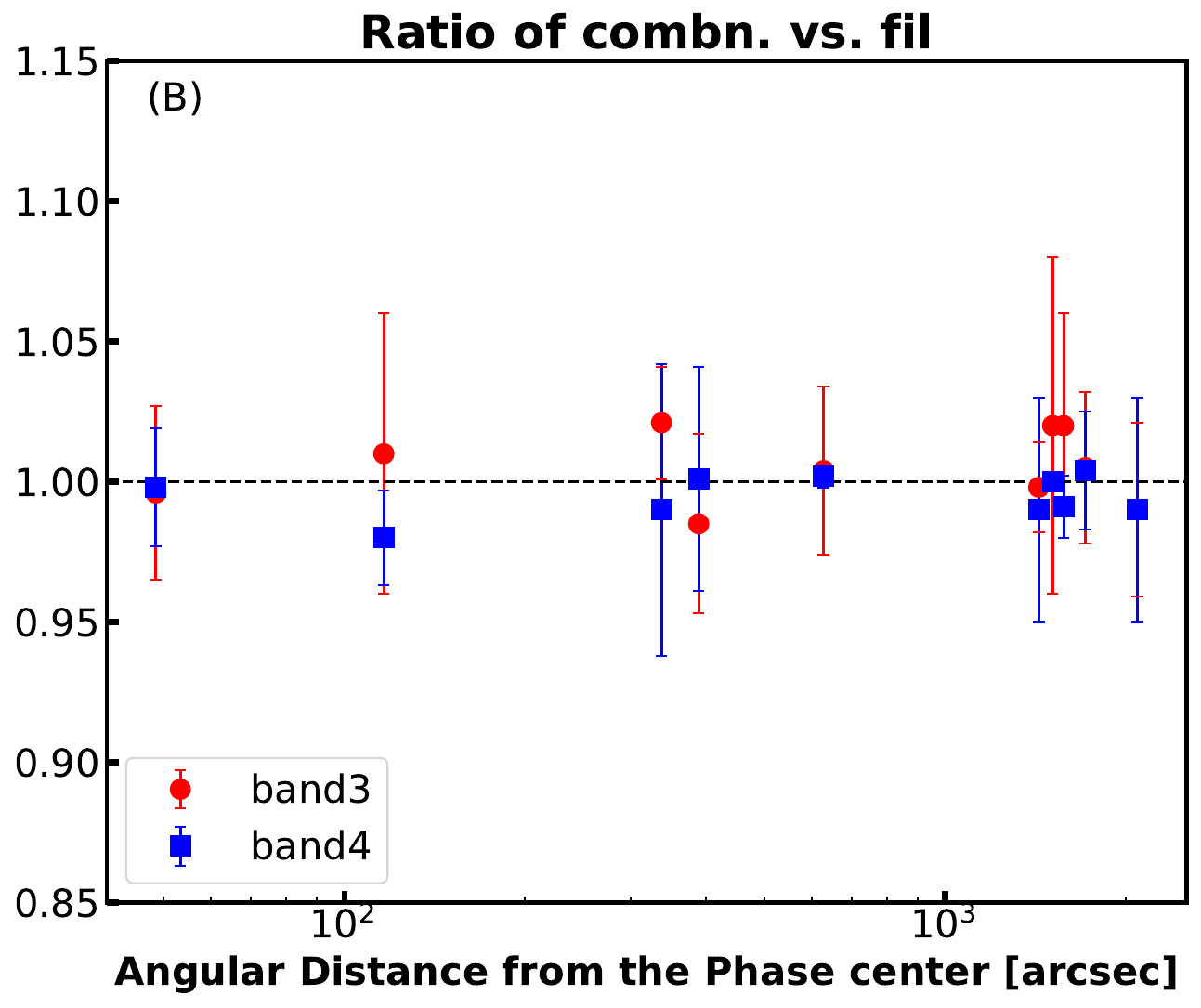}
        \includegraphics[width=.33\textwidth]{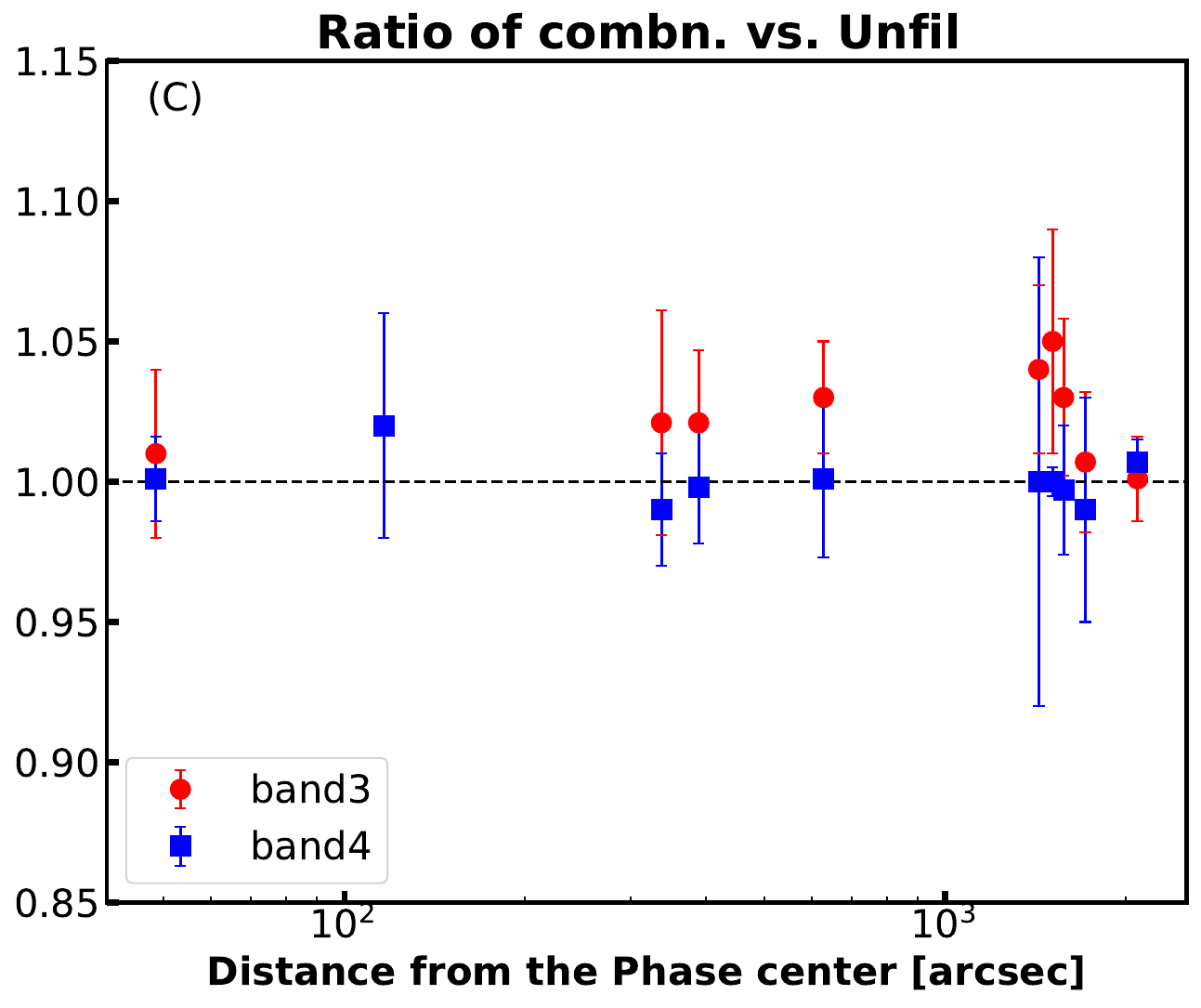}
        \caption{\textit{Left:} Ratio of the flux densities between the filtered and unfiltered data sets for the different sources are shown as a distance from the phase center. The dotted line at 1, indicates the similarity of flux densities for two cases. The error bars are purely statistical and contain the 10\% flux density error and measurement error.       
        \textit{Middle:} The similar image is shown here, but for the ratio of flux density between the combined and filtered data set. The values are not very much scattered around the 1 at both frequencies.  
        \textit{Right:} Here we have shown the same plot for the ratio between the combined and unfiltered one. Regions distant from the phase center show the variations}
        \label{img:21a}
        \label{img:21b}
        \label{img:21c}
\end{figure*}

\begin{table}
\centering
\caption{ List of the discrete unresolved sources in the radio halo and relic region. The positions are reported in the units of degree and flux densities are in units of mJy. The first block reports the sources in the radio halo, the second block is in relic R2, and the third block is for the sources in relic R1. Most of the sources are compact and unresolved.}
 \begin{tabular}{@{}ccccccc@{}}
 \hline
 Name &RA & Dec & F$_{\nu, 400 \rm MHz}$ & F$_{\nu, 650 \rm MHz}$ \\
\hline\hline 
J0454-1012a & 73.585  & -10.207 &0.91$\pm$0.03&0.54$\pm$0.02 \\

J0454-1012b & 73.598 & -10.211 & 0.48$\pm$0.02&0.22$\pm$0.02  \\
 
 J0454-1012c & 73.597 & -10.210 &0.38$\pm$0.02&0.12$\pm$0.01   \\
 
 J0454-1013a &73.598 & -10.217 &0.57$\pm$0.04&0.30$\pm$0.02   \\
 
  J0454-1014a & 73.597 & -10.235 & 0.49$\pm$0.03&0.27$\pm$0.02    \\
 
  J0454-1012d& 73.566 & -10.215 &0.4$\pm$0.02 &0.28$\pm$0.01   \\
 
  J0454-1013b& 73.557 & -10.224 &0.27$\pm$0.01 &0.20$\pm$0.02  \\
 
  J0454-1014b& 73.562 & -10.234 &1.36$\pm$0.08 &1.143$\pm$0.01  \\

  J0454-1015a& 73.557 & -10.261 &0.47$\pm$0.02&0.32$\pm$0.02\\
 
  J0454-1015b & 73.556  &-10.253 &0.38$\pm$0.08&0.2$\pm$0.02\\
  J0454-1015c& 73.566 & -10.251 &0.59$\pm$0.04&0.25$\pm$0.02\\
 
  J0454-1015d& 73.563 & -10.234 &0.5$\pm$0.06&0.32$\pm$0.01 \\
  J0454-1013c& 73.557 & -10.224 &0.45$\pm$0.03&0.16$\pm$0.01\\

 J0454-1013d & 73.541& -10.223 &0.8$\pm$0.04&0.15$\pm$0.01\\
  
  J0454-1013e& 73.527& -10.221 & 2.58$\pm$0.3&1.13$\pm$0.18\\
  
 J0454-1014c& 73.508 & -10.235 &0.45$\pm$0.01&0.26$\pm$0.01\\
  
  J0454-1016a& 73.510 & -10.269 &0.34$\pm$0.02&0.22$\pm$ 0.02\\
  
  J0454-1016b& 73.520 &  -10.267&0.33$\pm$0.02&0.14$\pm$0.01\\
  
  J0454-1015e& 73.526 & -10.266 &0.5$\pm$ 0.01&0.3$\pm$0.01\\
  
  J0454-1014d& 73.548 & -10.242 &0.47$\pm$ 0.02&0.38$\pm$ 0.01\\
  
  J0454-1015f& 73.538 & -10.252 &2.24$\pm$0.07&1.94$\pm$0.02\\
  
  J0454-1011a&73.541 & -10.186 &1.22$\pm$ 0.04&0.68$\pm$0.02\\
  
  J0454-1011b& 73.536 & -10.188 &0.78$\pm$0.05 & 0.5$\pm$0.01\\
  
  J0454-1011c& 73.504 & -10.187 &0.54$\pm$0.03& 0.21$\pm$0.02\\

  J0454-1012e& 73.514  & -10.216 &1.02$\pm$0.08 &0.94$\pm$0.05\\
  
 J0454-1013f& 73.502 & -10.218 &0.45$\pm$0.05& 0.21$\pm$ 0.02\\
  
  J0454-1016c& 73.520 &-10.267 &0.26$\pm$ 0.02 &0.15$\pm$0.01\\
  
  J0453-1014e & 73.491 & -10.240 &0.48$\pm$ 0.02 & 0.17$\pm$0.01\\
  
  J0453-1014f & 73.486 & -10.242 &0.72$\pm$ 0.02&0.33$\pm$0.01\\
  
  J0453-1015g & 73.489 & -10.262 &0.46$\pm$ 0.08& 0.25$\pm$0.02\\
  
  J0453-1016d &73.494&-10.273 &0.41$\pm$ 0.03&0.2$\pm$0.01\\
  
 J0454-1016e & 73.510& -10.269 &0.41$\pm$ 0.03&0.12$\pm$0.02 \\
  
 J0454-1016f &73.516 &-10.282  &0.41$\pm$ 0.02&0.2$\pm$0.01  \\
 J0454-1016g &73.494 &-10.274  &0.37$\pm$ 0.01&0.2$\pm$0.01 \\
\hline
J0453-1012& 73.487& -10.211 & 1.3 $\pm$ 0.04 &0.87$\pm$0.02 \\
J0453-1014& 73.491 & -10.201 & 1.1 $\pm$ 0.06 & 0.55 $\pm$ 0.06\\
\hline

J0454-1015a& 73.594& -10.265 &1.19$\pm$ 0.05& 0.92$\pm$0.01\\
J0454-1015b& 73.585& -10.255& 0.51$\pm$ 0.03&0.31$\pm$0.02\\
J0454-1017a& 73.593& -10.286&1.54$\pm$ 0.09&1.08$\pm$0.03\\
J0454-1016a & 73.587& -10.279&0.9$\pm$ 0.09&0.53$\pm$0.02\\
J0454-1018& 73.572 & -10.301 & 0.53 $\pm$ 0.01 &0.13$\pm$0.01\\
J0454-1017b& 73.590 & -10.284& 0.43 $\pm$ 0.05&0.15$\pm$0.01\\ 
J0454-1017c& 73.585& -10.286 & 0.40 $\pm$ 0.03  &0.13$\pm$0.01\\
 
 \hline

 \label{table5}
\end{tabular}
\end{table}

\begin{table*}
\centering
\caption{Comparison of the flux densities for the point sources}
\begin{tabular}{@{}cccccccc@{}}
\hline
\multicolumn{0}{c}{RA} & \multicolumn{0}{c}{DEC} & \multicolumn{3}{c|}{F$_{\nu, 400 \rm MHz}$(mJy)}&\multicolumn{3}{c|}{F$_{\nu, 650 \rm MHz}$(mJy)} \\
\cline{3-8}
 & &unfil. & Fil. &Combined  & Unfil. & Fil. &Combined \\
\hline
 04:54:09&-10:15:08&1.99$\pm$ 0.05 &2.1$\pm$0.04 &2.08$\pm$0.05 &1.52$\pm$0.01 &1.54$\pm$0.03&1.52$\pm$0.02 \\

04:54:03 &-10:12:59 & 1.06$\pm$0.06&1.08$\pm$0.07 &1.09$\pm$0.07 & 0.68$\pm$0.04& 0.71$\pm$0.03&0.7$\pm$0.04 \\

04:53:54&-10:24:07& 5.97$\pm$ 0.08& 6.14$\pm$0.08& 6.40 $\pm$ 0.08 & 3.90$\pm$0.08& 4.05$\pm$0.06&4.01$\pm$ 0.08 \\

04:52:26& -10:38:20& 4.98$\pm$0.13& 5.25$\pm$0.13& 5.12$\pm$ 0.12 & 0.99$\pm$0.04 & 1.05$\pm$0.04&1.06$\pm$ 0.04 \\

04:53:37& -09:48:51& 1.57$\pm$0.02&1.65$\pm$0.02 &1.66$\pm$0.02 & 0.54$\pm$ 0.007& 0.54$\pm$0.01& 0.54$\pm$ 0.007 \\

04:54:21&-9:50:41& 4.66$\pm$ 0.03& 4.84$\pm$ 0.05&4.82$\pm$ 0.05 &1.50$\pm$ 0.017 & 1.62 $\pm$ 0.02 &1.61$\pm$ 0.017 \\

04:54:28 & -10:17:23 &0.85$\pm$ 0.04 & 1.02$\pm$ 0.04 & 1.04$\pm$ 0.04 &0.87$\pm$0.03 & 0.88$\pm$ 0.02& 0.87$\pm$ 0.03 \\

04:55:38 & -10:00:01 &0.83$\pm$ 0.02 & 0.89$\pm$ 0.02& 0.91 $\pm$ 0.03 & 0.23 $\pm$0.01 & 0.23 $\pm$ 0.01& 0.23 $\pm$ 0.01 \\

04:53:46 & -10:11:09 & 1.7 $\pm$ 0.03 & 1.7 $\pm$ 0.03&  1.72 $\pm$ 0.03 & 0.92 $\pm$ 0.02& 0.95$\pm$ 0.01& 0.91 $\pm$ 0.01 \\

04:52:38 & -9:56:12 & 3.39 $\pm$ 0.03& 3.55 $\pm$ 0.08& 3.51 $\pm$ 0.08  & 1.3 $\pm$ 0.04& 1.4 $\pm$ 0.03 & 1.39 $\pm$ 0.04\\

\hline
\label{table9}
\end{tabular}
\end{table*}

\appendix

\section{Effect of the online RFI filtering}\label{app:A}
Broadband (BB) RFI (narrow high spikes in the time domain) occurs due to the power-line discharge and these are very strong at low frequencies. uGMRT has developed and implemented a real-time RFI filter \citep[see for further details][]{2022JAI....1150008B, 2023JApA...44...37B} to mitigate the BB RFI in the pre-correlation domain (for each antenna and each polarization).  
We have presented a comparison between the RFI-filtered (hereafter, filtered) and unfiltered data sets at bands 3 and 4. On average, $2-3\%$ of the raw voltage samples were flagged and replaced by the filter at both bands. Both the filtered and unfiltered data sets had similar calibration strategies. The flagging in the filtered data was lower (at the central antennas) than in the unfiltered data. The parameters for the imaging steps were similar for the two data sets. We have shown the unfiltered (left), filtered (middle), and combined (right) images (grey scales) for both band 3 and band 4, overlaid with the contours of each image in Figure~\ref{img:20a}. Compared to unfiltered data, the centrally located radio halo is well recovered in the filtered data.

\begin{figure}
	\includegraphics[width=\columnwidth]{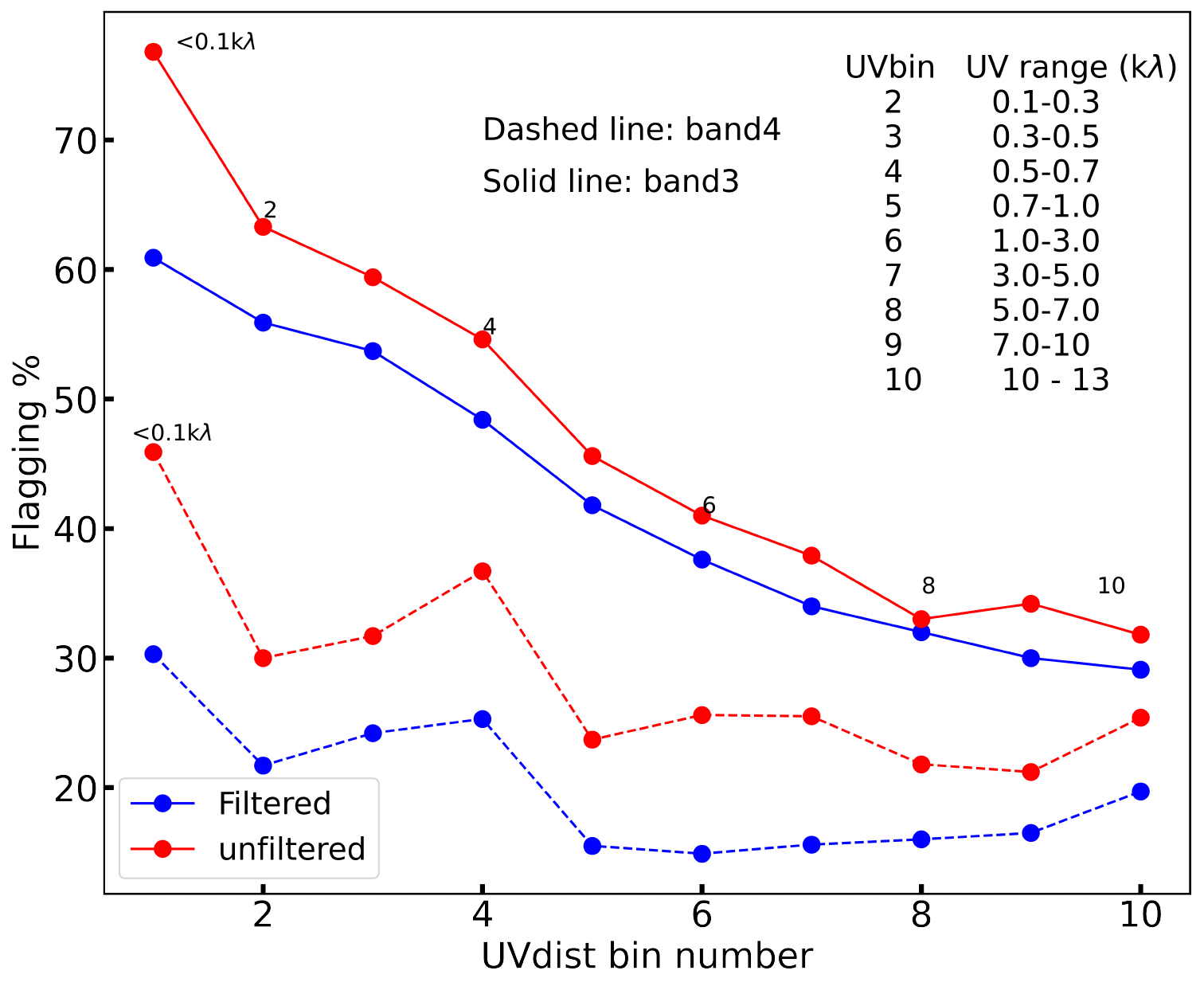}
    \caption{The flagging percentage as a function of the baseline is shown. The central baselines are less flagged in the filtered data, in both frequencies, leading to a better recovery of extended emissions.}
    \label{flagging}
\end{figure}

The ratio of flux densities for the various point sources as a function of their distances from the cluster center is shown in Figure~\ref{img:21a}. The ratio of flux densities of point sources between the filtered and unfiltered data sets is shown in the left image, where the variation in the flux densities is not high. The variation is seen clearly in the regions away from the phase center. A similar pattern can be seen in the other cases. In the filtered and unfiltered data sets, the flux density of the point sources has not changed drastically however, the extended emission is better recovered in the filtered data set. Therefore, the less flagging at central square antennas (Figure~\ref{flagging}) due to the application of the RFI filter is significant.

\section{Point source Image of the A521}\label{app:b}

\begin{figure*}[t!]
        \includegraphics[width=8.9cm,height =8.9cm]{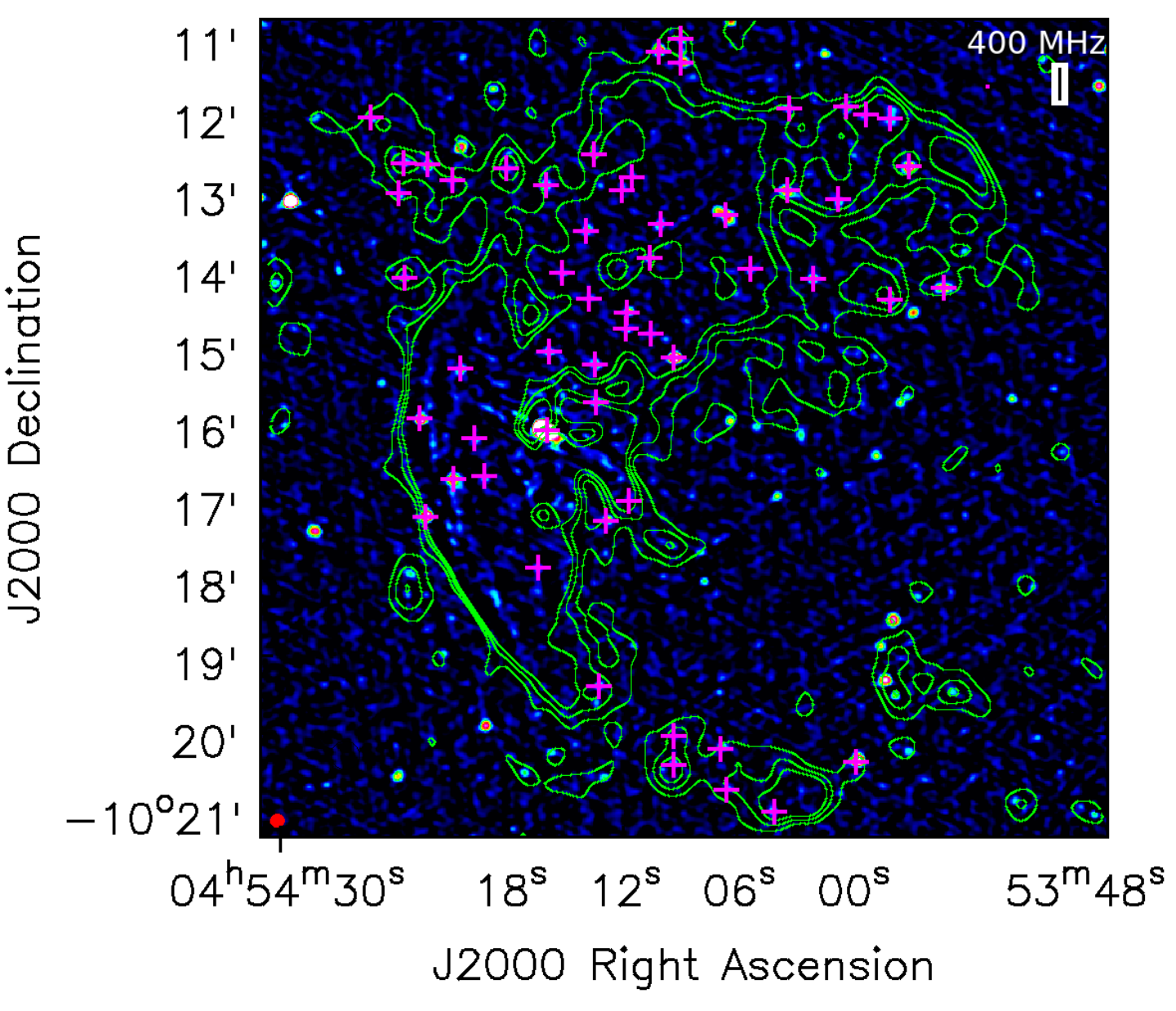}
        \includegraphics[width=8.9cm,height =8.9cm]{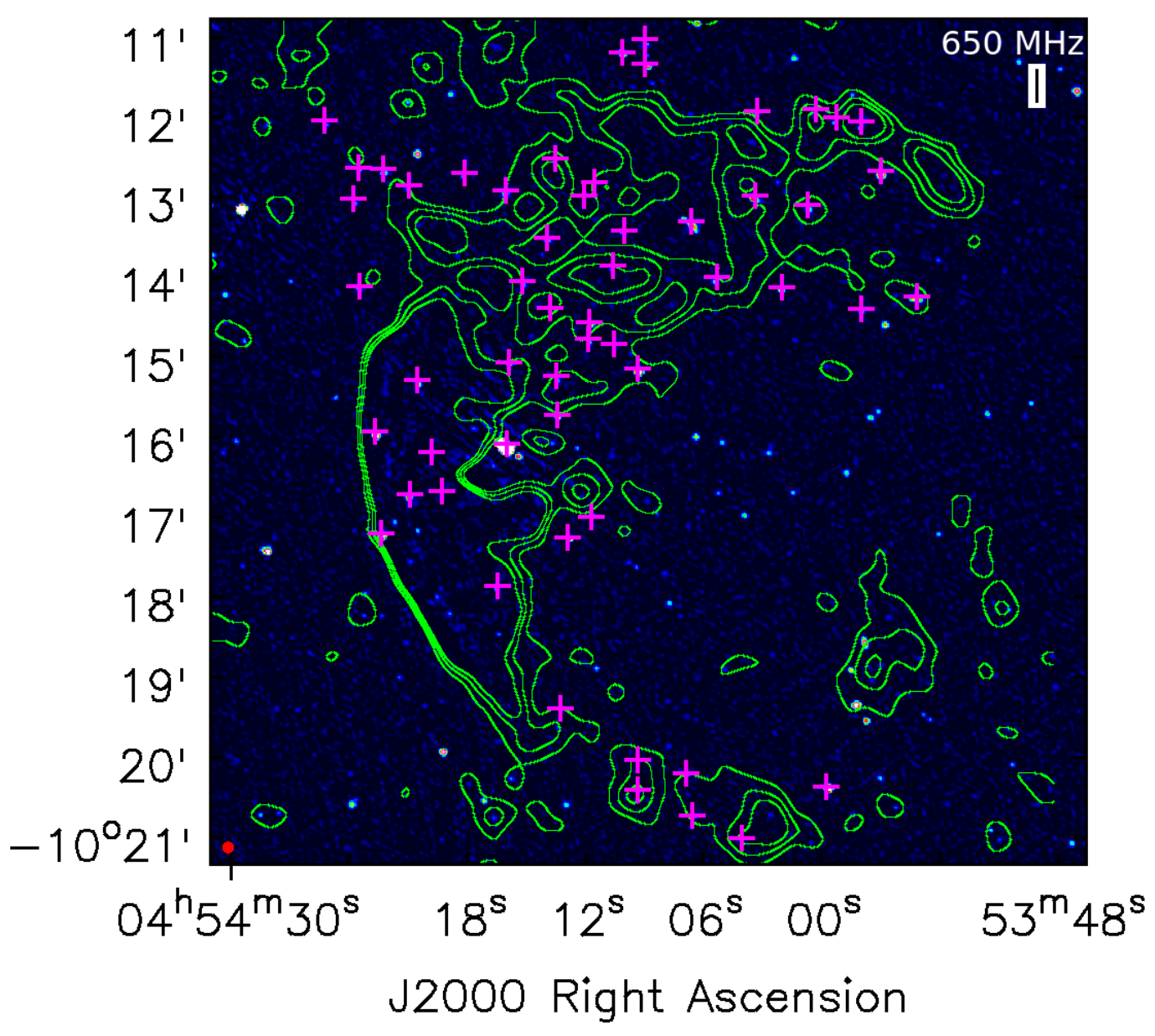}
        \caption{\textit{Left:} We have shown the point source image (color) at 400 MHz and the corresponding point source subtraction low-resolution image (green contours). The magenta plus signs show the unresolved sources that have been removed in the central region. The contours are drawn at 3$\sigma_{\rm rms}$ $\times$ [1,2,3], with $\sigma_{\rm rms} = 45 \mu$ Jybeam$^{-1}$. \textit{Right:} Here the similar is shown at 650 MHz. The green contours start with similar values, with a $\sigma_{\rm rms} = 30 \mu$ Jybeam$^{-1}$. The point sources are listed in the Table~\ref{table5}.}
        \label{img:22a}
       \label{img:22b}        
\end{figure*}

We have shown the point source image of the A521 cluster field in Figure~\ref{img:22a}.  Table~\ref{table5} we have presented the flux densities (both band 3 and 4) of the point sources in the central region (specifically in the central radio halo and peripheral relic region). In table.~\ref{table9}, the reported values of the flux densities are used in Fig~\ref{img:21a}. \cite{2006NewA...11..437G} had listed the radio sources in the 30$'$ $\times$ 30$'$ region. We have cross-checked their table with ours. At the same frequency, our image is 5 times more sensitive than theirs.

\bibliography{sample631}{}
\bibliographystyle{aasjournal}



\end{document}